\documentclass[usenatbib,a4paper,times,fleqn]{mnras}
\usepackage{graphicx,natbib,times,amssymb,amsmath,pstricks}
\usepackage{aas_macros}
\usepackage{bm,url}
\usepackage{textcomp}
\usepackage{breakurl}
\usepackage{subfig}

\newcommand {\grtsim} {\ {\raise-.5ex\hbox{$\buildrel>\over\sim$}}\ }
\newcommand {\aplt} {\ {\raise-.5ex\hbox{$\buildrel<\over\sim$}}\ }

\title[Gap opening in dusty discs]{An opening criterion for dust gaps in protoplanetary discs}
\author[Dipierro \& Laibe]{Giovanni Dipierro$^{1,2}$\thanks{giovanni.dipierro@leicester.ac.uk} and Guillaume Laibe$^{3,4}$ \\%, Daniel J. Price$^{3}$ and Giuseppe Lodato$^{1}$ \\
$^{1}$Dipartimento di Fisica, Universit\`a degli Studi di Milano, Via Celoria, 16, Milano, I-20133, Italy \\
$^{2}$Department of Physics and Astronomy, University of Leicester, Leicester, LE1 7RH, UK \\
$^{3}$Univ Lyon, Univ Lyon1, Ens de Lyon, CNRS, Centre de Recherche Astrophysique de Lyon UMR5574, F-69230, Saint-Genis-Laval, France\\
$^{4}$School of Physics and Astronomy, University of St. Andrews, North Haugh, St. Andrews, Fife KY16 9SS, UK 
}
\pagerange{\pageref{firstpage}--\pageref{lastpage}} \pubyear{2016}
\defcitealias{nakagawa86a}{NSH86}
\defcitealias{dipierro16a}{DLPL16}
\date{}
\begin{document}
\label{firstpage}
\bibliographystyle{mnras}
\maketitle

\begin{abstract}
We aim to understand under which conditions a low mass planet can open a gap in viscous dusty protoplanetary discs. For this purpose, we extend the theory of dust radial drift to include the contribution from the tides of an embedded planet and from the gas viscous forces. From this formalism, we derive i) a grain size-dependent criterion for dust gap opening in discs, ii) an estimate of the location of the outer edge of the dust gap and iii) an estimate of the minimum Stokes number above which low-mass planets are able to carve gaps which appear only in the dust disc. These analytical estimates are particularly helpful to appraise the minimum mass of an hypothetical planet carving gaps in discs observed at long wavelengths and high resolution. We validate the theory against 3D SPH simulations of planet-disc interaction in a broad range of dusty protoplanetary discs. We find a remarkable agreement between the theoretical model and the numerical experiments.
\end{abstract}

%----------------------------------------------------------------------------------------------------------------
\section{Introduction}
Dust rings and gaps-like structures have been recently revealed by high-resolution observations in both young and evolved protoplanetary discs \citep{alma-partnership15a,canovas16a,van-der-plas16a,de-boer16a,ginski16a,van-boekel16a,andrews16a,isella16a,fedele17a}. Various mechanisms have been proposed to explain the origin of these structures. A first category of models invokes discs that are dynamically young and in which planets have not yet formed. In those, rings may originate from self-induced dust pile-ups \citep{gonzalez15a,gonzalez17a}, zonal flows \citep{flock15a,bethune16a}, rapid pebble growth around condensation fronts \citep{zhang15a}, aggregate sintering \citep{okuzumi16a}, large scale instabilities due to dust settling \citep{loren-aguilar16a} or secular gravitational instabilties \citep{takahashi16a}. 

The alternative and more natural explanation is to interpret the rings as an observational signature of embedded planets. The tides generated by planets of sufficient masses overcome the gap-closing contributions induced by the pressure gradient and the viscous spreading of the gas. As a result, the planet carves a gap in its vicinity by pushing material away from its orbit \citep[e.g.][]{goldreich79a,goldreich80a,lin86a,lin93a,rafikov02a,kley12a,baruteau14a}. In the prototypal case of the disc around HL Tau, this explanation is supported by some observational features, such as the increase of the gap eccentricity at large orbital radii, as well as spectral index variations between dark and bright rings which suggests that dark rings are regions of low dust density \citep{alma-partnership15a}.  
So far, the structures observed in this disc \citep{alma-partnership15a} and HD163296 \citep{isella16a} have been better reproduced by assuming the presence of planets \citep[e.g.][]{dipierro15a,dong15a,jin16a,isella16a}. This scenario would be consistent with the increasing number of extrasolar planets detected \citep{laughlin15a}, but requires Saturn mass planets within a few million years at most, challenging the scenario of planet formation through core accretion.

A criterion on the minimum mass required for a planet to open a gap in a gas disc was derived by \citet{crida06a}. They considered the balance between the gap-opening tidal torque and the gap-closing viscous torque, taking into account the non-local deposition of angular momentum by the density waves excited by planets \citep[e.g.][]{goodman01a,rafikov02a}. Recent investigations have refined this analysis, showing that less massive planets may open gaps as well \citep{dong11a,duffell12a,duffell13a,duffell15b,zhu13a}, but more massive planets may not \citep{malik15a}.

However, observations of protoplanetary discs probe mostly the dust content of the disc, not the gas, and the two phases are not necessarily coupled. Recently, numerical investigations have shown that gap opening is more effective in the dust than in the gas \citep{paardekooper04a,paardekooper06a,fouchet07a,fouchet10a,ayliffe12a,gonzalez12a,zhu14a,picogna15a,dipierro15a,dipierro16a,rosotti16a}. 

Dust is usually modelled as an almost collisionless fluid in protoplanetary discs. As such, the dust intrinsic pressure and viscosity are not effective at closing the gap. Moreover, the tidal torque is amplified by geometrical effects due to dust settling. In the vicinity of the planet, the dust dynamics depends strongly on the size of the grains. When the planet is massive enough to carve a gap in the gas, micron-sized grains couple to the gas viscous flow and enter the gap \citep{rice06a}, whereas drag produces a pile up of large grains into the pressure maxima at the gap edges \citep[e.g.][]{pinilla15a}. This radial size-sorting produces well defined features in the mm and scattered light emission, consistent with recent observations \citep[e.g.][]{follette13a,canovas16a}. %Moreover, for dust grains with $\mathrm{St}\aplt 1$, a shallow gap can be carved in the dust for planets able to slightly affect the gas without creating pressure maxima \citep{rosotti16a}. 

Recently, the disc around TW Hydrae has been observed by ALMA and SPHERE, probing the dust continumm emission at $850\,\mu m$ and the scattered light in the H-band at $1.6\,\mu m$ respectively \citep{andrews16a,van-boekel16a}.  A comparison of SPHERE and ALMA images reveals that the gaps at $\sim 37$ and $\sim 43$ au observed with ALMA are absent in the SPHERE image (see Fig.~7 of \citealt{van-boekel16a}). This mismatch between the distribution of large dust grains probed by ALMA and of small dust grains (expected to be well mixed with the gas) probed by SPHERE indicates that large dust grains may be more susceptible to gap formation than gas.

Surprisingly, the minimum planet mass required to open a gap in a dusty disc has not been clearly identified theoretically yet. Recently, \citet{rosotti16a} have found that a shallow gap can be carved in the dust for planets able to slightly affect the local gas structure without creating pressure maxima. Moreover, \citet{dipierro16a} have shown numerically that even lower mass planets could open gaps in the dust only, without any perturbation in the radial pressure gradient at the planet location. 
In this case, the creation of gap results from the competition between the tidal torque and the drag torque outside the planet orbit since the planet is not able to affect the gas structure.  The drag torque acting on dust is negative all through the disc, whereas the tidal torque exerted by the planet is positive outside its orbit and negative inside. The balance between these two torques outside the planet orbit determines if the planet is able to carve a gap in the dust or not. 
In this paper,  we propose a theory to model this mechanism, with the aim of deriving a grain size-dependent criterion for dust gap opening by non-migrating planets in protoplanetary discs. To this purpose, we extend the formalism of dust drift introduced in \citet{nakagawa86a} to include viscous forces and the disc-planet tidal interactions. 
We also infer the radial location of the outer edge of the dust gap and the minimum Stokes number above which low mass planet are able to carve gap only in the dust.
The results of our analysis are thoroughly tested against 3D Smoothed Particle Hydrodynamics (SPH) gas and dust simulations of different disc models. 

The paper is organised as follows: in Sect.~\ref{sec:gapopenig} we describe the dust dynamics in disc hosting a non-migrating planet under the action of tidal and drag forces. In Sect.~\ref{sec:criterion} we apply the formalism developed in Sect.~\ref{sec:gapopenig} to derive a simple gap opening criterion for the dust in the regime where the planet does not significantly alter the gas disc. In Sect.~\ref{sec:testcriterion}, we perform a set of simulation to test our model. In Sect.~\ref{sect:discussion} we discuss how our criterion can be used in practice to interpret observations and, finally, in Sect.~\ref{sec:conclusion} we summarise our findings and conclusions.

%---------------------------------------------------------------------------------------------------------------------
\section{Dust dynamics in a viscous disc with a planet}
\label{sec:gapopenig}

\subsection{Disc-planet tidal interaction}
\label{sec:discplanetinteraction}
\subsubsection{Gas disc}
\label{sec:tidalgas}

A planet transfers angular momentum to its surrounding disc through the excitation of spiral density waves at specific locations called Lindblad resonances. %As a consequence of the waves dissipation, the disc
%driving material into mean-motion resonances  
\citet{goldreich79a,goldreich80a} derived an analytic expression for the tidal torque per unit mass $\Lambda$ (the \emph{excitation} torque density) exerted by a fixed planet on an elementary ring of a pressureless disc:
\begin{equation}
\Lambda(r)=\mathrm{sgn}(r-r_{\mathrm p})f \frac{(\mathcal{G} M_{\mathrm{p}})^2}{\Omega_{\mathrm{p}}^2} \frac{1}{\Delta^4} ,
\label{eq:torquegrav}
\end{equation}
where $M_{\mathrm p}$ is the mass of the planet, $\mathcal{G}$ is the gravitational constant, $r_{\mathrm p}$ is the planet location, $\Omega_{\mathrm p}=(\mathcal{G}M_{\star}/r_{\mathrm p}^3)^{1/2}$ is the Keplerian angular velocity at $r_{\mathrm p}$ where $M_{\star}$ is the mass of the central star, $\Delta \equiv \left |r-r_{\mathrm p}\right |$, and $f$ is a constant of order unity (e.g. $f \sim 0.4$ - \citealt{goldreich79a}, $f \sim 0.15$ - \citealt{lin79a}, $f \sim 0.1$ - \citealt{rafikov12a}). While the key features of the tidal torque equation, such as the $\Delta^4$ dependence, have been widely accepted \citep{lin86a,bryden99a,bate03a,varniere04a,dangelo08a}, the proportionality coefficient $f$ is mostly inferred by analyzing the shape of the gap carved by the planet \citep[e.g.][]{armitage02a}. %In particular, \citet{armitage02a} found that the gap extension carved by a planet is consistent with $f\sim10^{-2}$.
The $\mathrm{sgn}$ factor in front of the right-hand side of Eq.~\ref{eq:torquegrav} shows that a planet tends to push material outside of its orbit. In a disc with non-zero pressure, the exchange of torque between the planet and the disc is a two-step mechanism. 

The planet first excites density waves in the gas. The amount of initial torque density stored in these density waves is the one given by Eq.~\ref{eq:torquegrav}. Then, a fraction of this torque (the \emph{deposition} torque) is transferred from the waves to the disc by virtue of  damping processes such as viscosity or shocks (\citealt{takeuchi96a,goodman01a,rafikov02a}). %, whereas the remaining part (the \emph{pressure} torque, \citealt{crida06a}) is carried away by the waves until being deposited in a more distant region of the disc. 
This pressure-supported transport modifies also the effective location of the Lindblad resonance. An important consequence is that the deposition torque is essentially zero in the vicinity of the planet, since high order Lindblad resonances are shifted away from the planet by a typical length of the order of the scale height of the disc $H$ \citep{goldreich80a,artymowicz93a}. The linear theory (valid for low mass planets, e.g. \citealt{dong11a}) shows that there is no Lindblad resonances in the region $\left| r - r_{\mathrm p} \right| \lesssim 2H/3$. This effect, called torque cut-off, prevents the tidal torque to diverge close to the planet orbit. Finally, no torque is deposited by Lindblad resonances in the co-orbital region of the planet, where the gravity of the planet dominates over the one of the star and particles experience horseshoe orbits \citep{bate03a,dangelo08a}. This region  typically extends over a distance $r_{\mathrm{H}}$ from the planet's orbit, where $r_{\mathrm{H}}$ denotes the Hill radius of the planet:
\begin{equation}
r_{\rm H} \equiv r_{\mathrm p}\left( \frac{M_{\rm p}}{3 M_{\star}}\right)^{1/3} .
\label{eq:rhill}
\end{equation}
To include the torque cut-off due to the pressure and the corotation region, the deposition toque is often calculated from the expression given by Eq.~\ref{eq:torquegrav}, but with the following modified $\Delta_{\rm g}$
\begin{equation}
\Delta_{\rm g} = \mathrm{max}(\left |r-r_{\mathrm p} \right |,H,r_{\mathrm{H}}) .
\label{eq:deltag}
\end{equation}
The prescription given by Eq.~\ref{eq:deltag} has been extensively used in a variety of contexts, such as to model the disc-planet interaction \citep[e.g.][]{bryden99a} or the evolution of supermassive black hole binaries embedded in a disc \citep[e.g.][]{lodato09a}. 

In this work, we use Eq.~\ref{eq:torquegrav} to treat the tidal torque density by a prescription with is analytically tractable. This expression assumes a few simplification that should be kept in mind. Firstly, Eq.~\ref{eq:torquegrav} gives the tidal torque under the form of a smooth function. This regularity property originates from the fact that individual modes of the tidal potential interact with large regions of the disc rather than narrow rings centred over Lindblad resonances \citep{rafikov12a}. %This condition is always satisfied close to the planet. 
Secondly, Eq.~\ref{eq:torquegrav} neglects effects due to the non-linear propagation of the density waves excited by the planet, some details of the wave damping processes and the recently discovered negative torque correction. About the latter, in a disc of uniform surface density, the tidal torque density changes sign at a radial separation $\gtrsim 3 H$ from the planet \citep{dong11a,rafikov12a}. However, we do not expect this correction to play a major role since we study processes developing in regions of size $\lesssim H$ around the planet (see Sect.~\ref{sec:limits}), where the tidal torque exerted by the planet on the gas is always positive in the outer disc and negative in the inner disc. Moreover, we focus here on low mass planets embedded in viscous discs, where density waves are expected to be strongly damped close to the Lindblad resonsances. In this case, the exact role of the negative torque phenomenon remains an open question (Sect.~7 of \citealt{rafikov12a}). Hence, we choose to restrain our theory to a minimum but sufficient level of refinement. %We verify in Sect.~\ref{sec:testcriterion} that this approximation is legitimate.

\subsubsection{Dust disc}

From the discussion in Sect.~\ref{sec:tidalgas}, we expect the tidal torque to be more effective in the dust than in the gas for three reasons:
\begin{enumerate}
\item Dust is a pressureless fluid, where density waves cannot propagate far from the planet. The deposition torque equals therefore the excitation torque.
\item For the same reason, there is no torque cut-off in the dust, and the effective $\Delta_{\rm d}$ is
\begin{equation}
\Delta_{\rm d} = \mathrm{max}(\left |r-r_{\mathrm p} \right |, r_{\mathrm{H}}) .
\label{eq:deltad}
\end{equation}
Close to the planet, angular momentum can be deposited in the dust, but not in the gas. In detail, if $r_{\mathrm{H}}<H$, the torque exerted at the Lindblad resonances in the region between $r_{\mathrm{H}}$ and $H$ is effective in the dust, while in the gas is suppressed due to pressure effects.
The ratio between the maximum torque density in the gas and in the dust is of order $\sim (r_{\mathrm H}/H)^4=(r_{\mathrm p}/H)^4 (M_{\mathrm p}/3M_{\star})^{4/3}$. 
\item For large grains, the thickness of the dust layer $H_{\mathrm d}$ is smaller than $H$, as a result from the competition between settling and turbulent stirring \citep{dubrulle95a}. For dust layers with $H_{\mathrm d}\geq r_{\mathrm H}$ this results in an enhancement of the tidal torque due to local geometrical effects. 
\end{enumerate}

In absence of gas, the use of the smooth functional form given by Eq.~\ref{eq:torquegrav} is not appropriate. Indeed, the tidal torque exerted on the pressureless dust phase concentrates at Lindblad resonances, where particles eccentricities are effectively excited \citep{ayliffe12a,zhu14a}. In this case, the planet-disc interaction is better described by Hill's equations \citep{hill78a}, additional effects related to dust pressure induced by mutual collisions and velocity fluctuations must be taken into account \citep{henon86a,petit87a,petit87b,rafikov01a}. However, for a disc containing a low-mass planet, two arguments support the use of Eq.~\ref{eq:torquegrav} for dust as well. Firstly, grains experience gas drag (see Sect.~\ref{sec:dragdynamics}), a force which dominates the dynamics of the particles at the resonances as long as they are not too large (Fig.~18, top-center panel of \citealt{zhu14a}). In this case, orbits of dust grains shrink around the planet and the ability for opening a dust gap is enhanced \citep{ayliffe12a,zhu14a}. Secondly, with a low-mass planet, $r_{\mathrm{H}}<H$, we will show that the width of the dust gap $\Delta_{\mathrm{gap}}$ satisfies $r_{\mathrm{H}}\lesssim\Delta_{\mathrm{gap}}\lesssim H$ (see Sect.~\ref{sec:limits}). In this region, high-order Lindblad resonances of order $m$ are highly concentrated ($r_{\mathrm{p}}/H\aplt m \aplt r_{\mathrm{p}}/r_{\mathrm{H}}$) and degenerate into a continuum. This region is also sufficiently far away from low-order resonances (i.e.1:2, 2:3, 3:2 and 2:1), where eccentricity pumping is effective and can not be neglected \citep{zhu14a,ayliffe12a}. Thus, we use Eq.~\ref{eq:torquegrav} to model the tidal torque density in the dust as well, and test this assumption with numerical simulations.

%\subsection{Dust-gas aerodynamical coupling}
\subsection{Equations of motion}% for low-mass planet}
\label{sec:dragdynamics}

The motion of dust particles in protoplanetary discs is affected by the aerodynamical interaction with the gas and \textit{vice-versa}. The differential motion between the two phases gives rise to a drag force that  damp this velocity difference. In discs hosting planets, an additional velocity difference can be produced 
due to the different tidal interaction of the planet with the gas and dust (Sect. \ref{sec:discplanetinteraction}).
%since the interactions of the planet with the gas differs from the ones with the dust (Sect. \ref{sec:discplanetinteraction}).

We assume a thin, non-magnetic, non-self-graviting, dusty viscous and vertically isothermal protoplanetary disc hosting a non-migrating planet. We treat the dust phase as a continuous pressureless and viscousless fluid. The equations of motion for the gas and the dust are
\begin{eqnarray}
%\frac{\partial \rho_{\rm g}}{\partial t} + \nabla \cdot \left( \rho_{\rm g} \mathbf{v}_{\rm g} \right) & = & 0 , \label{eq:gene1}\\
%\frac{\partial \rho_{\rm d}}{\partial t} + \nabla \cdot \left( \rho_{\rm d} \mathbf{v}_{\rm d} \right) & = & 0 , 
%\label{eq:gene2}\\
\frac{	\partial \mathbf{v_{\mathrm g}}}{\partial t}+(\mathbf{v_{\mathrm g}}\cdot \mathbf{\nabla})\,\mathbf{v_{\mathrm g}} & = & \frac{K}{\rho_{\mathrm g}}(\mathbf{v_{\mathrm d}}-\mathbf{v_{\mathrm g}}) -\mathbf{\nabla} (\Phi + \Phi_{\mathrm p}) \nonumber \\
&& -\frac{1}{\rho_{\mathrm g}}(\mathbf{\nabla} P  - \mathbf{\nabla} \cdot \sigma), 
\label{eq:gene3}\\
\frac{	\partial \mathbf{v_{\mathrm d}}}{\partial t}+(\mathbf{v_{\mathrm d}}\cdot \mathbf{\nabla})\,\mathbf{v_{\mathrm d}} & = & -\frac{K}{\rho_{\mathrm d}}(\mathbf{v_{\mathrm d}}-\mathbf{v_{\mathrm g}})-\mathbf{\nabla} (\Phi + \Phi_{\mathrm p}), \label{eq:gene4}
\end{eqnarray}
where the indices g and d refer to the gas and the dust phases, $v$ and $\rho$ denote the velocities and the densities, $\Phi$ and $\Phi_{\mathrm p}$ denote the gravitational potentials of the star and the planet. $P$ and $\sigma$ denote the pressure and the viscous tensor of the gas. $K$ denotes the drag coefficient, whose expression depends on the local values of the parameters of the grain and of the disc (e.g. \citealt{laibe12a}). $K$ is related to the stopping time of the mixture $t_{\rm s}$ by 
\begin{equation}
%K\equiv\frac{\rho_{\mathrm d} \rho_{\mathrm g}}{t_{\mathrm s} (\rho_{\mathrm g} + \rho_{\mathrm d})} .
K\equiv\frac{\rho_{\mathrm{d}}}{t_{\mathrm{s}} (1+\epsilon)} ,
\end{equation}
where $\epsilon=\rho_{\mathrm d}/\rho_{\mathrm g}$ is the dust-to-gas density ratio. Instead of $K$, one uses generally the Stokes number $\mathrm{St}  \equiv \Omega_{\mathrm k} t_{\mathrm s} $, defined as the ratio of the stopping time to the local dynamical timescale. In typical discs, the mean free path of the gas molecules is smaller than the dust particle size $s_{\mathrm{grain}}$. For this so-called Epstein regime, the Stokes number is given by \citep[e.g.][]{price15a}
\begin{equation}
\mathrm{St} = \frac{ \rho_{\mathrm{grain}} s_{\mathrm{grain}} \Omega_{\mathrm{k}}}{(1+\epsilon) \rho_{\mathrm{g}} c_{\mathrm{s}}}  \sqrt{\frac{\pi\gamma}{8}},
\label{eq:ts}
\end{equation}
where $\rho_{\mathrm{grain}}$ is the intrinsic grain density, $c_{\mathrm s}$ is the sound speed and $\gamma$ is the adiabatic index.

To reduce Eqs.~\ref{eq:gene3} -- \ref{eq:gene4} to a system of equations that describes a steady steady solution for the gas and the dust, we follow the approach introduced in \citet{nakagawa86a} (hereafter \citetalias{nakagawa86a}) and make two approximations. Firstly, we assume that the orbits have circularised after a transient regime occurring over a time $t_{\rm s}$ \citep{adachi76a}. Secondly, the long term evolution of the gas surface density profiles due to viscous effect is neglected. To include the tides from the planets, we only consider the azimuthally averaged contribution of the tidal torque by replacing the source term $\mathbf{\nabla} \Phi_{\mathrm p}|_\theta$ by $\Lambda / r $, where $\Lambda$ is given by Eq.~\ref{eq:torquegrav}, adopting the prescription expressed in Eqs.~\ref{eq:deltag} - \ref{eq:deltad} for each phases. For low mass planets, i.e. $r_{\mathrm{H}}<H$, the values of $\Lambda_{\rm g}$ and $\Lambda_{\rm d}$ are \textit{a priori} different (the cut-off is at $H$ for the gas and at $r_{\rm H}$ for the dust). We assume classical shear viscosity and $\eta = \nu \rho_{\mathrm g}$ to denote the dynamical viscosity of the gas. This approach is commonly used to model angular momentum transport driven by turbulence generated e.g. by magneto-rotational or gravitational instabilities \citep{balbus91a,lodato04a,rafikov15a}. However, \citet{rafikov17a} has recently shown that accretion might not proceed viscously in protoplanetary discs, but may be driven non-diffusively by magnetohydrodynamic winds or spiral density waves \citep{rafikov02a,bai16a,fung17a}. In this paper, we assume that accretion is mediated by a viscous-like mechanism \citep{lynden-bell74a}. The only non-zero component of $\sigma$ for the axisymmetric sheared flow of the disc is
\begin{equation}
\sigma_{r\theta}=\eta r \frac{\partial}{\partial r}\biggr(\frac{v_{\mathrm g,\theta}}{r}\biggl)   ,
\label{streesstensor}
\end{equation}
where 
\begin{equation}
\frac{v_{\mathrm g,\theta}^{2}}{r} = \frac{v_{\rm k}^{2}}{r} + \frac{1}{\rho_{\rm g}} \frac{\partial P}{\partial r} + \mathcal{O}\left( H / r \right)^{2}.
\label{eq:kep_dev}
\end{equation}
For low mass planets, the deviation of $\sigma_{r\theta}$ induced by the sub-Keplerian rotation of the unperturbed pressure profile is only of order $(H/R)^2 \ll 1$ and can be neglected. 
 
Under these assumptions, we perform a perturbative expansion relative to the Keplerian velocity $\mathbf{v_{\mathrm k}}=(0,r\Omega_{\mathrm k},0)$ in both phases, and look for stationary solutions assuming an axisymmetric disc. %(e.g. \citet{laibe12b} for the technical details of the expansion). 
Note that from here, the notation $v_{\theta}$ will be used to refer to the \textit{perturbed} azimuthal velocities for sake of simplicity. The equations of motion for the perturbed velocities can be expressed in polar coordinates as
\begin{align}
\frac{	\partial v_{\mathrm{g},R}}{\partial t} &=\frac{K}{\rho_{\mathrm g}}(v_{\mathrm{d},r}-v_{\mathrm{g},r})  -\frac{1}{\rho_{\mathrm g}}\frac{\partial P}{\partial r}+2\Omega_{\mathrm k} v_{\mathrm g,\theta},
\label{eq:1} \\
\frac{	\partial v_{\mathrm{g},\phi}}{\partial t}&=\frac{K}{\rho_{\mathrm g}}(v_{\mathrm{d},\theta}-v_{\mathrm{g},\theta})  -\frac{\Omega_{\mathrm k}}{2}v_{\mathrm{g},r} +\frac{1}{\rho_{\mathrm g}}\nabla \cdot \sigma|_\theta+\frac{\Lambda_{\mathrm{g}}}{r}, \\
\frac{	\partial v_{\mathrm{d},R}}{\partial t}&=-\frac{K}{\rho_{\mathrm d}}(v_{\mathrm{d},r}-v_{\mathrm{g},r})  +2\Omega_{\mathrm k} v_{\mathrm d,\theta},
\label{eq:3} \\
\frac{	\partial v_{\mathrm{d},\phi}}{\partial t}&=-\frac{K}{\rho_{\mathrm d}}(v_{\mathrm{d},\theta}-v_{\mathrm{g},\theta})  -\frac{\Omega_{\mathrm k}}{2} v_{\mathrm{d},r}+\frac{\Lambda_{\mathrm{d}}}{r} .
\label{eq:4}
\end{align}
Eqs.~\ref{eq:1} -- \ref{eq:4} consist of a system of non-homogeneous differential equations of the form $\mathbf{X}'(t) + \mathbf{A}\mathbf{X}(t) = \mathbf{B}$, where $\mathbf{A}$ and $\mathbf{B}$ are two constant matrices. Its steady state is $\mathbf{X}_{\rm stat}=\mathbf{A}^{-1}\mathbf{B}$. The system relaxes towards this stationary regime in the typical time $\min \left|  \Re \left( \sigma_{\rm A} \right) \right|^{-1}  = t_{\mathrm{s}}$, where $\sigma_{\rm A}$ are the eigenvalues of matrix $\mathbf{A}$ and $t_{\mathrm{s}}$ is the the stopping time of the mixture defined in in Eq.~\ref{eq:ts} in a dimensionless form. 

\subsubsection{Steady-state solution}
The stationary solution of the linear system Eqs.~\ref{eq:1} -- \ref{eq:4} is %provides the four perturbative velocities as
\begin{align}
v_{\mathrm{g},r}\,=\,&- \frac{1}{1+\epsilon}\left\lbrace \frac{\epsilon \,\Delta v}{\mathrm{St}+\mathrm{St}^{-1}}   - \left(1 + \epsilon \frac{ \mathrm{St}^{2}}{1+\mathrm{St}^{2} } \right) v_{\mathrm{visc}} \right\rbrace \nonumber \\ 
 &+ \frac{2\Lambda_{\mathrm{d}}}{v_{\mathrm k}} \frac{\epsilon}{(1+\epsilon)(1+\mathrm{St}^2)}
+\frac{2\Lambda_{\mathrm{g}}}{v_{\mathrm k}}\frac{1+\mathrm{St}^2(1+\epsilon)}{(1+\epsilon)(1+\mathrm{St}^2)}, \label{eq:vrad} \\
%&+\frac{2\Lambda_{\mathrm{d}}}{v_{\mathrm k}}
%+\frac{2(\Lambda_{\mathrm{g}}-\Lambda_{\mathrm{d}})}{v_{\mathrm k}}\frac{1+\mathrm{St}^2(1+\epsilon)}{(1+\epsilon)(1+\mathrm{St}^2)} , \label{eq:vrad} \\
v_{\mathrm{d},r}\,=\,& \frac{1}{1+\epsilon}\left\lbrace \frac{ \Delta v}{\mathrm{St}+\mathrm{St}^{-1}} + \frac{ v_{\mathrm{visc}}}{1+\mathrm{St}^{2}}\right\rbrace \nonumber \\
&+ \frac{2\Lambda_{\mathrm{d}}}{v_{\mathrm k}} \frac{\epsilon \left(1+\mathrm{St}^2\right) + \mathrm{St}^{2}}{(1+\epsilon)\left(1+\mathrm{St}^2 \right)} + \frac{2\Lambda_{\mathrm{g}}}{v_{\mathrm k}(1+\epsilon)(1+\mathrm{St}^2)}, \label{eq:vradust} \\
%&+\frac{2\Lambda_{\mathrm{d}}}{v_{\mathrm k}} + \frac{2(\Lambda_{\mathrm{g}}-\Lambda_{\mathrm{d}})}{v_{\mathrm k}(1+\mathrm{St}^2)(1+\epsilon)} \label{eq:vradust} \\
v_{\mathrm{g},\theta}\,=\,&  \frac{1}{2(1+\epsilon)}\left\lbrace \left(1 + \epsilon \frac{ \mathrm{St}^{2}}{1+\mathrm{St}^{2} } \right)\Delta v+\frac{\epsilon}{\mathrm{St}+\mathrm{St}^{-1}}v_{\mathrm{visc}}\right\rbrace \nonumber \\
& + \frac{\epsilon(\Lambda_{\mathrm{g}}-\Lambda_{\mathrm{d}})}{v_{\mathrm k}(1+\epsilon)(\mathrm{St}+\mathrm{St}^{-1})} , \label{eq:vthetag}\\
v_{\mathrm{d},\theta}\,=\,& \frac{1}{2(1+\epsilon)}\left\lbrace \frac{\Delta v}{1+\mathrm{St}^{2}}   - \frac{ v_{\mathrm{visc}}}{\mathrm{St}+\mathrm{St}^{-1}} \right\rbrace \nonumber \\
&- \frac{\Lambda_{\mathrm{g}}-\Lambda_{\mathrm{d}}}{v_{\mathrm k}(1+\epsilon)(\mathrm{St}+\mathrm{St}^{-1})} , \label{eq:vthetad}
\end{align}
where 
\begin{equation}
\Delta v  \equiv  \frac{1}{\rho_{\mathrm g} \Omega_{\mathrm k}}\frac{\partial P}{\partial r} ,
\end{equation}
is the typical optimal drift velocity derived in \citetalias{nakagawa86a}, and 
\begin{eqnarray}
v_{\mathrm{visc}} & \equiv & \frac{2}{\Omega_{\mathrm k}\rho_{\mathrm g}}\nabla \cdot \sigma|_\theta \nonumber \\
& = &  \frac{1}{r \rho_{\mathrm g} \frac{ \partial }{ \partial r}\left(   r v_{\mathrm k} \right)  }\frac{\partial}{\partial r} \left(\eta r^3 \frac{\partial \Omega_{\mathrm k}}{\partial r}\right) ,
\end{eqnarray}
is the viscous velocity derived by \citet{lynden-bell74a}, since for low mass planet, surface density gradients develop over large scales (no gas gap). From Eqs.~\ref{eq:vthetag} -- \ref{eq:vthetad}, the differential azimuthal velocity between the gas and dust is
\begin{eqnarray}
v_{\mathrm d,\theta}-v_{\mathrm g,\theta}&=& -\frac{\Delta v}{2 (1+\mathrm{St}^{-2})}  -\frac{v_{\mathrm{visc}}}{2( \mathrm{St}+\mathrm{St}^{-1})}\nonumber \\
&-&\frac{\Lambda_{\mathrm{g}}-\Lambda_{\mathrm{d}}}{v_{\mathrm k}(\mathrm{St}+\mathrm{St}^{-1})} ,
\end{eqnarray}
which is independent on the dust-to-gas ratio. The specific drag torque exerted by the gas phase on an elementary dust ring si therefore
\begin{eqnarray}
\Lambda_{\mathrm{g}\rightarrow \mathrm{d}}&= &-r\frac{K}{\rho_{\mathrm d}}\left (v_{\mathrm d,\theta}-v_{\mathrm g,\theta} \right)\nonumber \\&=& \frac{v_{\mathrm k}}{2 (1 + \epsilon)} \left\lbrace \frac{\Delta v}{ \mathrm{St}+\mathrm{St}^{-1}}  + \frac{v_{\mathrm{visc}}}{ 1 + \mathrm{St}^{2}}  \right\rbrace\nonumber\\
&-& \frac{\Lambda_{\mathrm{g}}-\Lambda_{\mathrm{d}}}{(1+\epsilon)(1 + \mathrm{St}^{2})} ,
\label{eq:draggd}
\end{eqnarray}
while the back reaction drag torque from the dust to the gas is 
\begin{equation}
\Lambda_{\mathrm{d}\rightarrow \mathrm{g}} =r\frac{K}{\rho_{\mathrm g}}\left (v_{\mathrm d,\theta}-v_{\mathrm g,\theta} \right)=-\epsilon \Lambda_{\mathrm{g}\rightarrow \mathrm{d}}.
\label{eq:dragdg}
\end{equation}

%\begin{eqnarray}
%\Lambda_{\mathrm{d}\rightarrow \mathrm{g}}  &= & -\frac{\epsilon v_{\mathrm k}}{2 (1 + \epsilon)} \left\lbrace \frac{v_{\mathrm p}}{ \mathrm{St}+\mathrm{St}^{-1}}  + \frac{v_{\mathrm{visc}}}{ 1 + \mathrm{St}^{2}}  \right\rbrace \nonumber \\
%&+& \frac{\epsilon(\Lambda_{\mathrm{g}}-\Lambda_{\mathrm{d}})}{1 + \mathrm{St}^{2}}.
%\label{eq:dragdg}
%\end{eqnarray}
%
\subsubsection{Physical interpretation}
In the limit of the dust grains being perfectly decoupled from the gas ($\mathrm{St} = + \infty$), particles orbit the star with Keplerian velocity ($v_{\mathrm{d},\theta} = 0$) and are pushed outside of the planet orbit by the tidal torque at the constant velocity $v_{\mathrm{d},r} = 2 \Lambda_{\rm d} / v_{\rm k}$. The gas orbits the star at the sub-Keplerian velocity $\Delta v / 2 < 0$, while the gas radial velocity is $v_{\mathrm{visc}} + 2 \Lambda_{\rm g} / v_{\rm k}$. When drag couples the two phases, the dust motion is dominated by the gas when $\epsilon \ll 1$ and $1/\left( 1 + \epsilon \right) \sim 1$. When $\epsilon \gg 1$ the gas motion is dominated by the dust. %  and $\epsilon/\left( 1 + \epsilon \right) \sim 1 $. 
Both the gas and the dust strongly feel the other phase when the dust-to-gas ratio is of order unity \citep{gonzalez17a}.

As it is known in in absence of a planet ($\Lambda_{\rm g} =\Lambda_{\rm d} =0$), differential dynamics due to gas pressure and viscosity are communicated from one phase to the other via drag. The term proportional to $\Delta v$ in Eq.~\ref{eq:vradust} implies that particles drift radially as a result of the residual differential orbital velocity between the two phases (Eq.~2.11 of \citetalias{nakagawa86a}). In the absence of any small scale pressure perturbation in the gas, $\Delta v < 0$ in a typical disc since the inner regions are denser and warmer. This results in a radial inward drift of particles. Drift is most efficient for $\mathrm{St} \sim 1$. For $\mathrm{St} \gg 1$ (resp. $\mathrm{St} \ll 1$) particles maintain a fixed Keplerian (resp. sub-Keplerian) orbit. In  Eq.~\ref{eq:vrad}, the term proportional to $\Delta v$ corresponds to the back-reaction term (Eq.~2.13 of \citetalias{nakagawa86a}): in discs with $\Delta v < 0$, the gas is pushed back in the regions of lower pressure by the dust to conserve the global angular momentum (this motion becomes significant when $\epsilon$ approaches unity). Futhermore, the radial viscous gas flow is weighted by a factor which depends on $\epsilon$ and $\mathrm{St}$ (second term of the right-hand side of Eq.~\ref{eq:vrad}). For small particles $\mathrm{St} \ll 1$, the viscous flow is reduces by a factor $\left( 1 + \epsilon \right)^{-1}$. For large particles ($\mathrm{St} \gg 1$), $v_{\mathrm{g},r}=v_{\mathrm{visc}}$, i.e. dust does non affect the viscous gas motion regardless the value of $\epsilon$.

 In Eq.~\ref{eq:vradust}, the term proportional to $v_{\mathrm{visc}}$ shows that additionally, drag makes very small grains stick to the gas. They are carried radially by the viscous flow. Hence, dust evolves ``viscously'' under the indirect effect of the viscous evolution of the gas, with a so-called ``drag-induced dust viscosity''
\begin{equation}
\nu_{\mathrm{d, eff}} \equiv \frac{1}{1+\epsilon} \frac{1}{1+\mathrm{St}^{2}} \nu.
\label{eq:viscositydragind}
\end{equation}
This contribution dominates over the pressure drift only when grains are tiny, i.e. $\mathrm{St} < \alpha$. Note that although Eq.~\ref{eq:viscositydragind} resembles to the dust diffusivity derived in \citet{youdin07a}, the drag-induced dust viscosity does not describe the motion induced by turbulence over dust grains, but models how the viscous evolution of the gas affects the radial dynamics of grains. Importantly, when $\mathrm{St} \sim 1$ and $\epsilon \geq 1$, the gas dynamics is dominated by the back-reaction from the dust and not by the viscosity \citep{gonzalez15a,taki16a}. 

In the general case, the differential motion between the gas and the dust induced by the different tides only affect the motion of smaller grains, i.e. $\mathrm{St}\aplt 1$. This concerns particularly the region $r_{\rm H} < \left| r - r_{\rm p}\right| < H$, where $\Lambda_{\rm d} \ne \Lambda_{\rm g} = 0 $. In this region, for a gas-dominated dynamics with $\epsilon \ll 1$, only large dust grains, i.e. $\mathrm{St}\grtsim 1$, are pushed away from the planet orbit, since they experience a lower drag related to the different tidal torque between the two phases (see the third term in the right hand side of Eq.~\ref{eq:draggd}). In other words, for smaller grains, the motion induced by the tidal torque in this region is damped by the drag torque that tends to reduce the velocity difference with the unperturbed gas flow. As a result, small grains are forced to stick to the fixed gas. 
On the other hand, it can be noticed that the gas is not affected by the tides in this region if $\epsilon \ll1$ (see the third and fourth terms in the right hand side of Eq.~\ref{eq:vrad}).
Sufficiently far away from the planet, where $\left| r - r_{\rm p}\right| > H$ and thus $\Lambda_{\rm g} = \Lambda_{\rm d}$, there is no differential tidal torque, and the tidal barycentric velocity is spread over the two phases proportionally to the respective density of each phase. 

%From this analysis, it is reasonably expected that planets which are not able to carve gaps in the gas might be able to open gaps in the dust structure only for larger grains, i.e. $\mathrm{St}\grtsim1$. 

\subsubsection{Orders of magnitude}
\label{sec:odg}

We compare the orders of magnitude of the different velocity terms in Eq.~\ref{eq:vradust}, related to the radial pressure gradient, the viscous and the tidal contribution. Until specified, we shall not restrict our analysis to the case of an unperturbed gas density profile. 
We obtain
\begin{eqnarray}
\left| - \frac{1}{\rho_{\rm g}\Omega_{\mathrm k}} \frac{\partial P}{\partial r} \right| & \sim &  \left( \frac{r}{l} \right) \left( \frac{H}{r} \right)^{2} v_{\rm k} ,\label{eq:odg_press}\\
\left|  \frac{1}{\rho_{\mathrm g}\Omega_{\mathrm k}}\nabla \cdot \sigma|_\theta \right| & \sim & \left( \frac{r}{l} \right)  \alpha\left( \frac{H}{r} \right)^{2} v_{\rm k} . \label{eq:odg_visc}
\end{eqnarray}
where $l$ denotes the typical length over which the gas surface density varies and where we have used the seminal Prandtl-like turbulent viscosity $\nu=\alpha c_{\rm s} H$ \citep{shakura73a}.
The prefactor in Eq.~\ref{eq:odg_visc} originates from the second derivative of the Keplerian deviation (Eq.~\ref{eq:kep_dev}). For an unperturbed gas profile, $l = r$ and the pre-factors in Eqs.~\ref{eq:odg_press} -- \ref{eq:odg_visc} equal unity. The maximal tidal contributions, obtained at the cut-off locations, are of order
\begin{equation}
\left| \frac{\Lambda}{v_{\rm k}} \right|_{\rm max} \sim \frac{r_{\rm H}^{6}}{r^{2} \Delta^{4}} v_{\rm k} .
\end{equation}
For low mass planets, i.e. $r_{\mathrm{H}}<H$, the maximum tidal torque density for the gas and the dust are given by 
\begin{eqnarray}
\left| \frac{\Lambda_{\rm g}}{v_{\rm k}} \right|_{\rm max} & \sim & \left( \frac{r_{\rm H}}{H} \right)^{6} \left( \frac{H}{r} \right)^{2} v_{\rm k} , \label{eq:odg_lambdag} \\
\left| \frac{\Lambda_{\rm d}}{v_{\rm k}} \right|_{\rm max} & \sim & \left( \frac{r_{\rm H}}{H} \right)^{2}\left( \frac{H}{r} \right)^{2} v_{\rm k} . \label{eq:odg_lambdad}
\end{eqnarray}
%
%
%\begin{equation}
%b = 
%\begin{cases}
%\frac{\mathrm{St}}{1 + \mathrm{St}^{2} } , & \text{if $\mathbf{a} = a \mathbf{e}_{r}$,} \\
%\frac{1}{1 + \mathrm{St}^{2} }  , & \text{if $\mathbf{a} = a \mathbf{e}_{\theta}$.}
%\end{cases}
%\end{equation}
%%
%$b$ ensures that large dust grains are decoupled from the gas, and that small dust grains drift inwards if and only if the gas does, as shown in Eq.~\ref{eq:draggd}. 
For the sake of clarity, we now assume $\epsilon \ll1$ and limit our analysis to large grains which are most affected by the tidal torque without being decelerated by the drag torque arising due to the differential tidal torque between the two phases. For larger grains, i.e. $\mathrm{St}\grtsim1$, the terms in Eq.~\ref{eq:vradust} proportional to $1/(1+\mathrm{St}^2)$ are negligible.
%Additionally, we assume that $\mathrm{St}> \alpha$, to limit our analysis to large grains and neglect the viscous forces. 
The remaining terms in Eq.~\ref{eq:vradust} are given by 

\begin{equation}
v_{\mathrm{d},r}|_{\Delta v}\sim \frac{\mathrm{St}}{1 + \mathrm{St}^{2} } \left( \frac{H}{r} \right)^{2} v_{\rm k},
\label{eq:contrpres}
\end{equation}
%
%\begin{equation}
%v_{\mathrm{d},r}|_{\mathrm{visc}}\sim \frac{\mathrm{1}}{1 + \mathrm{St}^{2} }\alpha \left( \frac{H}{r} \right)^{2} v_{\rm k}
%\label{eq:contrvisc}
%\end{equation}
%
\begin{equation}
v_{\mathrm{d},r}|_{\Lambda_{\rm d}}\sim \frac{\mathrm{St}^{2}}{1 + \mathrm{St}^{2} } \left( \frac{r_{\rm H}}{H} \right)^{2}\left( \frac{H}{r} \right)^{2} v_{\rm k}.
\label{eq:contrlamb}
\end{equation}
%
%\begin{equation}
%v_{\mathrm{d},r}|_{\Lambda_{\rm g}}\sim \frac{\mathrm{1}}{1 + \mathrm{St}^{2} } \left( \frac{r_{\rm H}}{H} \right)^{2}\left( \frac{H}{r} \right)^{2} v_{\rm k}
%\label{eq:contrlamb}
%\end{equation}
%
%\begin{equation}
%\Lambda_{\rm drag} \left( a_{\rm drift} \right) \sim \frac{\mathrm{St}}{1 + \mathrm{St}^{2} } \left( \frac{H}{r} \right)^{2} v_{\rm k}^{2} ,
%\label{eq:tq_d1}
%\end{equation}
%%
%\begin{equation}
%\Lambda_{\rm drag} \left( a_{\rm visc} \right) \sim \frac{1}{1 + \mathrm{St}^{2} } \alpha \left( \frac{H}{r} \right)^{2} v_{\rm k}^{2} ,
%\label{eq:tq_d3}
%\end{equation}
%%
%\begin{equation}
% \Lambda_{\rm drag} \left( a_{\rm tides,g} \right) \sim \frac{1}{1 + \mathrm{St}^{2}} \left( \frac{r_{\rm H}}{H} \right)^{6} \left( \frac{H}{r} \right)^{2} v_{\rm k}^{2} ,
% \label{eq:tq_d3}
%\end{equation}
%%
%\begin{eqnarray}
%\Lambda_{\rm d} + \Lambda_{\rm drag} \left( a_{\rm tides,d} \right) & \sim & \left(1 -  \frac{1}{1 + \mathrm{St}^{2} }\right) \left( \frac{r_{\rm H}}{H} \right)^{2} \left( \frac{H}{r} \right)^{2} v_{\rm k}^{2} \nonumber \\
%& = & \frac{\mathrm{St}^{2}}{1 + \mathrm{St}^{2} } \left( \frac{r_{\rm H}}{H} \right)^{2} \left( \frac{H}{r} \right)^{2} v_{\rm k}^{2} .
%\label{eq:tq_d4}
%\end{eqnarray}
%
%Eq.~\ref{eq:tq_d4} appears as a consequence of the differential tidal torque, which induces a relative motion between the dust and the gas. In return, this motion creates a drag torque which counterbalances the original torque. 
The term $\mathrm{St}^{2} / \left(1 + \mathrm{St}^{2} \right)$ in Eq.~\ref{eq:contrlamb} expresses that only large grains are entrained by the tides since small grains stick to the gas. For these grains the drag torque is dominated by the contributions from the pressure gradient and the tides.

%---------------------------------------------------------------------------------------------------------------------
\section{Gap opening in dusty discs}
\label{sec:criterion}

\subsection{The low mass planet regime}
The formalism derived in Sect.~\ref{sec:gapopenig} enables to study the gap opening process by a low massive planet embedded in dusty discs assuming that the pressure profile around the planet remains unperturbed. We now investigate under which condition this assumption is satisfied. 

A gap is carved in the gas when the tidal torque overpowers the viscous torque. %A condition for the tidal torque to be large enough is $r_{\rm H} \sim H$. The sixth power in Eq.~\ref{eq:odg_lambdag} ensures indeed that $\left( r_{\rm H} / H\right)^{6} \ll \alpha$ for smaller Hill radii. 
Assuming that the typical length over which the gas surface density varies by the tidal action of the planet is of order of $H$, we compare Eqs.~\ref{eq:odg_visc} and \ref{eq:odg_lambdag} by replacing $l$ by $H$ to estimate the condition for gap opening in the gas. We find
\begin{equation}
\frac{M_{\rm p}}{M_{\star}} \gtrsim \alpha^{1/2} \left( \frac{H}{r_{\mathrm p}} \right)^{5/2} .
\label{eq:gas_scaling}
\end{equation}
Although Eq.~\ref{eq:gas_scaling} provides an interesting scaling, more quantitative criteria are used in literature. The first one is based on the requirement that a strong shock forms within a scale height of the planet's orbit \citep{lin93a}, giving
\begin{equation}
\left( \frac{M_{\rm p}}{M_{\star}}\right)_{\mathrm{th}}\grtsim3\biggl(\frac{H}{r_{\mathrm p}}\biggr)^3.
\label{eq:cond1}
\end{equation}
according to which the planet Hill radius $r_{\mathrm H}$ must be greater than the vertical scale height of the disc $H$.
%Eq.~\ref{eq:cond1} corresponds to our first condition. 
However, recent 2D and 3D simulations of gas discs hosting planets have shown that planets with mass $M_{\mathrm{p}}\grtsim 0.2\,M_{\mathrm{p,th}}$ are able to create a pressure maximum outside the planetary orbit \citep{lambrechts14a,rosotti16a}. 
The second criterion is based on the requirement that gap opening should be faster than the viscous refilling of the gap, giving
\begin{equation}
\left( \frac{M_{\rm p}}{M_{\star}}\right)_{\mathrm{visc}}\grtsim \biggl(\frac{3 }{2f}\biggr)^{1/2} \biggl(\frac{H}{r_{\mathrm p}}\biggr)^{5/2}\alpha^{1/2} .
\label{eq:cond2}
\end{equation}
where $f$ is the constant in the tidal torque density formula (Eq.~\ref{eq:torquegrav}).
Moreover, Eq.~\ref{eq:cond2} is in agreement with the estimate derived in Eq.~\ref{eq:gas_scaling}. If the pressure (resp. the viscous) force dominates the gap closing mechanism, we expect $M_{\mathrm{p,th}}$ to be larger (resp. smaller) than $M_{\mathrm{p,visc}}$. 
In this work, we consider that the minimum mass able to create a pressure maxima in the outer disc is given by the maximum of all the masses predicted by the previously mentioned criteria,
\begin{equation}
M_{\mathrm{p, gap}}=\mathrm{max}(0.2\,M_{\mathrm{p,th}},M_{\mathrm{p,visc}}) .
\label{eq:mlimgas}
\end{equation}
However, \citet{rosotti16a} have recently found that planet of masses slightly lower than the one given by Eq.~\ref{eq:mlimgas} could create gap structures in the dust as well. In detail, if $M_{\mathrm{p}}\gtrsim 0.1\,M_{\mathrm{p,th}}$, the planet weakens the pressure gradient profile in its neighbourhood, which reduces the dust drift locally and leads to accumulation of particles. This traffic jam mechanism affects essentially marginally coupled particles ($\mathrm{St}\sim 1$) and lead to the formation of a dust gap.
%The gap-like structure produced in the dust by this mechanism is deeper for $\mathrm{St}\sim 1$. 
We consider this effect by assuming that the minimum planet mass able to affect the local gas profile is 
\begin{equation}
M_{\mathrm{p, lim}}=0.1\,M_{\mathrm{p,th}} .
\label{eq:mplim}
\end{equation}
In Sect.~\ref{sec:testcriterion}, we confirm the validity of this condition. Eventually, it should be noted that all the criteria above assume planets remains on a fixed orbit. However, \citet{malik15a} have shown that the migration may affect the ability of the planet to carve gaps, and the critical masses given by Eqs.~\ref{eq:cond1} and \ref{eq:cond2} might be underestimated.
%\citet{crida06a} have encompassed the two conditions Eqs.~\ref{eq:cond1} -- \ref{eq:cond2} into the folowing semi-analytical criterion
%\begin{equation}
%\frac{3}{4}\frac{H}{r_{\mathrm{p}}}+\frac{M_{\star}}{M_{\rm p}}\frac{50\nu_{\mathrm{p}}}{\Omega_{\mathrm{p}}r_{\mathrm{p}}^2}\aplt 1,
%\end{equation}
%where $\nu_{\mathrm{p}}$ is the kinematical viscosity at the planet's location. 

\subsection{Dust gap width}
\label{sect:dustgapwidth}

We now focus our analysis on planets not able to affect the local pressure structure, i.e. $M_{\mathrm{p}} \lesssim M_{\mathrm{p, lim}}$, embedded in standard discs ($\partial P / \partial r<0$).  The tidal interaction between the planet and the disc acts to carve the gap around the planetary orbit, whereas drag makes the grains drift inwards towards the central protostar. In particular, the flux of solids coming from the outside of the planet orbit tends to refill the dust depletion locally induced by the tides. In this case, the drag torque can be derived using the unperturbed pressure profile of the gas (Eq.~\ref{eq:draggd}).
\begin{figure}
\includegraphics[width=0.49\textwidth]{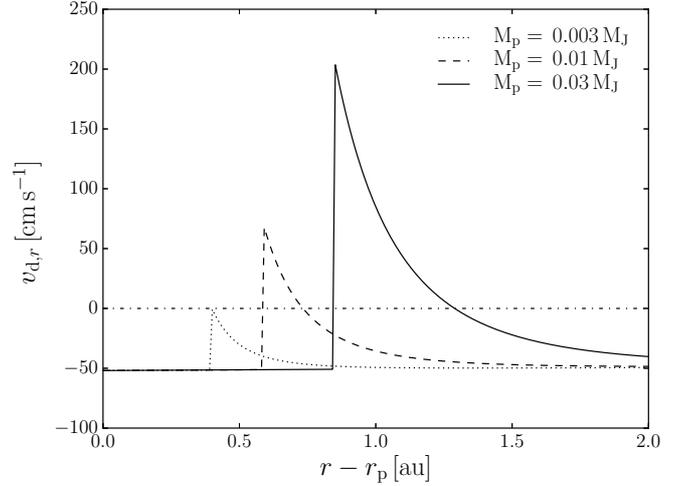}
\caption{Total radial dust velocity of millimetre grains outside the orbit of the planet for different planet masses: 0.003, 0.01 and 0.03 $M_{\mathrm{J}}$ adopting a disc model with $H/r=0.02$ at 1 au (corresponding to $H/r_{\mathrm{p}}\approx0.05$ at $r_{\mathrm{p}}$), $M_{\mathrm{p, th}}\sim 0.4\, M_{\mathrm{J}}$, $\alpha=0.005$, $\epsilon=0.01$, $p=1$, $q=1/2$, $\mathrm{St}\sim10$ at $r_{\mathrm{p}}$ and assuming the proportionality constant in front of the torque density prescription expressed in Eq.~\ref{eq:torquegrav} equal to the nominal value of $0.4$ introduced in \citet{goldreich79a}. The velocity peaks at $r-r_{\mathrm{p}}=r_{\mathrm{H}}$ and decreases with increasing distance from the planet. For planets with masses $\aplt 0.003 M_{\rm J}$, the tidal torque is not strong enough to halt the radial inflow induced by the drag torque ($v_{\mathrm{d},r}(r) < 0$).}
\label{fig:vdr} 
\end{figure}
We denote 
\begin{equation}
\zeta \equiv \left. \frac{\partial \log P}{\partial \log r} \right|_{r_{\rm p}} ,
\label{eq:zeta}
\end{equation}
the exponent which characterises the steepness of the pressure profile of the disc. If we assume power-law profiles for the surface density ($\Sigma \propto r^{-p}$) and the temperature ($T \propto r^{-q}$), we have $\zeta =-(p+q/2+3/2)$. For typical disc ($\,p=1$, $q=1/2$, \citealt{andrews07a,williams14a}), the power law exponent of the pressure profile is $\zeta \simeq -2.75 < 0$. With this notation, the azimuthally averaged radial dust velocity (Eq.~\ref{eq:vradust}) can be written as
\begin{eqnarray}
v_{\mathrm{d},r}&=& \frac{\zeta \mathrm{St}  - \left( 6 + 3\zeta \right)\alpha}{(1+\epsilon)(1+\mathrm{St}^2)}\frac{c_{\rm s}^2}{v_{\mathrm k}}
\nonumber \\
&& +\frac{2\Lambda_{\mathrm{d}}}{v_{\mathrm k}} + \frac{2(\Lambda_{\mathrm{g}}-\Lambda_{\mathrm{d}})}{v_{\mathrm k}(1+\mathrm{St}^2)(1+\epsilon)}.
\label{eq:outedge}
\end{eqnarray}
 As an example, Fig.~\ref{fig:vdr} shows the radial velocity of millimetre sized grains outside the orbit of the planet with different mass (lower than $M_{\mathrm{p, lim}}$) embedded in a typical disc model, assuming $f$ to be equal to the nominal value of $0.4$ introduced in \citet{goldreich79a}.  In this case, the planet is located at 40 au from the star, $M_{\mathrm{p, th}}\sim 0.4 \, M_{\mathrm{J}}$, $\alpha=0.005$, $p=1$, $q=1/2$, $H/r=0.02$ at 1 au (corresponding to $H/r_{\mathrm{p}}\approx0.05$ at $r_{\mathrm{p}}$) and $\epsilon=0.01$. The dust grains in the midplane have Stokes number $\mathrm{St}\sim 10$ at the planet location. %,  the aspect ratio of the disc at the planet location is $\sim 0.07$ and the exponent characterising the steepness of the pressure radial profile is $\zeta=-1.95$.
 
Fig.~\ref{fig:vdr} shows that in the presence of a planet of very low-mass, the tidal torque is not strong enough to halt the inward radial flow triggered by the drag torque (e.g. the case $\aplt 0.003 M_{\rm J}$, for which $v_{\mathrm{d},r} < 0$). Larger planet masses empower stronger tidal torques in the disc, and the balance between the tidal and the drag torques leads to an outward drift (Fig.~\ref{fig:vdr} shows regions where $v_{\mathrm{d},r} > 0$ for $M_{\rm p}  = 0.01 M_{\rm J}$ and $M_{\rm p}  = 0.03 M_{\rm J}$). 

The location of the outer edge of the gap $r_{\rm gap}$ can be estimated simply, by evaluating the distance to the planet where $v_{\mathrm{d},r}(r_{\rm gap}) = 0$, i.e. where the drift induced by the tides balances steadily the drift induced by the drag. We use $\Delta_{\mathrm{gap}}$ to denote the width of the dust gap outside the planetary orbit $r_{\rm gap}-r_{\rm p}$. 
To simplify the analysis, we assume that the temperature and surface density of the gas are uniform over the domain $\Delta_{\mathrm{gap}}$, i.e. the terms in Eq.~\ref{eq:outedge} are evaluated at $r_{\mathrm{p}}$ with exception of $\Delta$ in $\Lambda_{\mathrm{d}}$ and $\Lambda_{\mathrm{g}}$ (see Eq.~\ref{eq:torquegrav}). Corrections of order $\Delta_{\mathrm{gap}}/r_{\mathrm{p}}$ are negligible for our analysis (see Sect.~\ref{sec:results}).
The width of the dust gap is therefore given by
\begin{equation}
\frac{\Delta_{\mathrm{gap}}}{r_{\mathrm{p}}} \simeq \left( 2f\right)^{1/4} z^{-1/4} \, \mathrm{St}^{1/4} \left(\frac{H}{r_{\mathrm p}}\right)^{-1/2}  \left(\frac{M_{\rm p}}{M_{\star}}\right)^{1/2} ,
\label{eq:redgeapprox}
\end{equation}
where
\begin{equation}
z \! \left(\mathrm{St},\epsilon, \alpha \right) = 
\begin{cases}
\displaystyle \frac{-\zeta    + \left( 6 + 3\zeta \right)\alpha / \mathrm{St} }{\left( 1 + \epsilon \right) +\epsilon / \mathrm{St}^{2}} , & \text{$r_{\rm H} \le \Delta_{\mathrm{gap}} < H$,} \\ 
\displaystyle \frac{-\zeta    + \left( 6 + 3\zeta \right)\alpha / \mathrm{St} }{\left( 1 + \epsilon \right) +\left( 1 + \epsilon\right) / \mathrm{St}^{2}}, & \text{$ \Delta_{\mathrm{gap}} \ge H$.}
\end{cases}
\label{eq:def_f}
\end{equation}
%
%and 
%
%\begin{equation}
%\left( \frac{32 C}{81\pi}\right)^{1/4} \approx 0.95 .
%\end{equation}
%
Note that the difference between the two expressions in Eq.~\ref{eq:def_f} has no sensible effect on the value of $z$ in practice for larger grains. This suggests, as expected, that the different tidal torque experienced by the two phases does not cover the key role in the gap opening process. 
For $\mathrm{St} \gtrsim \alpha$, the second term of the numerator vanishes and for $\mathrm{St} \gtrsim 1+\epsilon$, Eq.~\ref{eq:def_f} reduces to
\begin{equation}
z \simeq \frac{-\zeta}{1 + \epsilon} .
\label{eq:f_approx}
\end{equation}
Eq.~\ref{eq:redgeapprox} is consistent with the fact that infinitely small grains at low dust densities follow the gas, in which no gap forms ($\Delta_{\mathrm{gap}} = 0$ for $\mathrm{St} = 0$ and $\epsilon = 0$). For large grains, $\Delta_{\mathrm{gap}} \propto \mathrm{St}^{1/4}$, an expression only weakly sensitive to the exact value of $\mathrm{St}$. The location of the outer edge of the dust gap is a weak increasing function of the Stokes number, since it corresponds to a weaker drag torque, as found numerically by \citet{dipierro16a} (see Sect.~\ref{sect:grainsize}). Moreover, this analysis shows that, since large grains experience a lower drag related to the differential tidal torque between the two phases, the criterion is mostly based on the balance between the tidal torque $\Lambda_{\mathrm{d}}$ and the drag torque related to pressure forces. The differential tidal torque between the two phases mostly influence the dynamics of small grains, forcing them to follow the unperturbed gas flow.
 For large Stokes number, i.e. $\mathrm{St}>\alpha$, the increase of the dust-to-gas ratio at the midplane due to settling affect the motion of dust and gas, as described in Sect.~\ref{sec:dragdynamics}. 
However, Eq.~\ref{eq:redgeapprox} shows that, since $\mathrm{St}\propto1/\left(1+\epsilon\right)$, the gap width does not depend on the dust-to-gas ratio when large grains are considered, i.e. when Eq.~\ref{eq:f_approx} is valid. %, and in the limit where the dust feedback does not affect the local gas structure. %Therefore, since the morphology of the dusty gap does not depend on the local value of the dust-to-gas ratio, it is not necessary to evaluate
The eventual modification of the gas surface density profile due to the dust back reaction is not included in our model (see discussion in Sect.~\ref{sec:limits}).

Finally, Eq.~\ref{eq:redgeapprox} shows that the distance between the planet and the outer edge of the gap is a decreasing function of the aspect ratio of the disc. Indeed the contribution of gas drag to gap closing goes as $\left (H/r_{\mathrm{p}} \right)^2$ (cf Eq.~\ref{eq:outedge}). Assuming power-law profiles for the surface density of the gas $\Sigma \propto r^{-p}$ and temperature $T \propto r^{-q}$, the location of outer gap edge scales as $\Delta_{\mathrm{gap}}/r_{\mathrm{p}} \propto r^{p+\left(q-1\right)/4}$. For a typical disc with $p=1$ and $q=1/2$, the outer gap edge is expected to increase with radius as $r^{0.87}$.% \citet{kanagawa16a} found that in the case of more massive planets, the width of the gap scales as $\left(H/r_{\mathrm{p}}\right)^{-3/4} \left(M_{\rm p}/M_{\star}\right)^{1/2} \alpha^{-1/4}$, a result similar to Eq.~\ref{eq:redgeapprox}.

 %In detail, it is reasonably expected that, taking into account the location of the maximum of the tidal torque in the gas (see Eq.~\ref{eq:torquegrav}), a gap of half width $\Delta_{\rm g} = \mathrm{max}(H,r_{\mathrm{H}})$ is roughly the smallest gap that can be opened in the gas phase \citep{duffell13a}. 
%For low-mass planets considered in our work, since $r_{\mathrm{H}}< H$, the gap width in the gas should be governed by the disc scale height, i.e. $\Delta_{\rm g}/r_{\mathrm{p}} \propto H/r_{\mathrm{p}}$ \citep{dong16f}. 

%back-reaction produces a remarkable effect when a population of small coupled grains becomes concentrated. Large values of $\epsilon$ imply weaker drag torques: dust concentration may create gaps of finite sizes unexpected for low dust-to-gas ratios. In appearance, this gap width depends on $\epsilon$. 
% It should be noted that $\epsilon$ results from the dust dynamics, and is therefore an indirect function of $\mathrm{St}$. As an example, the diffusive model of turbulent settling of \citet{dubrulle95a} provides 
%
%\begin{equation}
%\epsilon=\epsilon_0 \frac{H_{\mathrm{g}}}{H_{\mathrm{d}}}\simeq \epsilon_0\sqrt{1 + \frac{\mathrm{St}}{\alpha}}.
%\end{equation}
%

\subsection{Gap opening criterion}

\subsubsection{Orders of magnitude}
\label{sec:odgdust}
An order of magnitude for the minimum mass required for a planet to open a dust gap can be straightforwardly estimated from Sect.~\ref{sec:odg}. By equating Eqs.~\ref{eq:contrpres} and \ref{eq:contrlamb} with the definition given in Eq.~\ref{eq:rhill}, we obtain
\begin{equation}
\frac{M_{\rm p}}{M_{\rm \star}} \sim \mathrm{St}^{-3/2}  \left( \frac{H}{r_{\rm p}} \right)^{3}.
\label{eq:scaling}
\end{equation}
Eq.~\ref{eq:scaling} shows that the critical mass required to open a gap in the dust is lower for large Stokes numbers and large aspect ratios. More interestingly, the exponent $-3/2$ implies a sharp transition between small and large grains at $\mathrm{St} \simeq 1$. Fixing $H / r_{\rm p}$ and decreasing $\mathrm{St}$ in Eq.~\ref{eq:scaling} shows that $M_{\rm p}$ increases efficiently, up to reach the critical mass required to open a gap in the gas as well, i.e. $M_{\rm p}/M_{\rm \star}\sim \left(H/r_{\rm p}\right)^3$ (Eq.~\ref{eq:cond1}, \citealt{lin93a}). This suggests, as expected, that the condition $\mathrm{St} \grtsim 1$ should be fulfilled for a gap to be carved in the dust only (see Sect.~\ref{subsec:examples}).

\subsubsection{Necessary condition for dust gap opening}

Eq.~\ref{eq:redgeapprox} provides a simple way to estimate the minimum planet mass $M_{\rm p}$ able to halt the inward radial drift induced by the drag. Noting that the minimum radius of the outer edge of the dust gap is the Hill radius, $M_{\rm p}$ is the planet mass for which the radial dust velocity is zero at  $r=r_{\mathrm p}+r_{\mathrm H}$.  Assuming that the temperature and surface density of the gas are uniform over the domain $ \left | r-r_{\mathrm p}\right |\sim r_{\mathrm{H}}$ -- an approximation of order $r_{\rm H}  / r_{\rm p} \ll 1$ --  we obtain 
\begin{equation}
\frac{M_{\rm p}}{M_{\rm \star}} \ge \xi \, \left(\frac{z}{\mathrm{St}}\right)^{3/2}  \left(\frac{H}{r_{\rm p}}\right)^3 ,
\label{eq:necessarycond}
\end{equation}
with $z$ given by the first expression in Eq.~\ref{eq:def_f} and with
\begin{equation}
\xi = \frac{1}{9}\left(2f\right)^{-3/2}.% \approx 0.16,
%\xi = \left( \frac{81\pi}{32 C}\frac{1}{ 3^{4/3}}\right)^{3/2}% \approx 0.16,
\label{eq:defxi}
\end{equation}
Eq.~\ref{eq:necessarycond} provides the minimum mass for a planet  to produce a density depletion in the dust outside the planet orbit. For the typical disc described above, the minimum planet mass has a value of $\sim0.003 \, M_{\mathrm{J}}$, as expected from the previous analysis about the radial velocity (see dotted line in Fig.~\ref{fig:vdr}).
As expected, Eq. \ref{eq:necessarycond} is in agreement with the orders-of-magnitude estimate (see Eq.~\ref{eq:scaling}).

\subsubsection{Sufficient condition for dust gap opening}
\label{subsec:suffcond}
Eq.~\ref{eq:necessarycond} gives the minimum mass of a planet able to halt the radial inward drift of dust particles. However, for a gap to form, dust must not refill dust depletions as it explores different azimuths. Since refilling is a \textit{non-axisymmetric} process, its effects is not picked up by the previous analysis based on averaged axisymmetric torques. An alternative approach based on timescales estimates is therefore developed hereafter to take dust refilling into account.

The gap opening timescale is the time required to evacuate all the dust contained between $r_{\mathrm{p}}$ and  $r_{\mathrm{p}}+r_{\mathrm{H}}$. As previously mentioned, a gap of half width $r_{\mathrm{H}}$ is roughly the smallest gap that can be opened in the dust, since the Lindblad resonances are most effective at this distance from the planet. 
Focussing our analysis on the dust outside the planetary orbit, the angular momentum that must be removed to open the gap between $r_{\mathrm{p}}$ and  $r_{\mathrm{p}}+r_{\mathrm{H}}$ is
\begin{equation}
\Delta J=2\pi r_{\mathrm{p}} r_{\mathrm{H}} \Sigma_{\mathrm{d}}\left.\frac{\mathrm{d} l}{\mathrm{d} r}\right|_{r_{\mathrm{p}}}r_{\mathrm{H}}=\pi r_{\mathrm{p}} r_{\mathrm{H}}^2\Sigma_{\mathrm{d}} v_{\mathrm k},
\end{equation}
where $l$ denotes the specific angular momentum. The typical time to evacuate all the dust in this region is 
\begin{equation}
t_{\mathrm{open}}=\frac{\Delta J}{\left |\mathrm{d} J/\mathrm{d} t\right|},
\end{equation}
where $\left |\mathrm{d} J/\mathrm{d} t\right|$ is the one-sided total torque on the planet due to its interaction with dust outside the orbit. This total torque is the integral of the torque density given by Eq.~\ref{eq:torquegrav} over the entire outer disc, i.e. %weighted by the dust mass $2\pi r_{\mathrm{p}}\Sigma_{\mathrm{d}}$
\begin{equation}
\frac{\mathrm{d} J}{\mathrm{d} t}=\int_{r_{\mathrm{H}}}^{\infty} 2\pi r_{\mathrm{p}}\Sigma_{\mathrm{d}} \Lambda_{\mathrm{d}}(r-r_{\mathrm{p}}) \, \mathrm{d}(r-r_{\mathrm{p}}) .
\end{equation} 
The gap opening timescale is therefore given by
%
%\begin{equation}
%t_{\mathrm{open}}= \left(3\xi\right)^{2/3}\frac{(1+\epsilon)(1+\mathrm{St}^2)}{\mathrm{St}^2(1+\epsilon)+\epsilon}  \left( \frac{M_{\rm p}}{M_{\star}} \right)^{-1/3} \Omega_{\rm k}^{-1} .
%\label{eq:topengap}
%\end{equation}
%
\begin{equation}
t_{\mathrm{open}}= \left(3\xi\right)^{2/3}  \left( \frac{M_{\rm p}}{M_{\star}} \right)^{-1/3} \Omega_{\rm k}^{-1} .
\label{eq:topengap}
\end{equation}
%
%The tidal torque appears weighted by a factor that depends on $\mathrm{St}$ and $\epsilon$, taking into account the different grains dynamics in the various aerodynamical regimes. 
Setting $\Lambda_{\rm g} =\Lambda_{\rm d} = 0$ in Eq.~\ref{eq:outedge} and assuming $v_{\mathrm{d},r}$ constant over the domain $ \left | r-r_{\mathrm p}\right |\sim r_{\mathrm{H}}$, the closing time $t_{\rm close} = r_{\rm H} / v_{\mathrm{d},r}$ is
\begin{equation}
t_{\mathrm{close}}=  \frac{(1+\epsilon)(1+\mathrm{St}^2)}{- \zeta \mathrm{St}  + \left( 6 + 3\zeta \right)\alpha} \frac{v_{\rm k}}{c_{\rm s}^{2}}r_{\mathrm{H}}  .
\end{equation}
%\subsubsection{Sufficient condition for dust gap opening}
The critical mass ratio $M_{\mathrm{p}}/M_{\star}$ above which a planet sustains its gap in a dust disc is obtained by equating the opening and the closing timescale, which gives
\begin{equation}
\frac{M_{\rm p}}{M_{\rm \star}} \ge 3^{3/2} \xi \, \left(\frac{z }{\mathrm{St}}\right)^{3/2}  \left(\frac{H}{r_{\rm p}}\right)^3,
\label{eq:sufficientcondition}
\end{equation}
with $z$ given by the second expression in Eq.~\ref{eq:def_f}. As already mentioned, the two expressions in Eq.~\ref{eq:def_f} have the same value for our aims, i.e. for $\mathrm{St} \gtrsim \alpha$ and for $\mathrm{St} \gtrsim \epsilon$. Thus,  Eq.~\ref{eq:sufficientcondition} equals to $3^{3/2} \approx 5.2$ times the critical mass derived in Eq.~\ref{eq:necessarycond}. For the disc model described above the criterion gives a typical mass of $\sim0.015 \,M_{\mathrm J}$ to open a dust gap.
%. 
%The critical Stokes number over which the planet carve gaps only in the dust is close to unity.  
For large grains, Eq.~\ref{eq:sufficientcondition} reduces to
\begin{equation}
\frac{M_{\rm p}}{M_{\rm \star}} \ge \xi \, \left(\frac{-3\zeta}{1 + \epsilon}\right)^{3/2}  \mathrm{St}^{-3/2}\left(\frac{H}{r_{\rm p}}\right)^3 .
\label{eq:sufficientcondapprox}
\end{equation}
This criterion provides a good estimator for the minimum mass to open a gap in dusty disc. As a remark, Eq.~\ref{eq:necessarycond} originates from a balance of torques performed at steady-state (the gap is already opened), whereas Eq.~\ref{eq:sufficientcondition} originates from a balance of torques performed in a transient regime (the gap is not opened yet). This explains why the two conditions are not rigorously identical and differ by a factor of order unity. Note that, since $\mathrm{St}\propto1/\left(1+\epsilon\right)$, the gap opening criterion does not depend on the dust-to-gas ratio, assuming that the dust back reaction does not affect the local pressure profile. Therefore, our analysis shows that it is not necessary to estimate the local dust-to-gas ratio to derive the value of the minimum mass.

\begin{figure}
\begin{center}
\includegraphics[width=0.49\textwidth]{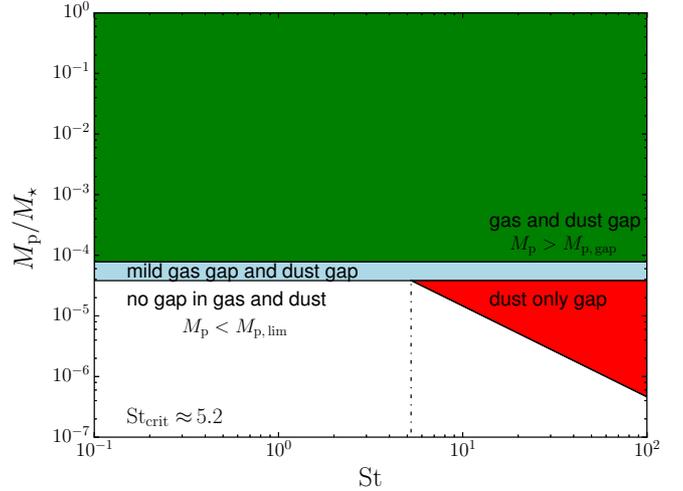}
 \caption{Sufficient condition for dust gap opening in a typical disc model with a local aspect ratio of 0.05 and $\zeta=-2.75$ for different Stokes numbers, adopting  $f=0.4$ \citep{goldreich79a}. The shaded areas indicates the range of planet masses and Stokes number for which a gap is carved in both phases (green) and only in the dust (red). The small blue area indicates the range of masses for which the local pressure profile is perturbed without creating a pressure maximum.
 The range of masses able to carve a gap in the dust only increases with the Stokes number.}
\label{fig:criterion}
\end{center}
\end{figure}

Fig.~\ref{fig:criterion} displays the minimum mass required for a planet to open a gap in the dust as a function of the Stokes number. This limit is calculated from Eq.~\ref{eq:sufficientcondition} in the disc model described above assuming $f=0.4$ \citep{goldreich79a}. %, $H / r_{\rm p} = 0.05$, $\epsilon=0.01$, $p=1$ and $q=1/2$, giving $\zeta=-2.75$. 
The red shaded area indicates the range of planet masses and Stokes numbers for which a gap is carved in the dust only. There is no ubiquitous lower mass for gap opening in the dust, since tides always overpower drag in the limit of very large and decoupled grains. The green area shows the domain for which the planet carves a gap in the gas as well, i.e. for $M_{\mathrm{p}}\geq M_{\mathrm{p, gap}}$. The small blue area indicates the range of masses $M_{\mathrm{p,lim}}\lesssim M_{\mathrm{p}}\lesssim M_{\mathrm{p,gap}}$ for which the local pressure profile is perturbed without creating a pressure maximum \citep{rosotti16a}. %A mild gap is produced in the gas and a gap structure in the dust under the combined action of the tidal torque and the reduced radial dust velocity \citep{rosotti16a}.}

As expected, the range of masses for which the planet is able to carve a gap in the dust only (red area) increases with increasing Stokes numbers, due to the reduced replenishment from the outer disc induced by the drag torque. For planets with masses inside the green area, the drag assists the gap opening in the dust, leading to an accumulation of dust particles at the pressure maximum and producing a well-defined dusty gap with a shape closely related to the Stokes number \citep[e.g.][]{pinilla15a}.

Eventually, the exact value of the minimum mass is related to the constant $f$ in front of the tidal torque density (Eq.~\ref{eq:torquegrav}). Eq.~\ref{eq:sufficientcondapprox} shows that the critical mass to open a gap in the dust is sensible to the actual value of $f$ as it varies as $f^{-3/2}$ (see Eq.~\ref{eq:defxi}), as long as the hypothesis of the low-mass planet regime is satisfied.  Measuring the critical mass provides therefore an effective way to measure $f$ in numerical simulations (see Sect.~\ref{sec:planet_mass}).

%The value is very close to what obtained using the sufficient condition (0.007 $M_{\mathrm J}$, Eq.~\ref{eq:cubic}).

%\begin{figure*}
%\begin{center}
%\includegraphics[width=0.35\textwidth]{criterion_honr_alpha0001new.eps}
%\includegraphics[width=0.3205\textwidth]{criterion_honr_alpha001new.eps}
%\includegraphics[width=0.3205\textwidth]{criterion_honr_alpha01new.eps}
%\caption{Sufficient condition of gap opening assuming different values of the viscosity $\alpha_{\rm SS}$ and aspect ratio at the planet location. In the viscous force dominates the gap closing mechanism, the value of the critical Stokes number decrease with increasing viscosity and decreasing aspect ratio (see Eq.~\ref{eq:critstokes}).}
%\label{fig:criterion_hover}
%\end{center}
%\end{figure*}

\subsection{Critical Stokes number}
\label{subsec:examples}
An important feature mentioned in Sect.~\ref{sec:odg} appears clearly: low mass planets can carve dust gaps only if the Stokes number is above a critical value of order unity. The value of the critical Stokes number is obtained by comparing the minimum mass given by Eq.~\ref{eq:sufficientcondapprox} to $M_{\mathrm{p, lim}}$, the minimum mass able to perturb the gas pressure profile. Since the value of the critical Stokes number is expected to be higher than unity, we assume $\mathrm{St} \gtrsim \alpha$ and $\mathrm{St} \gtrsim \epsilon$ and obtain
%
%\begin{equation}
%\mathrm{St_{crit}} =   
%\begin{cases}
%\displaystyle  3z\left(\frac{\xi}{0.3}\right)^{2/3} , & \text{$M_{\mathrm{p, lim}}=0.1\,M_{\mathrm{p,th}}$,} \\ 
%\displaystyle  2z\left(\frac{\xi^{2}}{\pi}\right)^{1/3} \alpha^{-1/3}\left( \frac{H}{ r_{\mathrm{p}}} \right)^{1/3}, & \text{$M_{\mathrm{p, lim}}=M_{\mathrm{p,visc}}$,}
%\end{cases}
%\label{eq:critstokes}
%\end{equation}
\begin{equation}
\mathrm{St_{crit}} \simeq   3\left(\frac{\xi}{0.3}\right)^{2/3}\left(  \frac{-\zeta}{1 + \epsilon} \right) .
\label{eq:critstokes}
\end{equation}
For a typical disc with $\zeta=-2.75$ and $\epsilon=0.01$, $\mathrm{St_{crit}} \sim 5.2$ (see Fig.~\ref{fig:criterion}).
%The small exponents $\pm 1/3$ involved make the the value of the critical Stokes number is very close to unity. When the gap opening criterion for the gas is dominated by pressure forces, i.e. $M_{\mathrm{p, lim}}=0.1\,M_{\mathrm{p,th}}$, 
Importantly, the value of the critical Stokes number does not depend on $H/r_{\mathrm{p}}$ since the gap opening criterion for the gas and the dust scales equally with the aspect ratio. Moreover, Eq.~\ref{eq:zeta} shows that $\mathrm{St_{crit}}$ scales linearly with the power-law exponent of the pressure profile and is proportional to $\left(1+\epsilon \right)^{-1}$. Since $\mathrm{St} \propto\left(1+\epsilon \right)^{-1}$ as well, the value of the critical grain size does not depend on $\epsilon$. %Eventually, planet migration would also increase the mass of the planet required to carve a gap in the gas and hence, decrease the value of the critical Stokes number.

\begin{figure*}
\begin{center}
\includegraphics[width=0.33\textwidth]{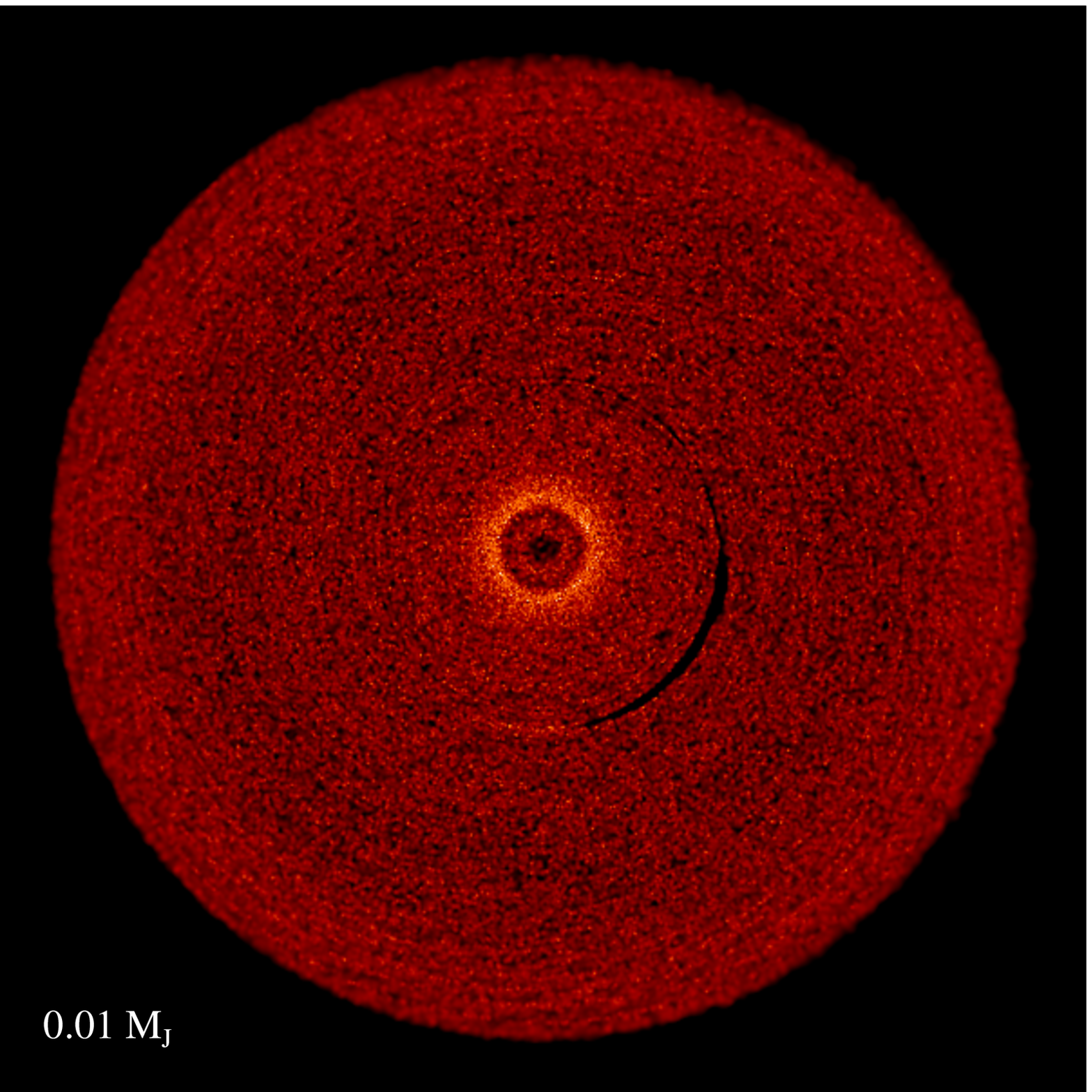}
\includegraphics[width=0.33\textwidth]{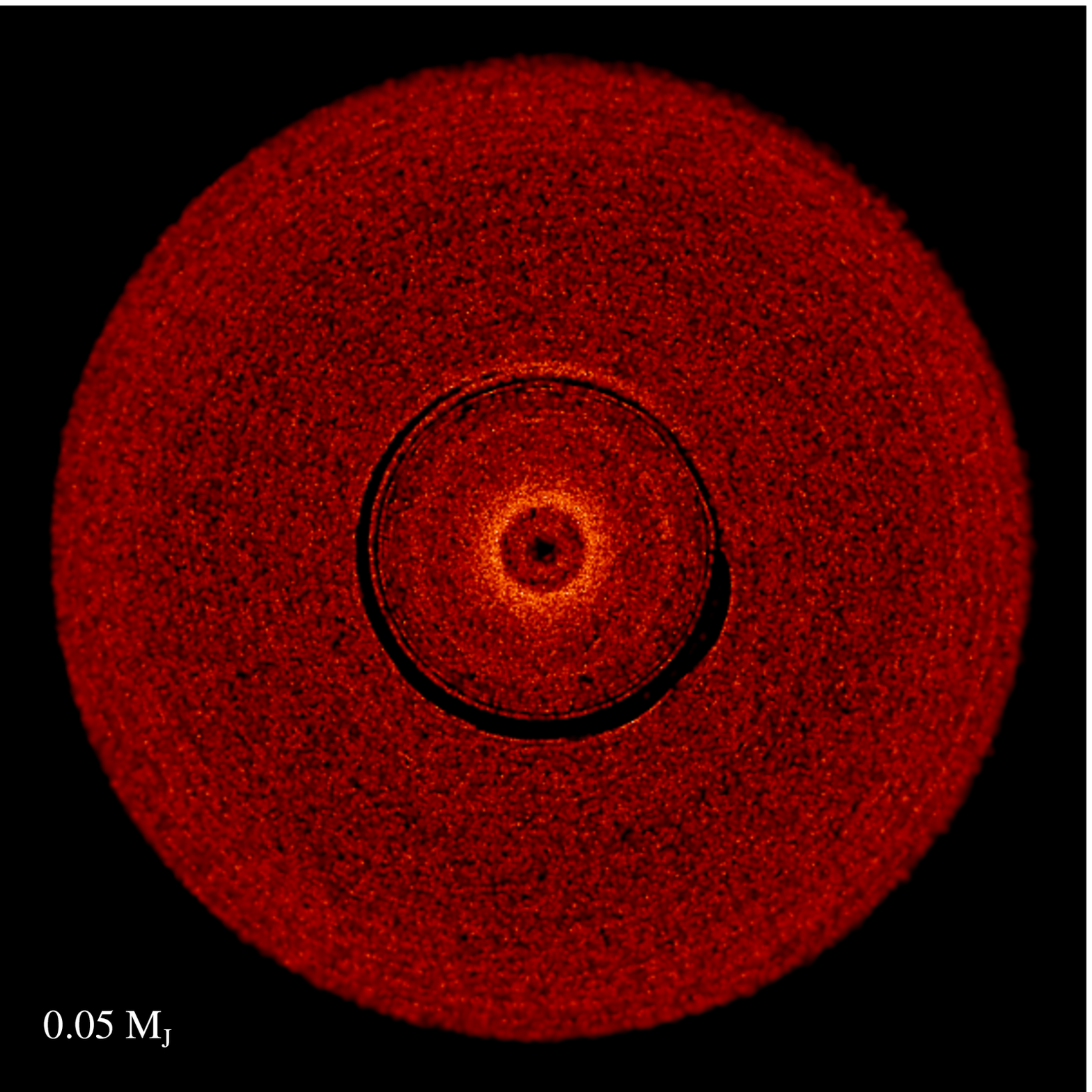}
\includegraphics[width=0.33\textwidth]{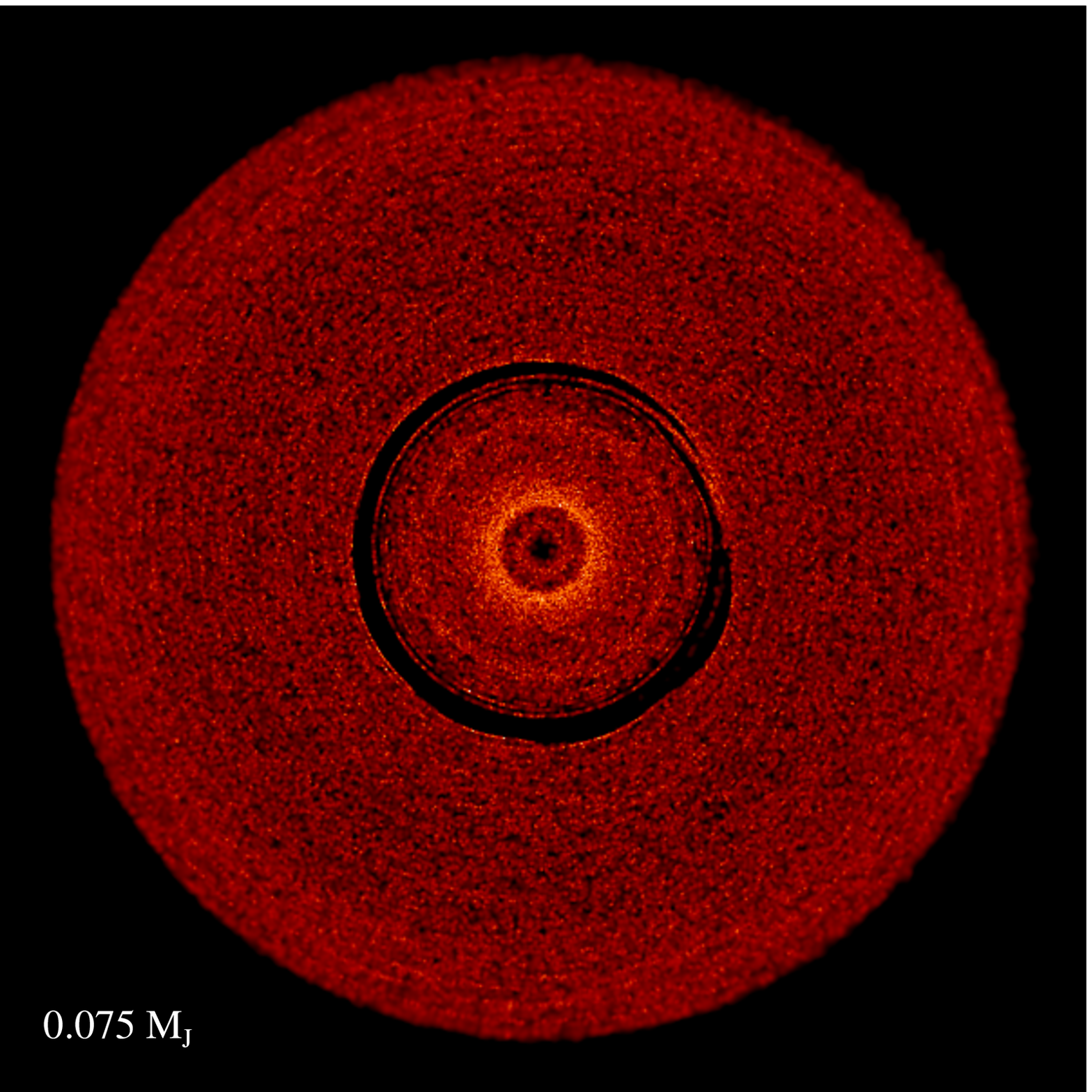}
\includegraphics[width=0.33\textwidth]{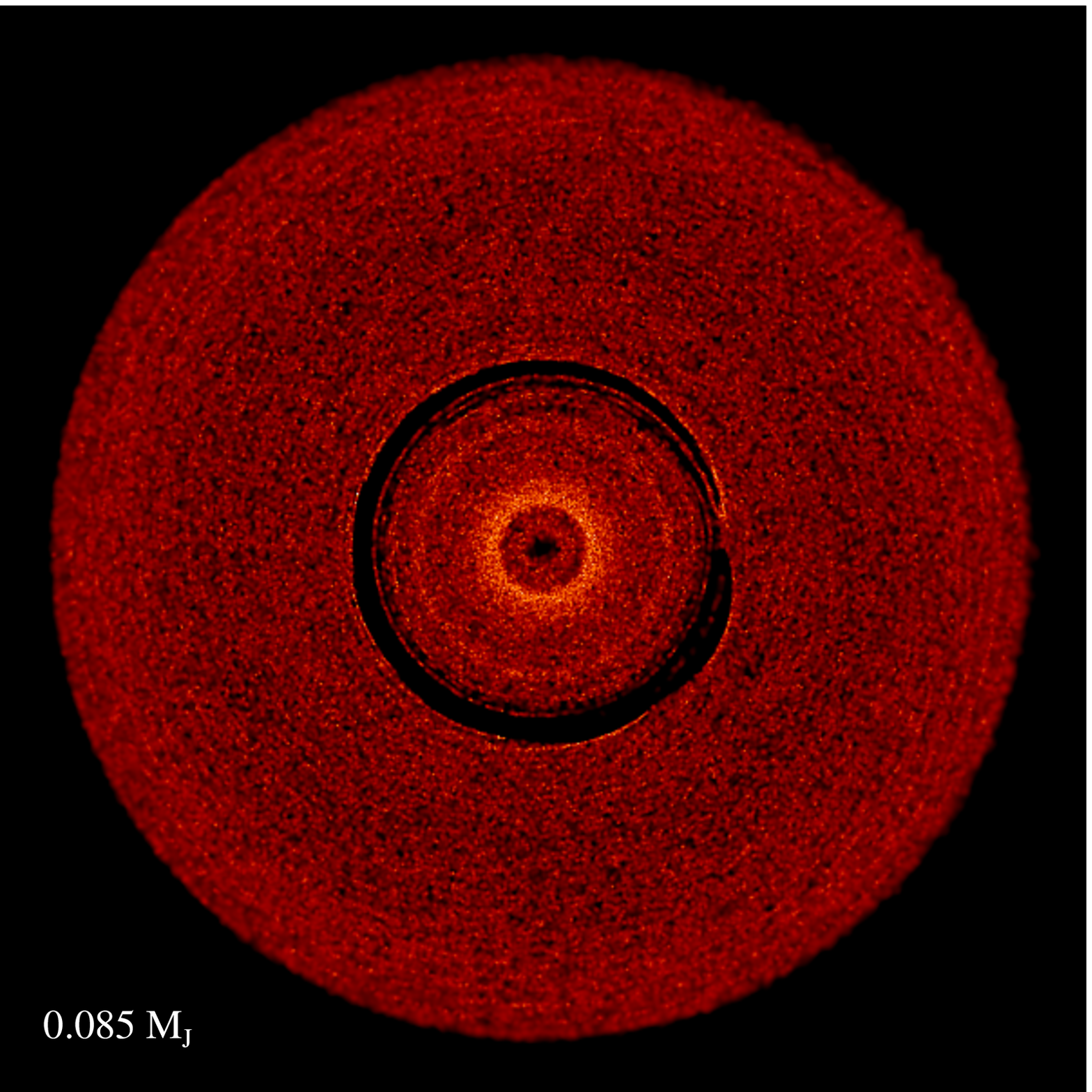}
\includegraphics[width=0.33\textwidth]{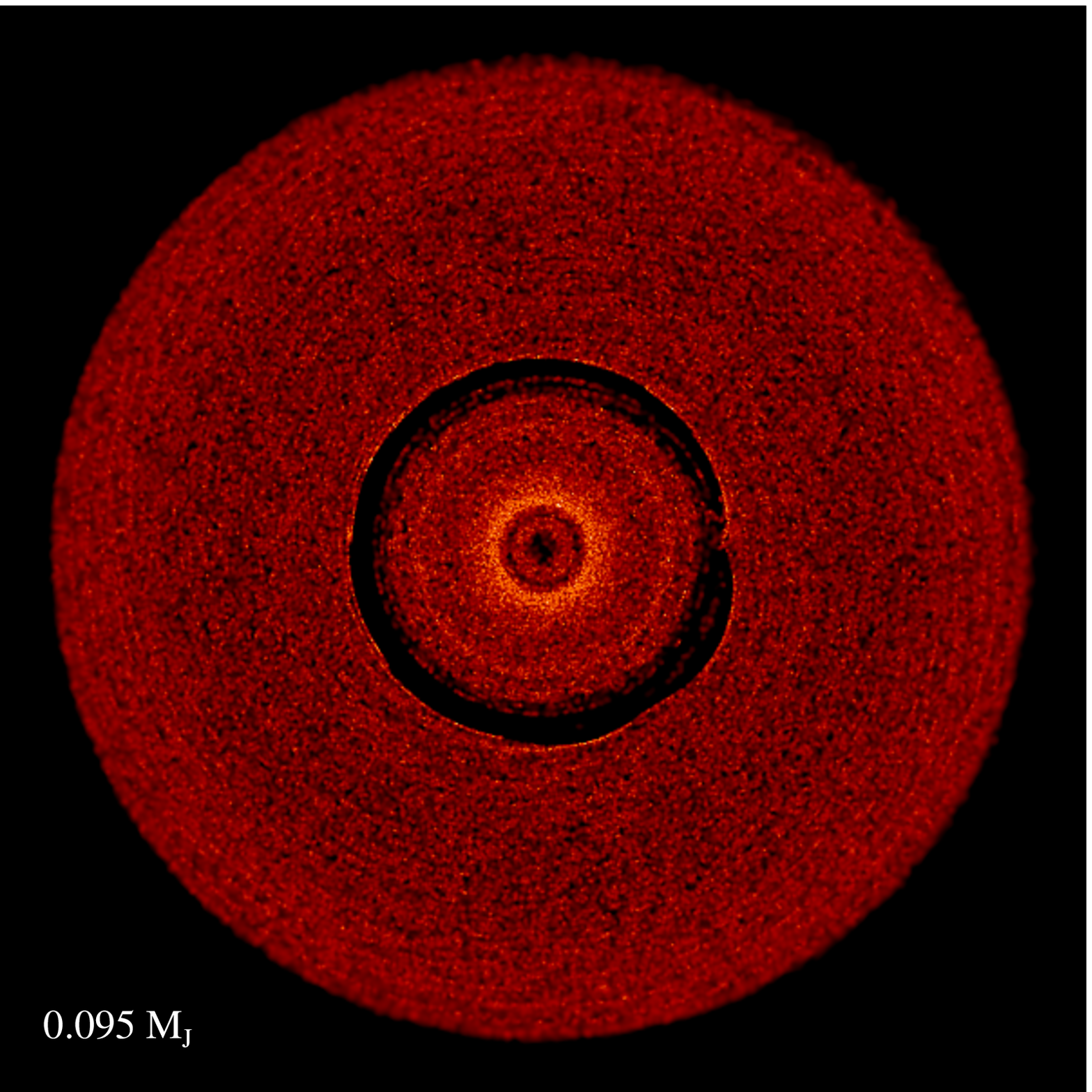}
\includegraphics[height=0.33\textwidth]{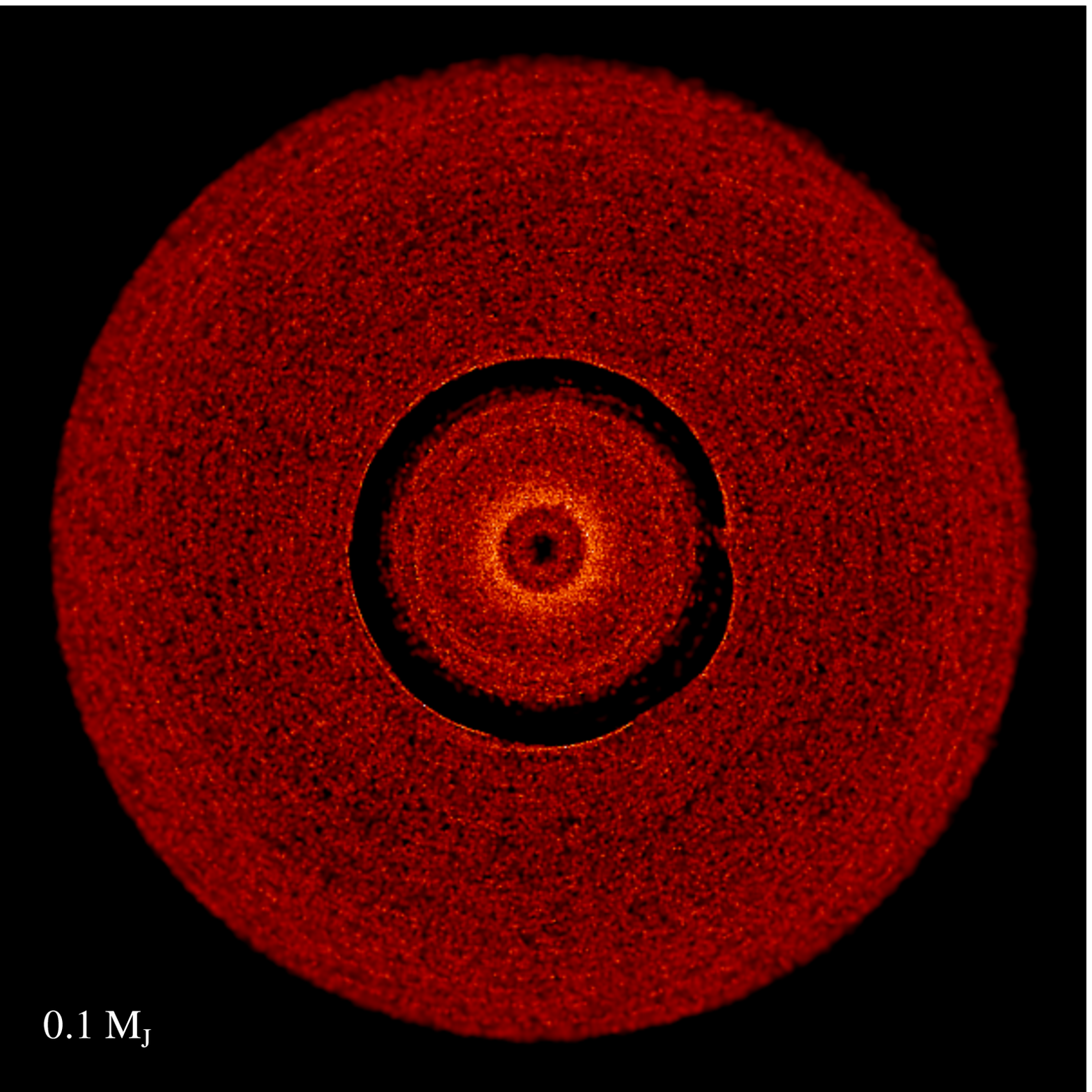}
\includegraphics[height=0.33\textwidth]{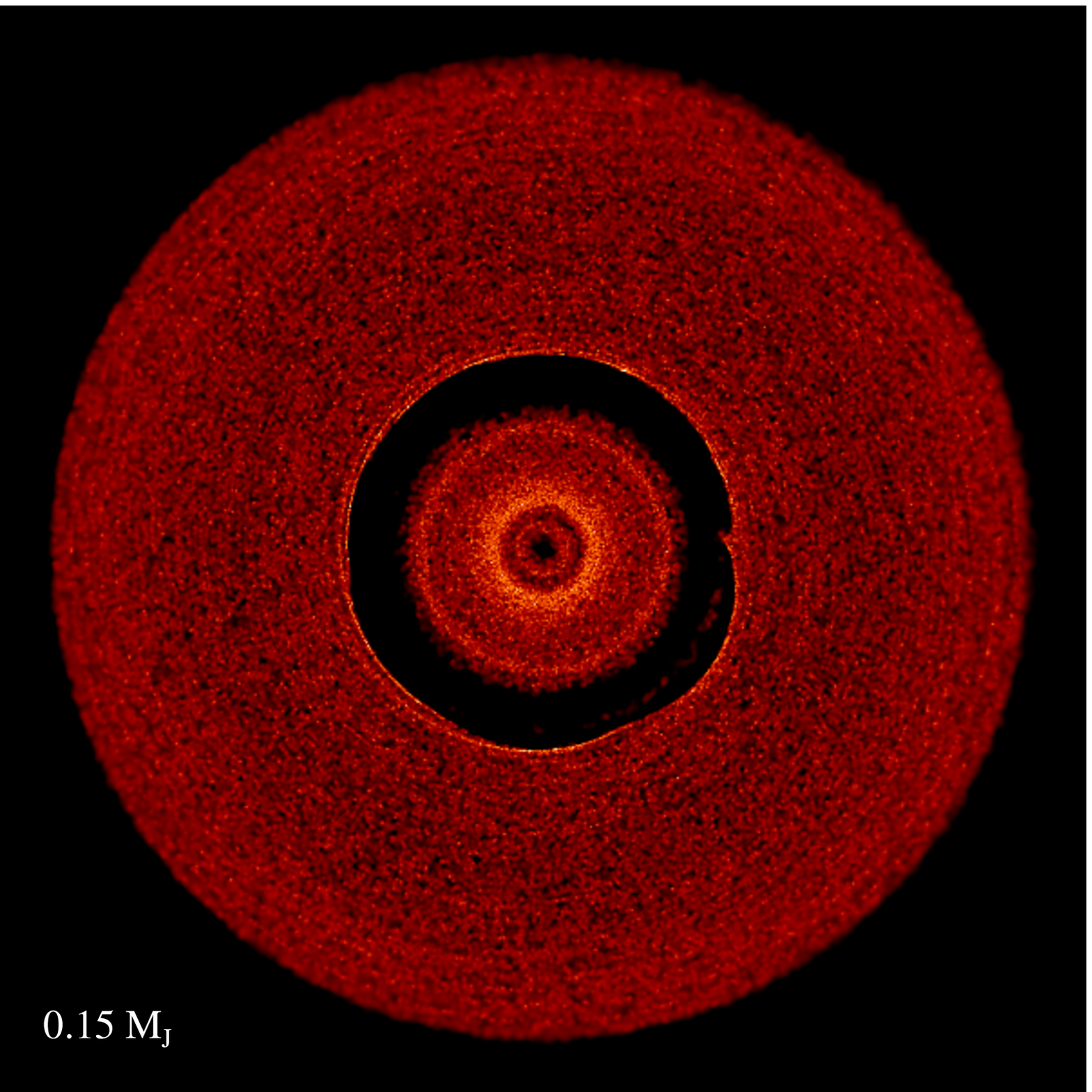}
\includegraphics[height=0.33\textwidth]{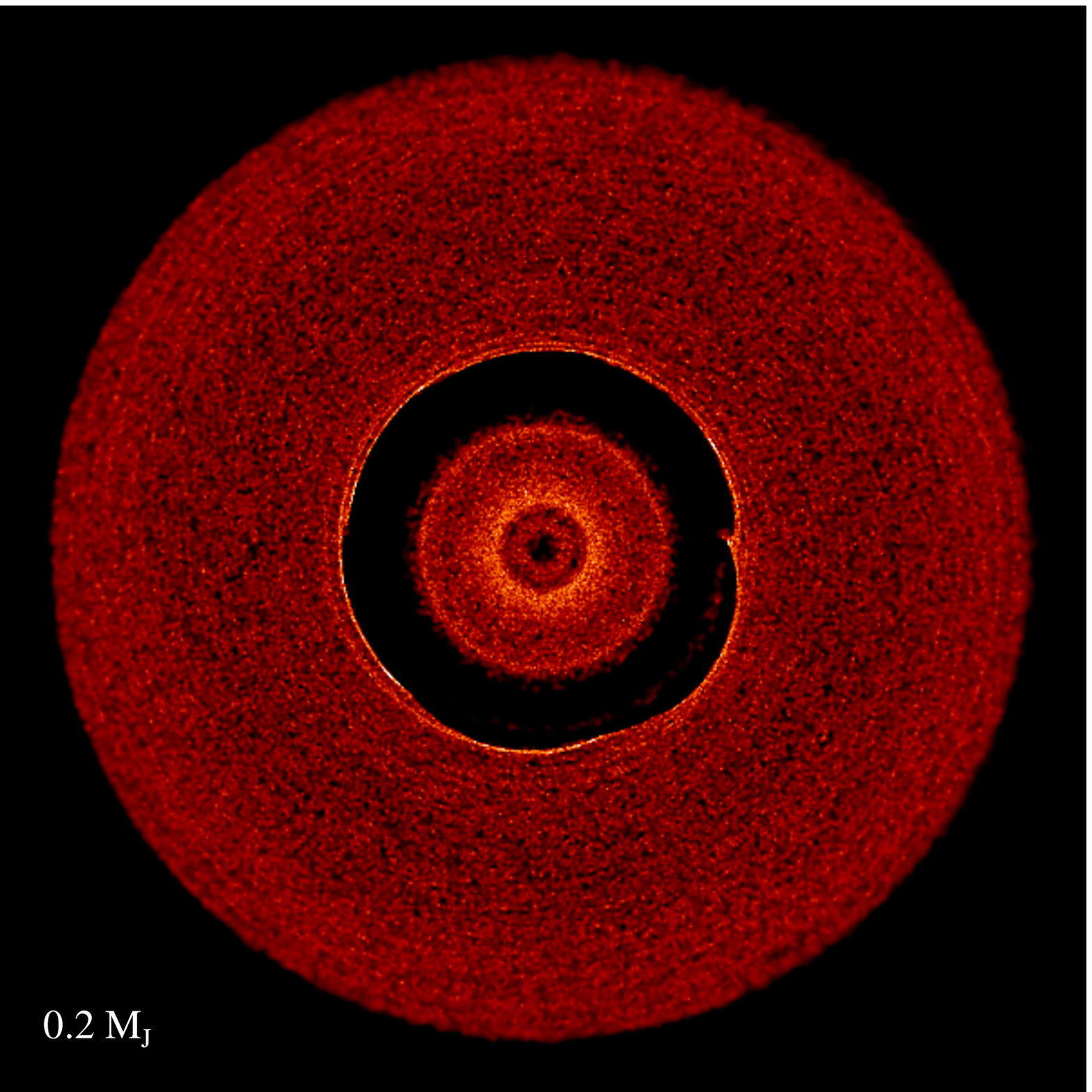}
\includegraphics[height=0.33\textwidth]{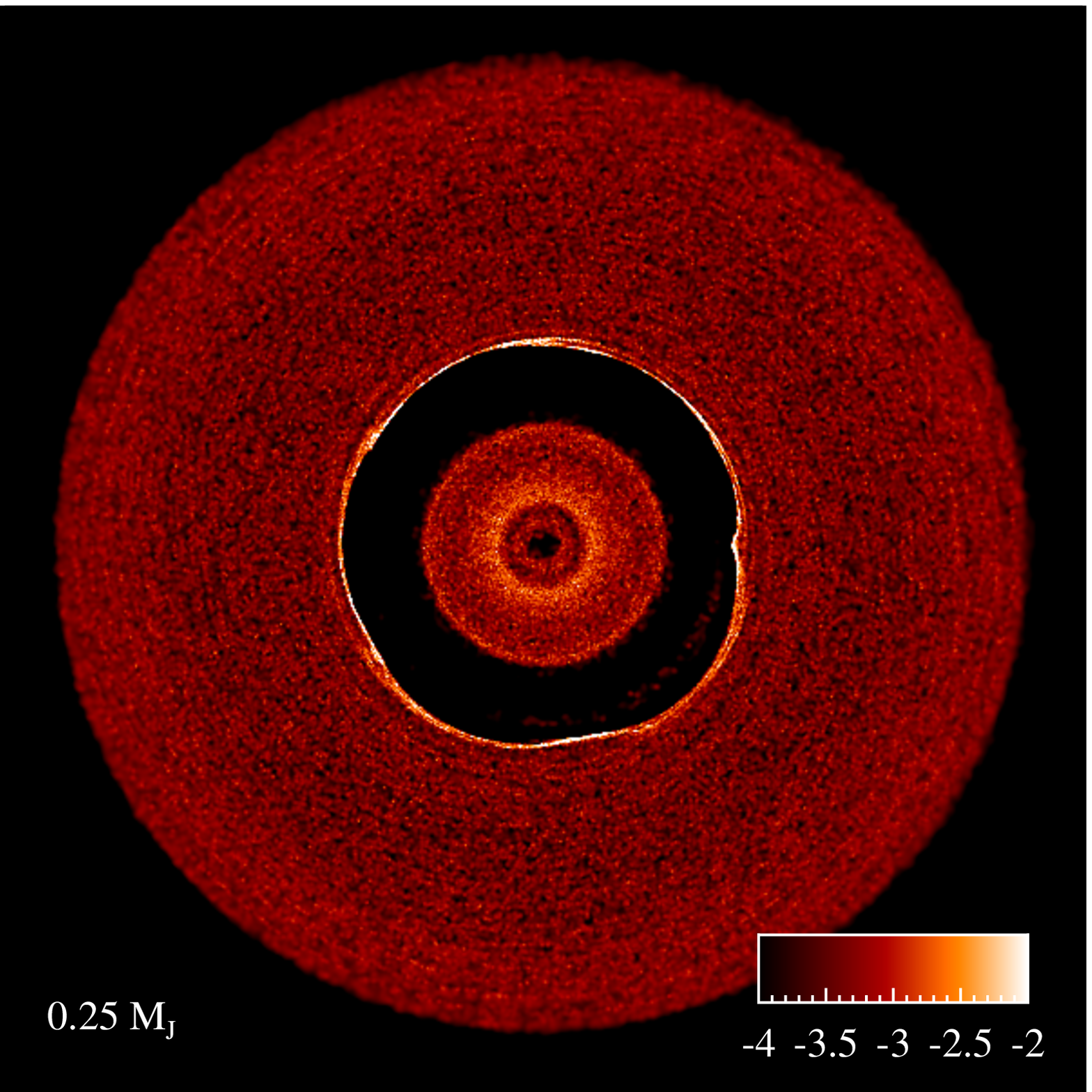}
\caption{Rendered images of the steady-state dust surface density of millimetre size grains for the disc model described in Sect.~\ref{subsect:initialcond} hosting embedded planets of mass 0.01 (top-left), 0.05 (top-center), 0.075 (top-right), 0.085 (mid-left), 0.095 (mid-center), 0.1 (mid-right), 0.15 (bottom-left), 0.2 (bottom-center) and 0.25 (bottom-right) $M_{\rm J}$ initially located at 40 au after 40 planetary orbits. A full-cleared gap is carved by a planet with a minimum mass of $\sim 0.09\,M_{\rm J}$. From Eq.~\ref{eq:sufficientcondapprox}, we infer $f=0.28$.}
\label{fig:changemass}
\end{center}
\end{figure*}
\begin{figure*}
\begin{center}
\includegraphics[width=0.48\textwidth]{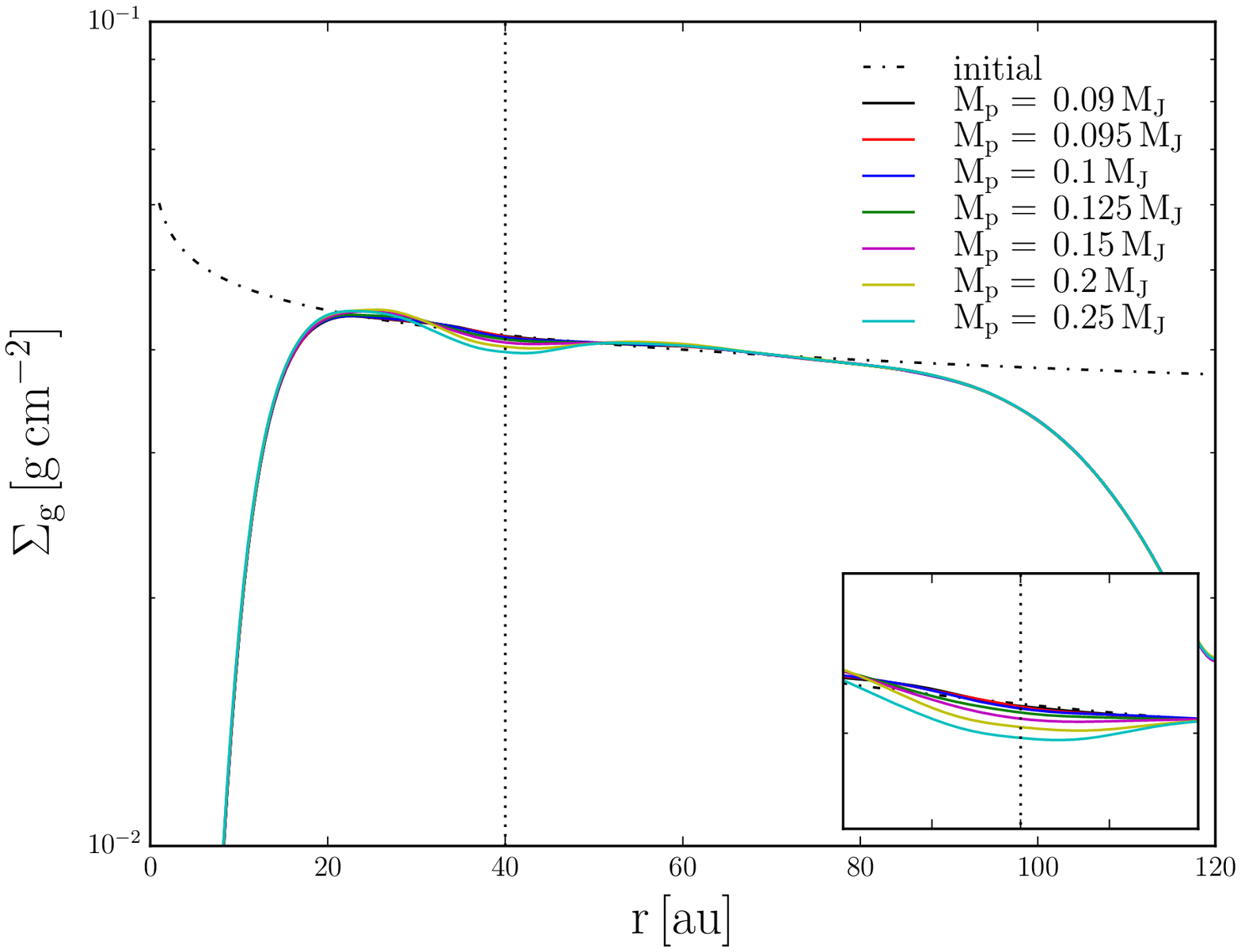}
\includegraphics[width=0.49\textwidth]{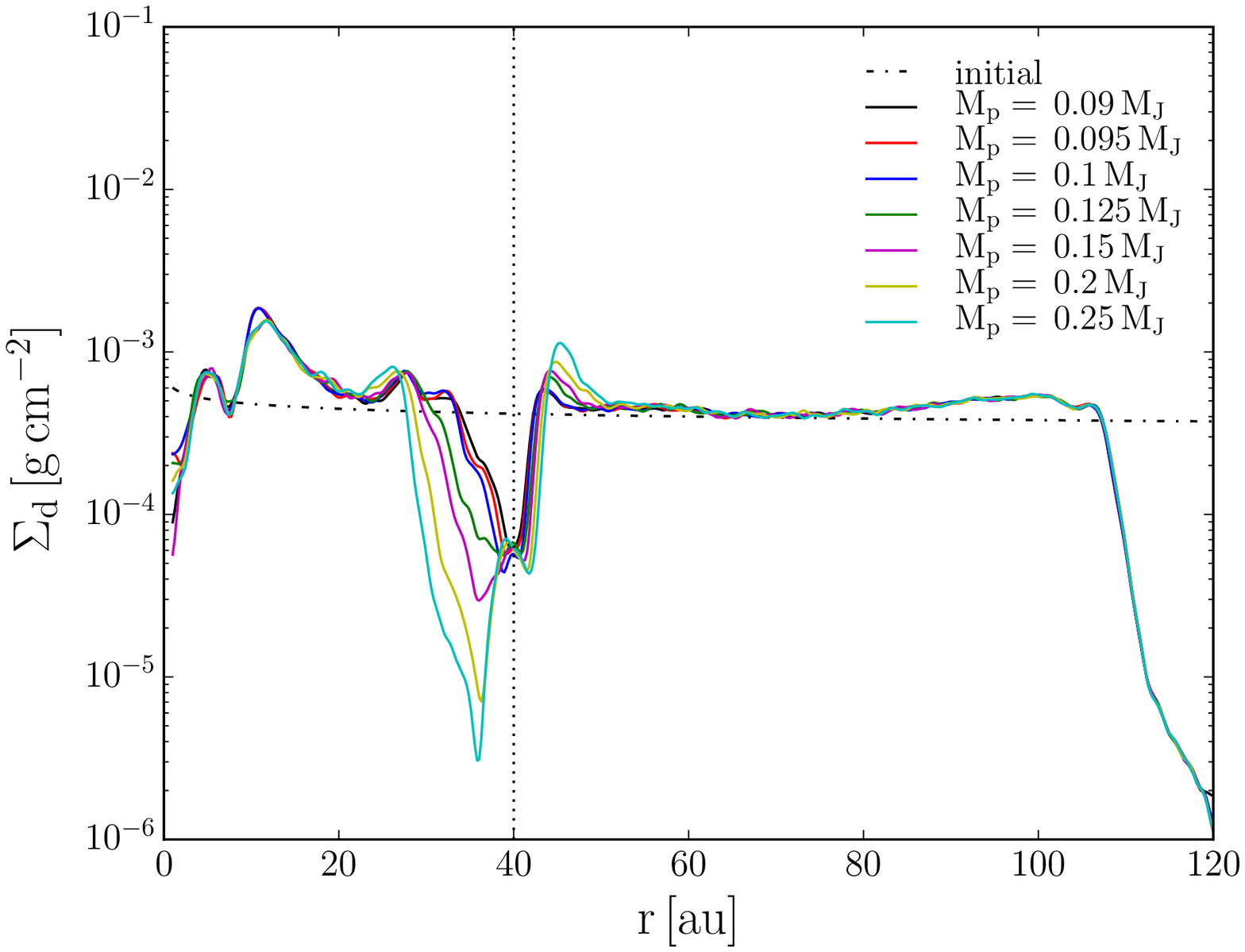}
\includegraphics[width=0.48\textwidth]{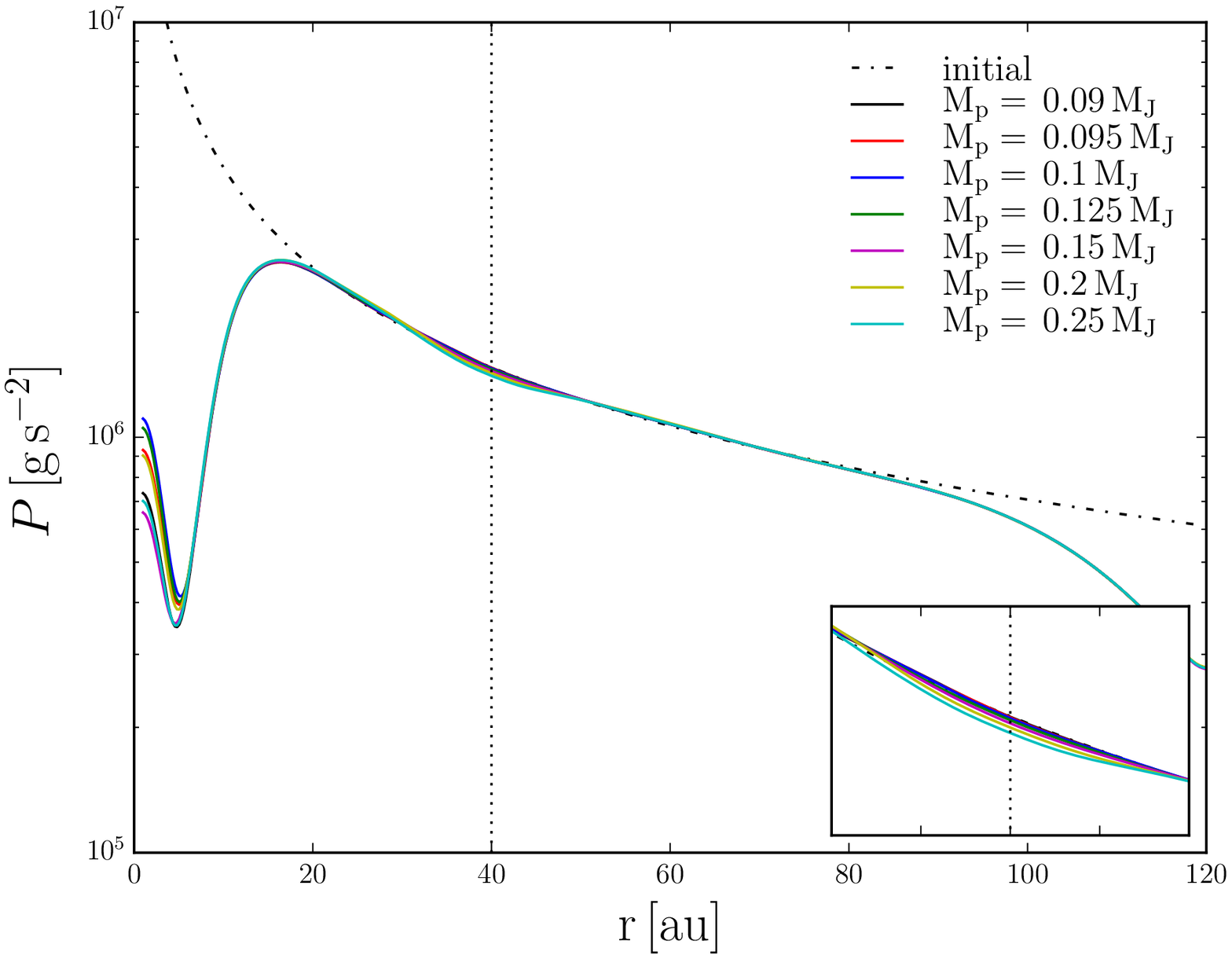}
\includegraphics[width=0.49\textwidth]{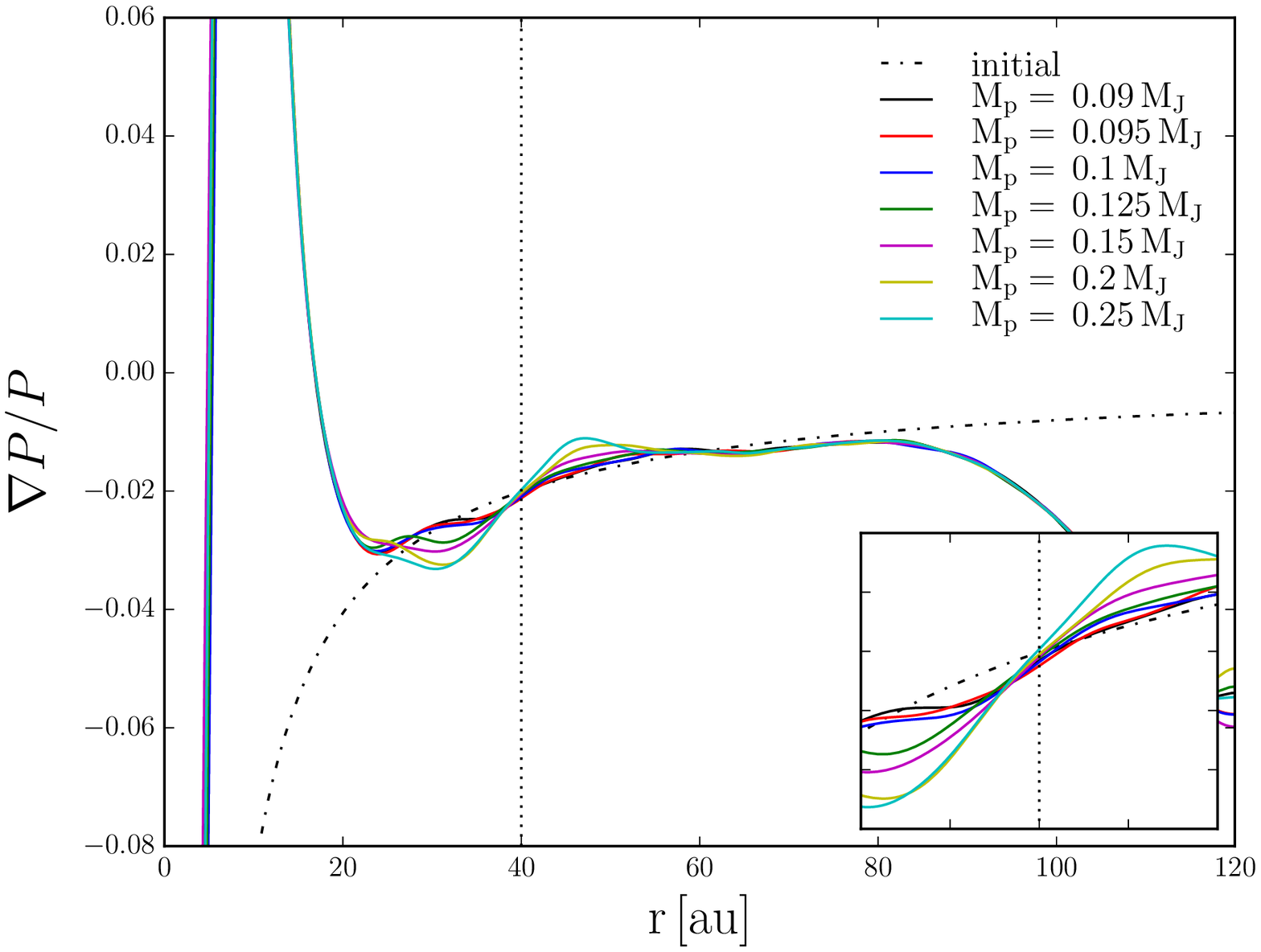}
\caption{Azimuthally averaged surface density radial profiles of (top-left) gas and dust (top-right)  for the disc hosting a planet with different mass. The bottom panels show the (left) pressure radial profile and (right) its gradient.
The dotted vertical line indicates the planet orbit. For planet of masses $0.09 M_{\mathrm{J}}\lesssim M_{\mathrm{p}}\lesssim 0.1 M_{\mathrm{J}}$ the pressure profile is not perturbed by the presence of the planet, while more massive planet perturb the local pressure profile, leading to a larger and deeper dust gap. }
\label{fig:profiles}
\end{center}
\end{figure*}

%
%----------------------------------------------------------------------------------------------------------------
\section{Numerical simulations}
\label{sec:testcriterion}

\subsection{Dust/gas simulations}
We use the SPH code \textsc{phantom} to perform 3D global simulations of gas/dust discs containing an embedded protoplanet \citep{price17a}. Importantly, every process involved in the physical problem (viscosity gravity and drag) is computed self-consistently. A calibrated non-zero viscosity is applied on each gas particle \citep{lodato10a}, to mimic the viscous transport of gas in disc described with an Prandtl-like model of turbulence \citep{shakura73a}. The star and the planet are treated as mobile point sources of mass. Their gravitational interactions with the gas/dust particles are computed using the sink particles approach \citep*{bate95a}, which allows the bodies to migrate from their interactions with the disc. Dust and gas particles are accreted onto the sinks when two conditions are fulfilled: i) the SPH particle is found to be gravitationally bound the sink, and ii) the divergence of the velocity field at the location of the particle is negative. The gravitational acceleration between the $n_{\mathrm{sinks}}$ sink particles and the $i$-th gas/dust SPH particle is computed according to
\begin{equation}
\frac{\mathrm{d} \mathbf{v}_{i}}{\mathrm{d}t}=-\sum_{j=1}^{n_{\mathrm{sinks}}}\frac{\mathcal{G} M_j}{\left( \left|\mathbf{r}_{ij}\right|^2+s_j^2 \right)^{3/2}} \mathbf{r}_{ij} ,
\end{equation}
where $\mathbf{r}_{ij} = \mathbf{r}_{i} -  \mathbf{r}_{j}$ denotes the differential location between the particles, $M_j$ is the mass of the $j$-the sink particle and $s_j$ is the usual softening parameter which prevents singularities at the sink locations. $s_{j}$ is also chosen to be the accretion radius of the sink particle. %We discuss how the choice of this value affects the planet-disc interaction in Sect.~\ref{sec:plan_acc}. 

The dust motion is computed using the two fluid algorithm described in \citet{laibe12a}. The drag force between a particle of one type and its neighbours of the other type is calculated in a pairwise manner, to conserve the linear and the angular momentum of both phases as well of the energy of the gas to machine precision. In particular, the drag from the dust onto the gas (sometimes referred as ``back-reaction'') is included self-consistently. A specific double-hump drag kernel ensures the accuracy of the interpolation. To ensure better resolution within the gap, the smoothing length of the gas is used to compute drag terms. Outside of the gap, results do not depend from this choice. The algorithm has been extensively benchmarked on simple test problems including waves and shocks in dust and gas mixtures \citep{laibe11a,laibe12a,price15a}. We model spherical, compact and uncharged grains of constant sizes. The drag coefficient is computed consistently, based on the local values of the Knudsen number, as well as the Reynolds and the Mach number of the relative flow between the two phases \citep{kwok75a,paardekooper06a,laibe12a}. %We additionally perform simulations where gravity from the planets in the dust phase and the dust-gas aerodynamical drag is artificially turned off. This allows to disentangle the contributions of the different torques.

\begin{figure}
\begin{center}
\includegraphics[width=0.48\textwidth]{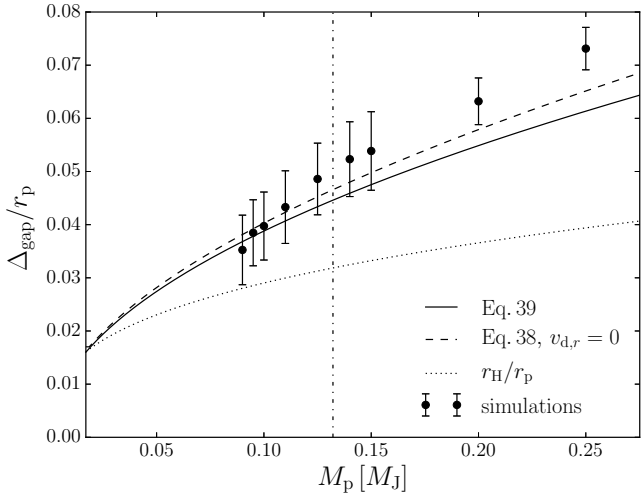}
\caption{Positions of the outer gap edge $\Delta_{\mathrm{gap}}=r_{\mathrm{gap}}-r_{\mathrm{p}}$ for different planet mass in our full sample. The bold dots indicates the estimates from the surface density profile computed by simulations, whereas the solid line are the predictions of the theory (Eq.~\ref{eq:redgeapprox}). The dashed lines are the numerically evaluation of the gap outer edge by calculating where the radial dust velocity is null taking into account the radial profiles of the surface density and temperature far from the planet by computing the null values of Eq.~\ref{eq:outedge}.
The vertical dashed line denotes the limit in planet mass ($M_{\mathrm{p,lim}}$, Eq.~\ref{eq:mplim}) below which the analysis is valid. The dotted line indicates the Hill radius that can be considered the minimum width of the gap that can be carved in the dust. Our analysis on the gap outer edge are consistent with the results of simulations for low massive planets, while for larger masses it gives an underestimate.}
\label{fig:reedgecommpl}
\end{center}
\end{figure}
\begin{figure*}
\begin{center}
\includegraphics[width=0.33\textwidth]{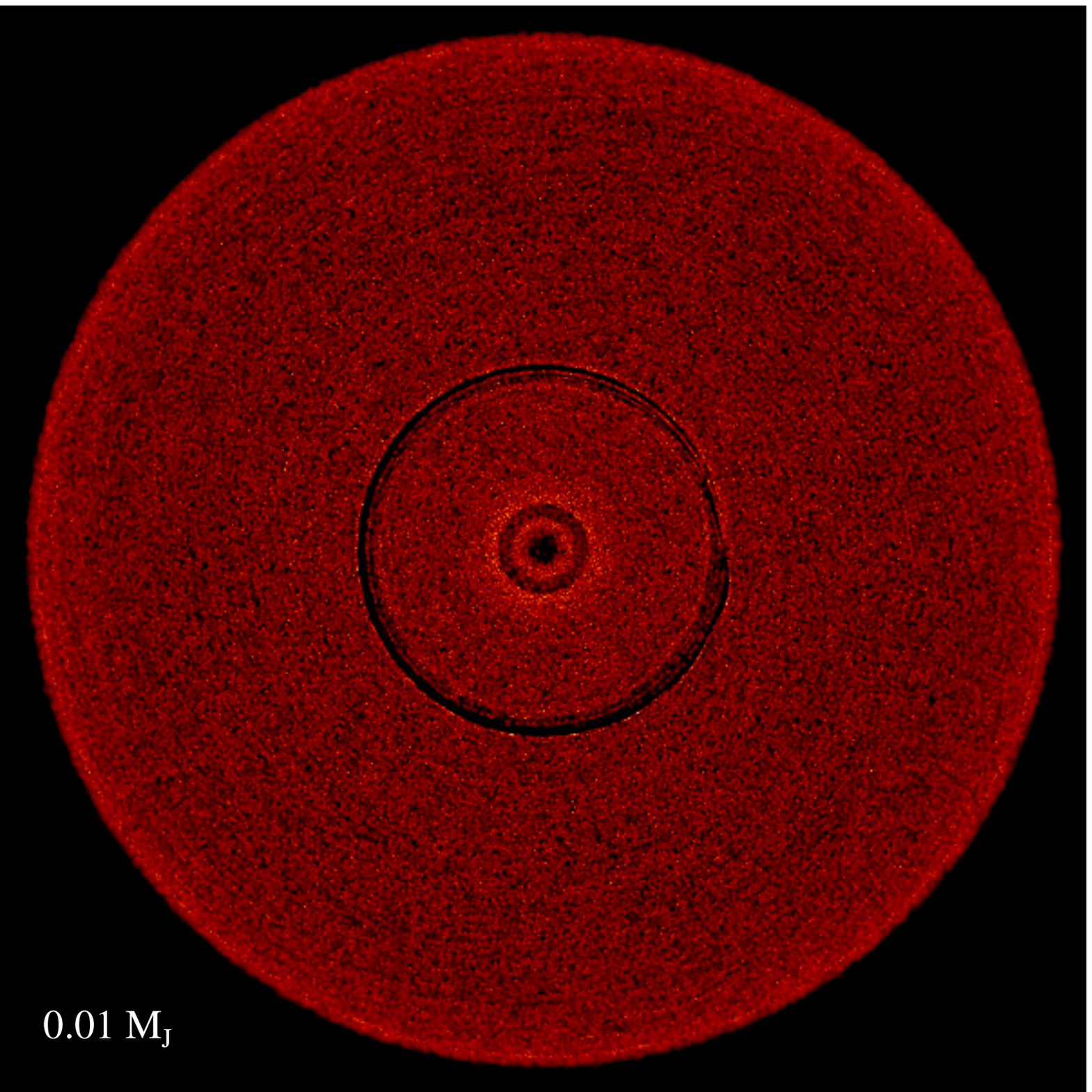}
\includegraphics[width=0.33\textwidth]{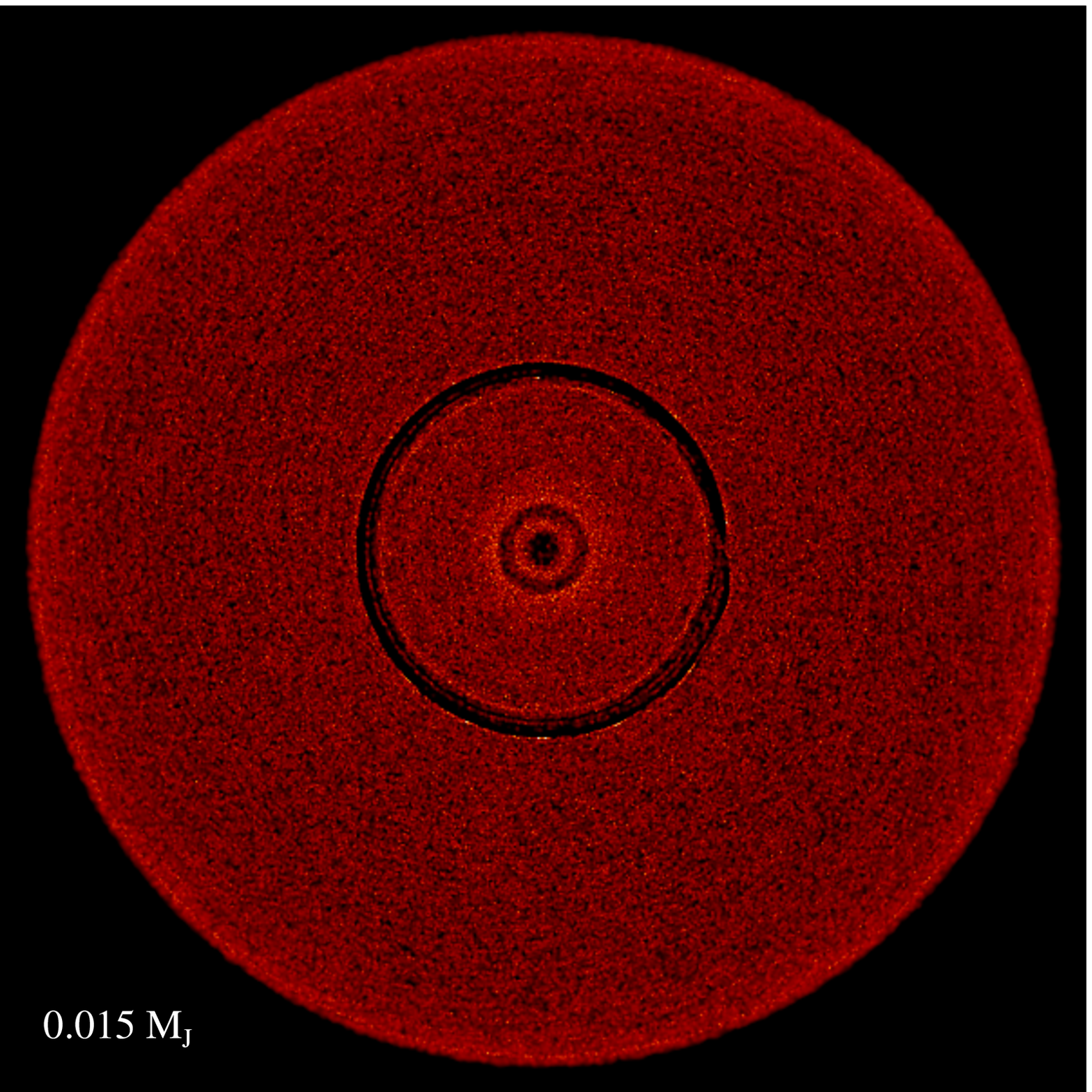}
\includegraphics[width=0.33\textwidth]{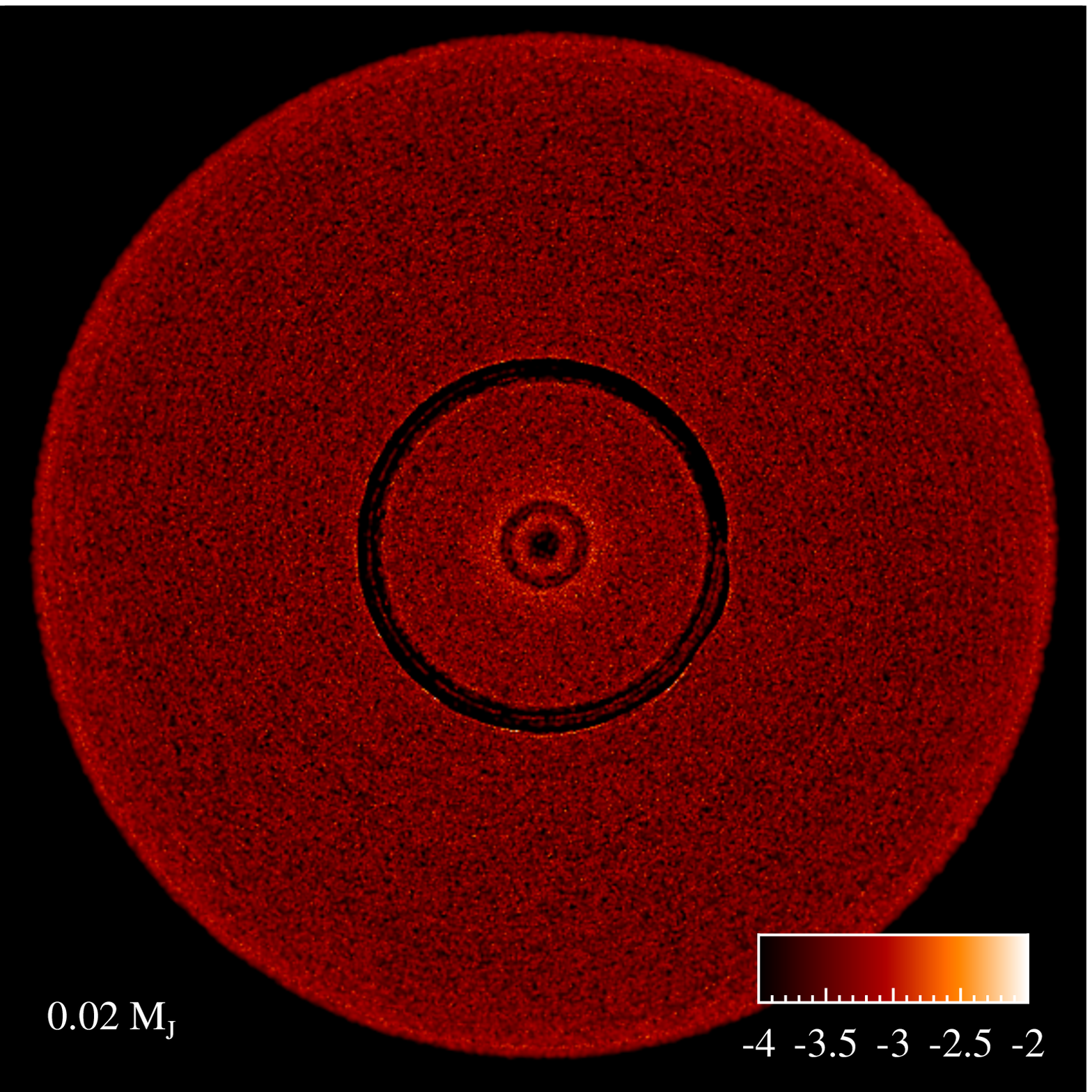}
\caption{Rendered images of dust surface density for the disc model described in Sect.~\ref{subsect:initialcond} but with an aspect ratio $H/r = 0.02$ at 1 au hosting a planet with mass 0.01 (left), 0.015 (center) and 0.02 (right) $M_{\rm J}$ initially located at 40 au after 40 planetary orbits. A planet with mass $\grtsim 0.01\, M_{\mathrm{J}}$ ($3\, M_{\oplus}$) is able to carve dust gaps, consistent with our analysis.}
\label{fig:changemasshr002}
\end{center}
\end{figure*}
\begin{figure*}
\begin{center}
\includegraphics[width=0.333\textwidth]{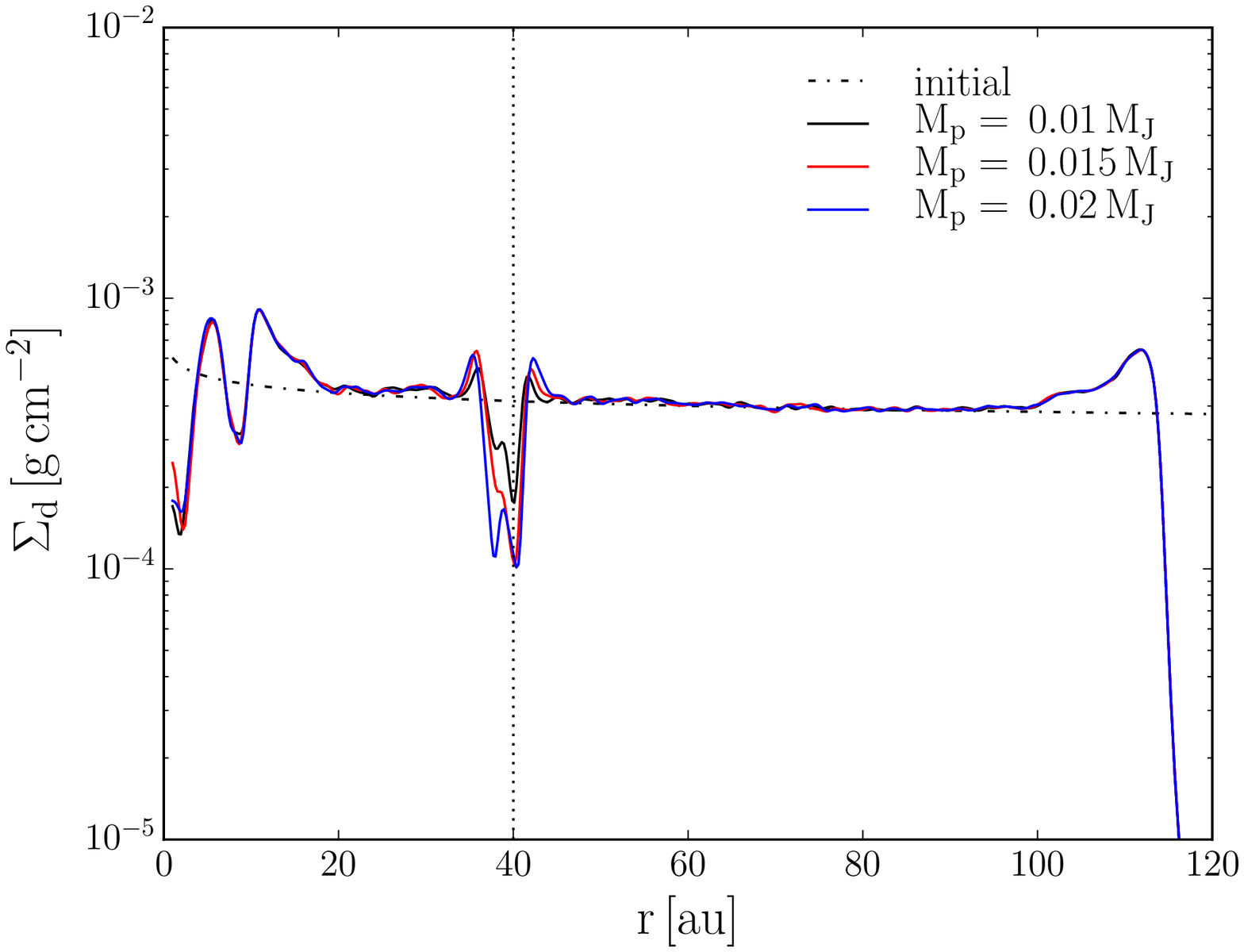}
\includegraphics[width=0.325\textwidth]{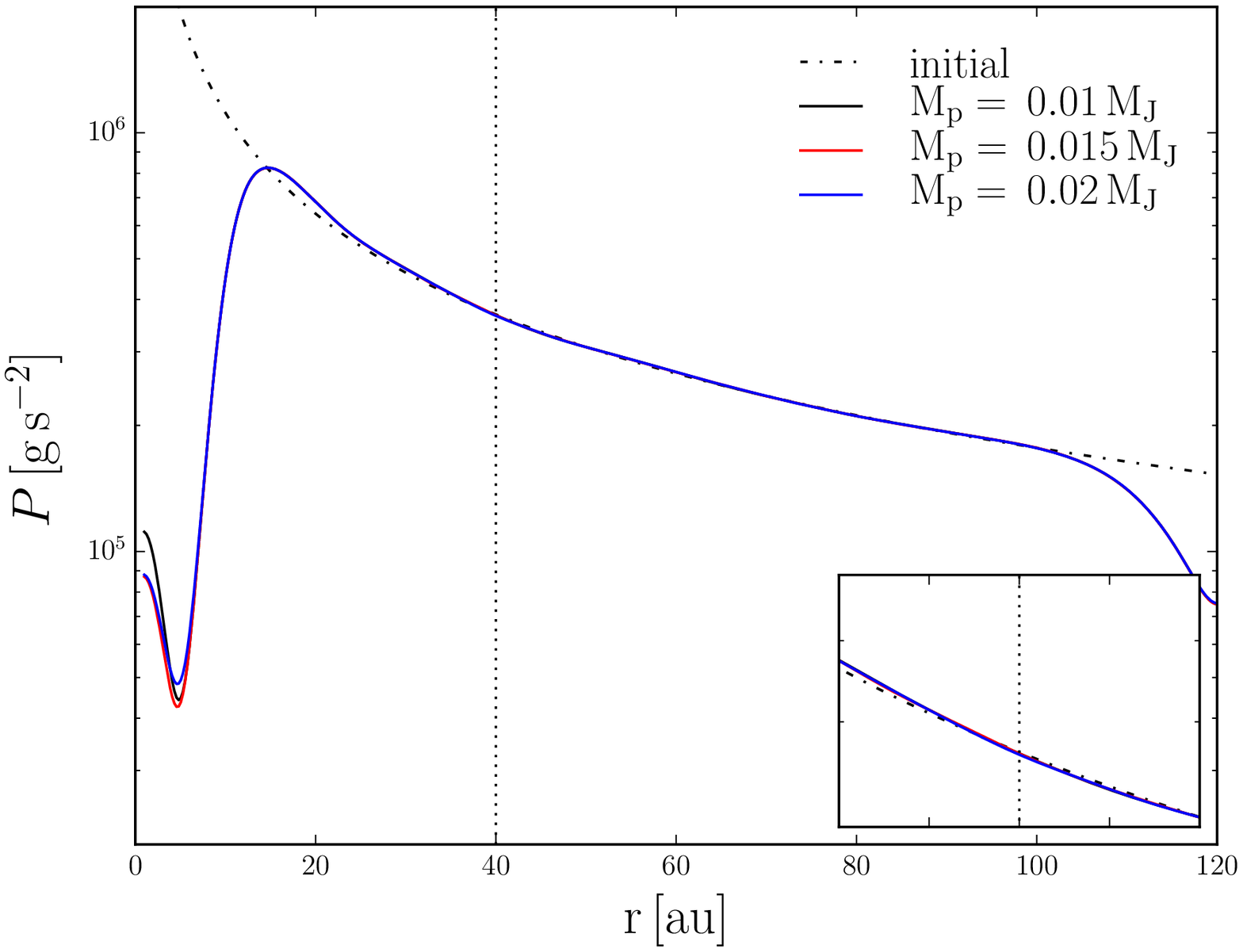}
\includegraphics[width=0.333\textwidth]{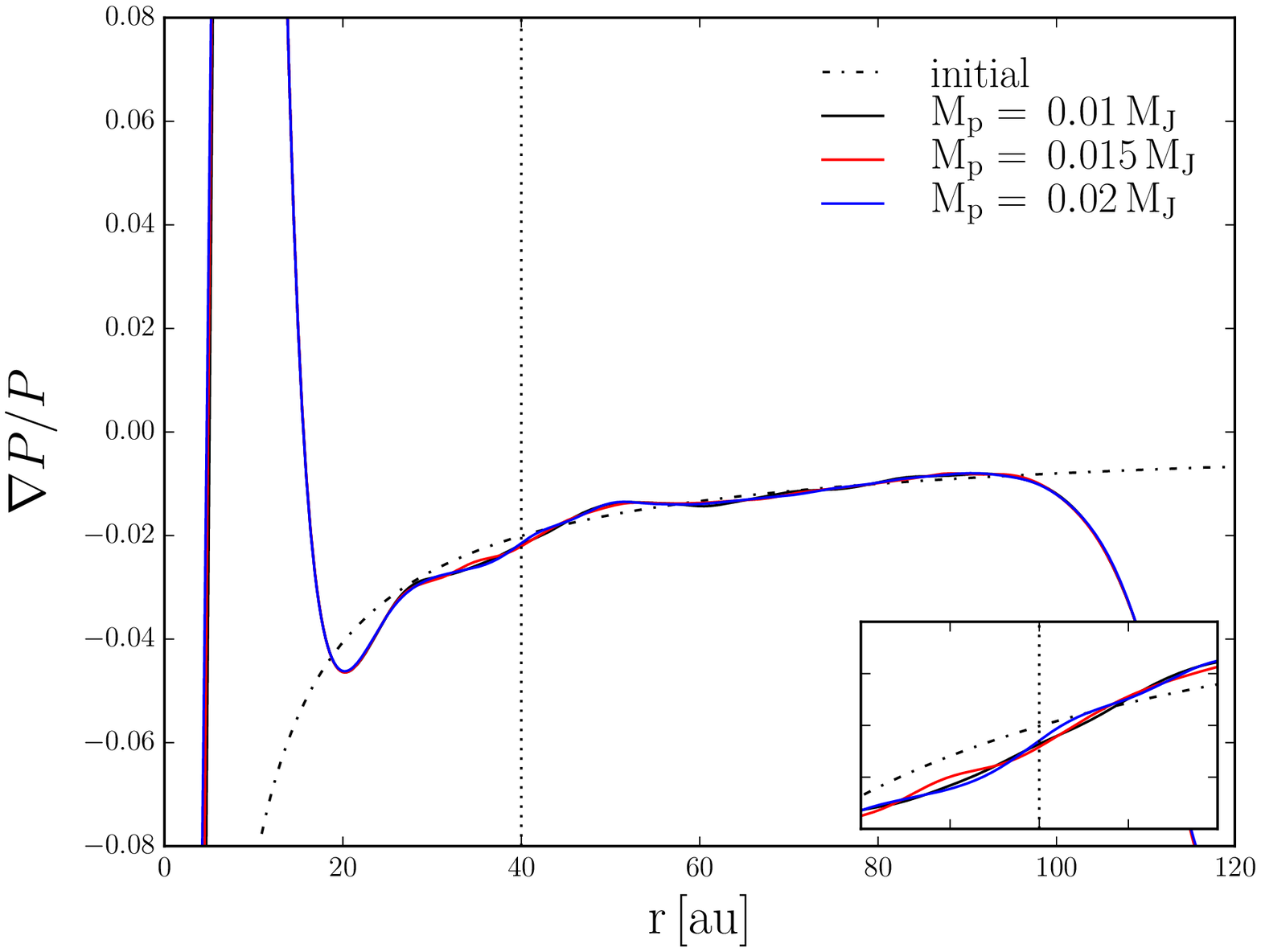}
\caption{Azimuthally averaged (left) dust surface density and (center) pressure profiles for a gap created by planets of various masses embedded in a disc with a disc aspect ratio equal to half of the one adopted in the reference case. The right panel show the radial pressure gradient.
The dotted vertical line indicates the planet orbit. For planet masses in our sample the pressure profile is not perturbed by the presence of the planet, in accordance with  our analysis. }
\label{fig:profileshonr}
\end{center}
\end{figure*}
\begin{figure}
\begin{center}
\includegraphics[width=0.48\textwidth]{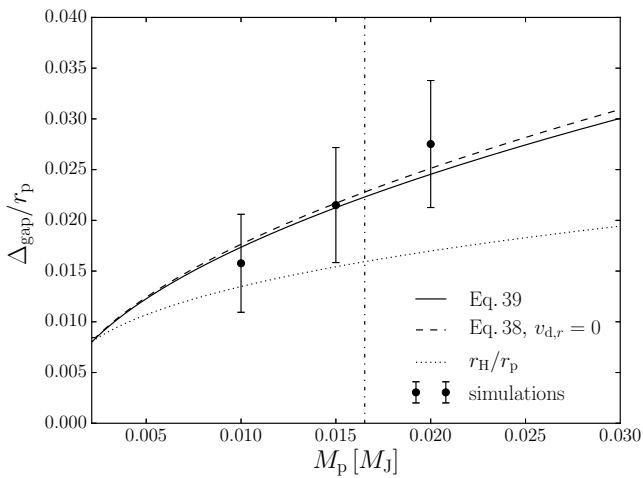}
\caption{Same as Fig.~\ref{fig:changemasshr002} but using a disc model with a disc aspect ratio two times lower than the value previously adopted. Our theoretically predictions on the gap outer edge are consistent with the results of simulations.}
\label{fig:reedgecomhonr}
\end{center}
\end{figure}
\begin{figure*}
\begin{center}
\includegraphics[width=0.33\textwidth]{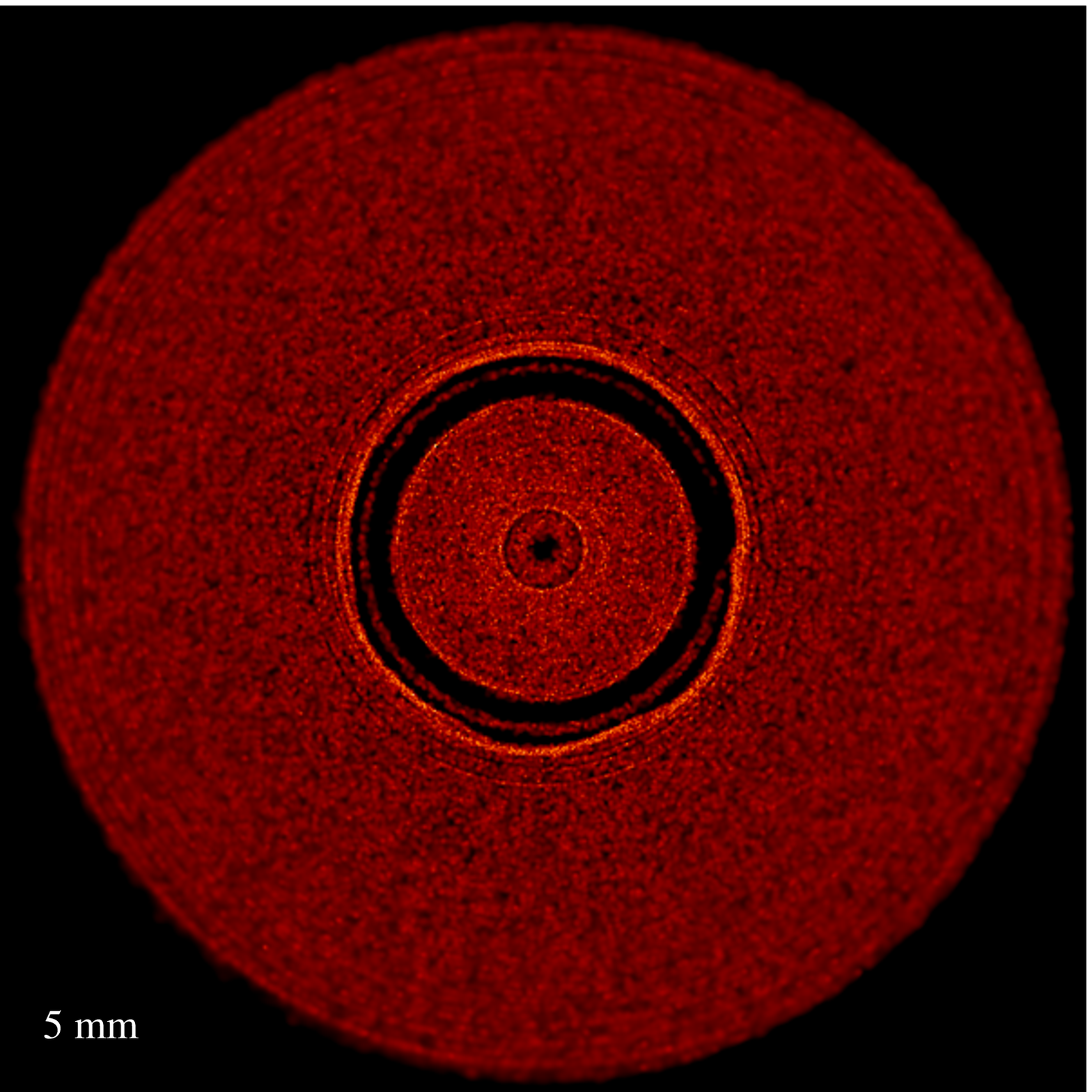}
\includegraphics[width=0.33\textwidth]{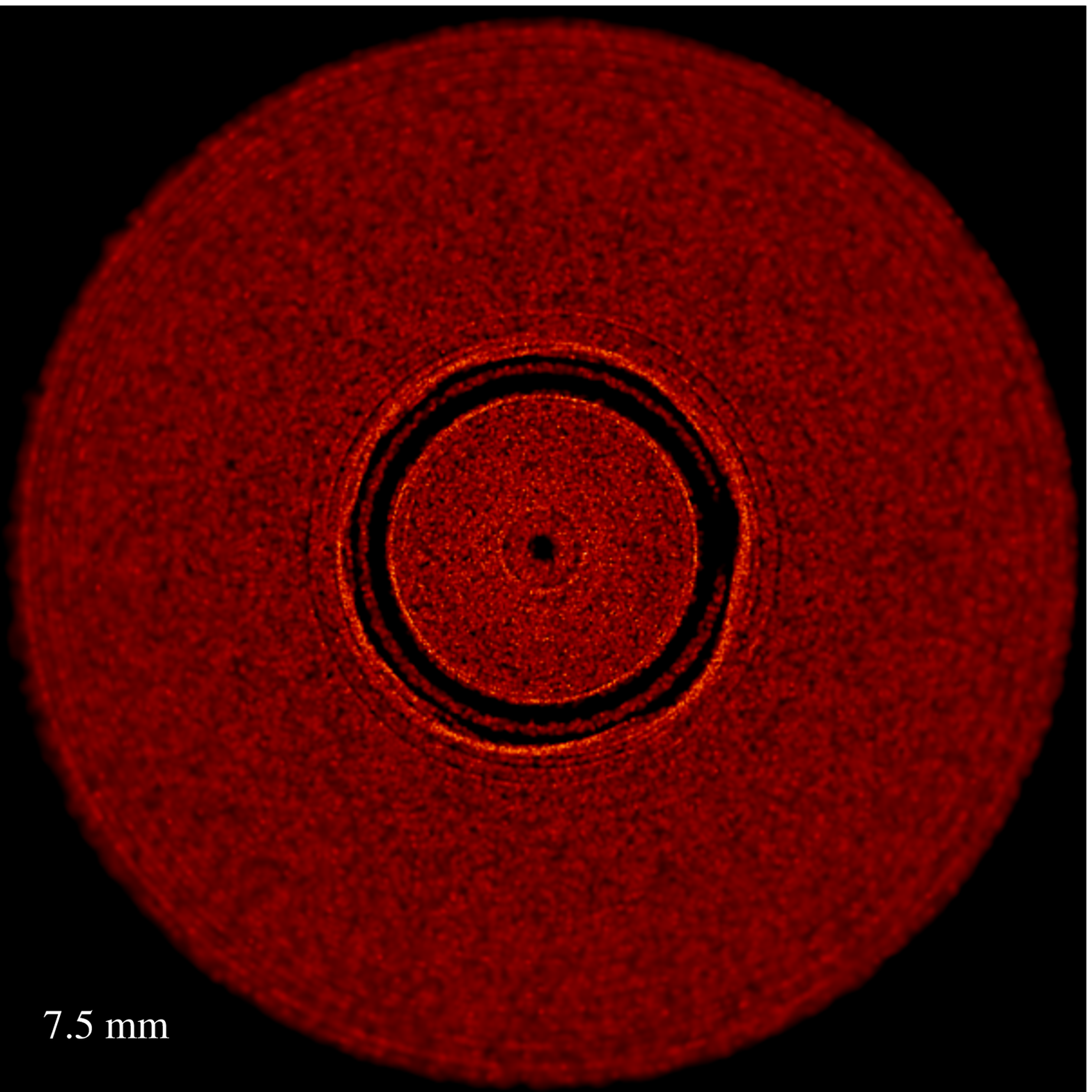}
\includegraphics[width=0.33\textwidth]{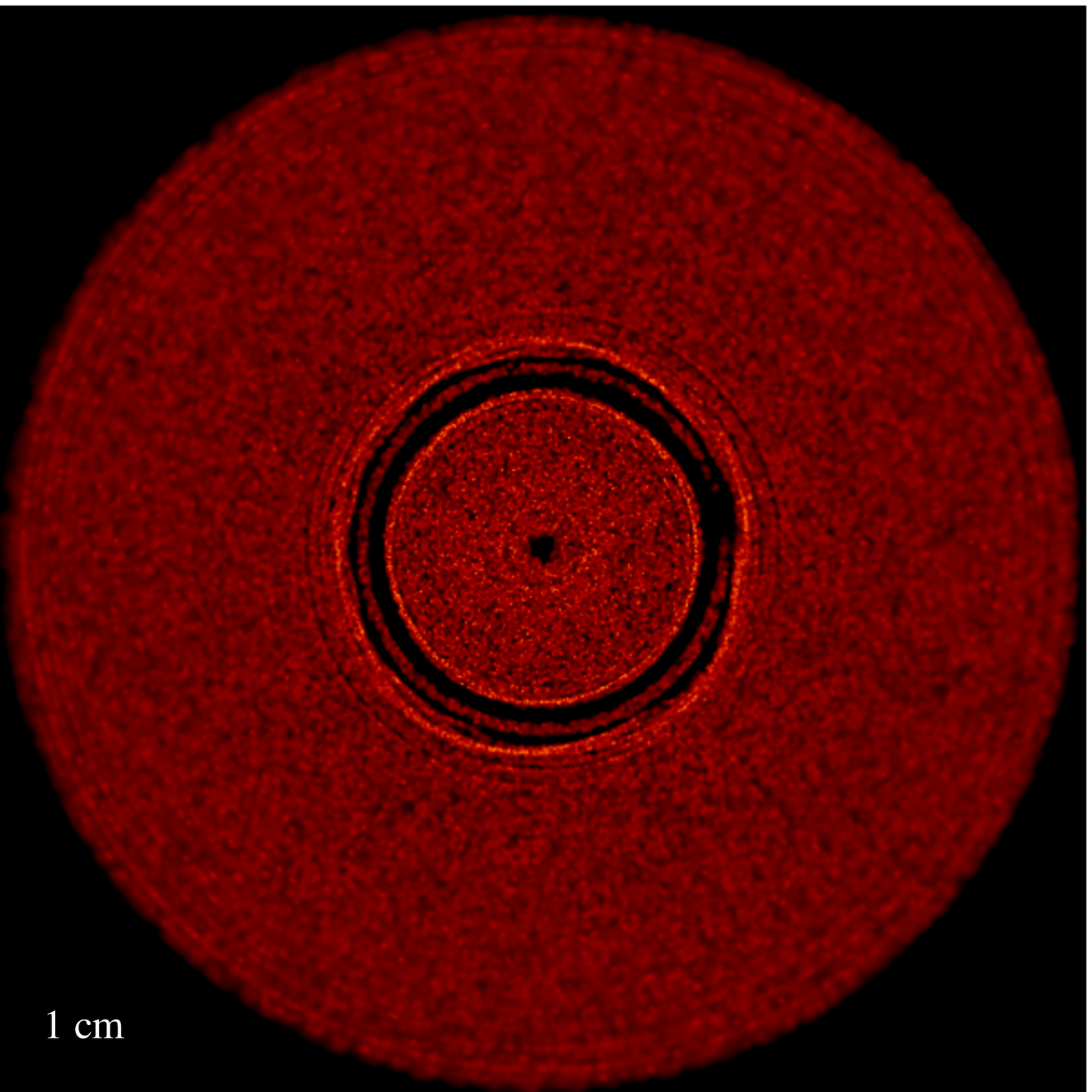}
\caption{Rendered images of dust surface density for the disc model described in Sect.~\ref{subsect:initialcond} hosting a planet with mass $0.1 \, M_{\rm J}$ at 40 au  after 100 planetary orbits adopting different grain sizes: (left) 5 mm, (center) 7.5 mm and (right) 1 cm. }
\label{fig:changestoke}
\end{center}
\end{figure*}
\begin{figure*}
\begin{center}
\includegraphics[width=0.48\textwidth]{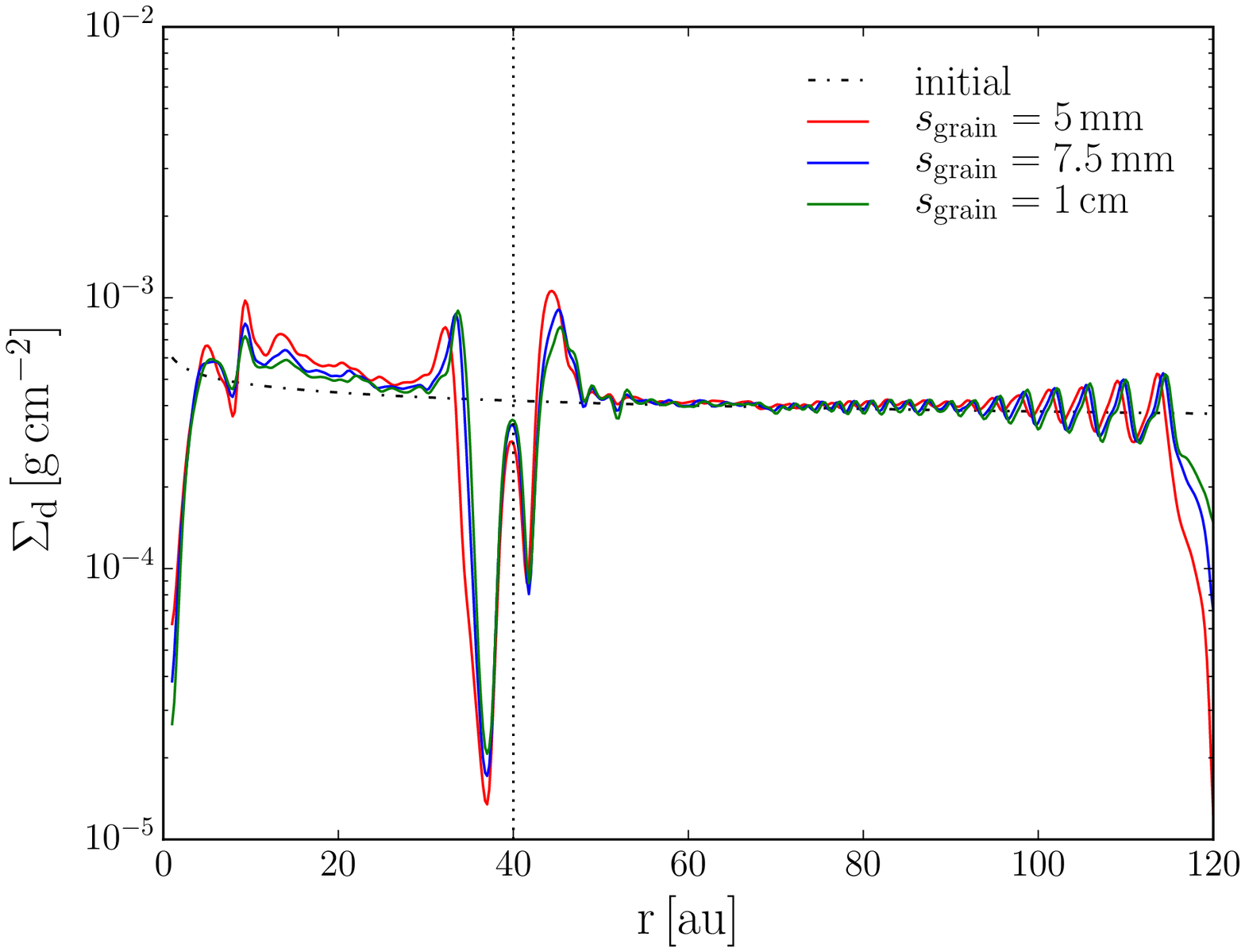}
\includegraphics[width=0.47\textwidth]{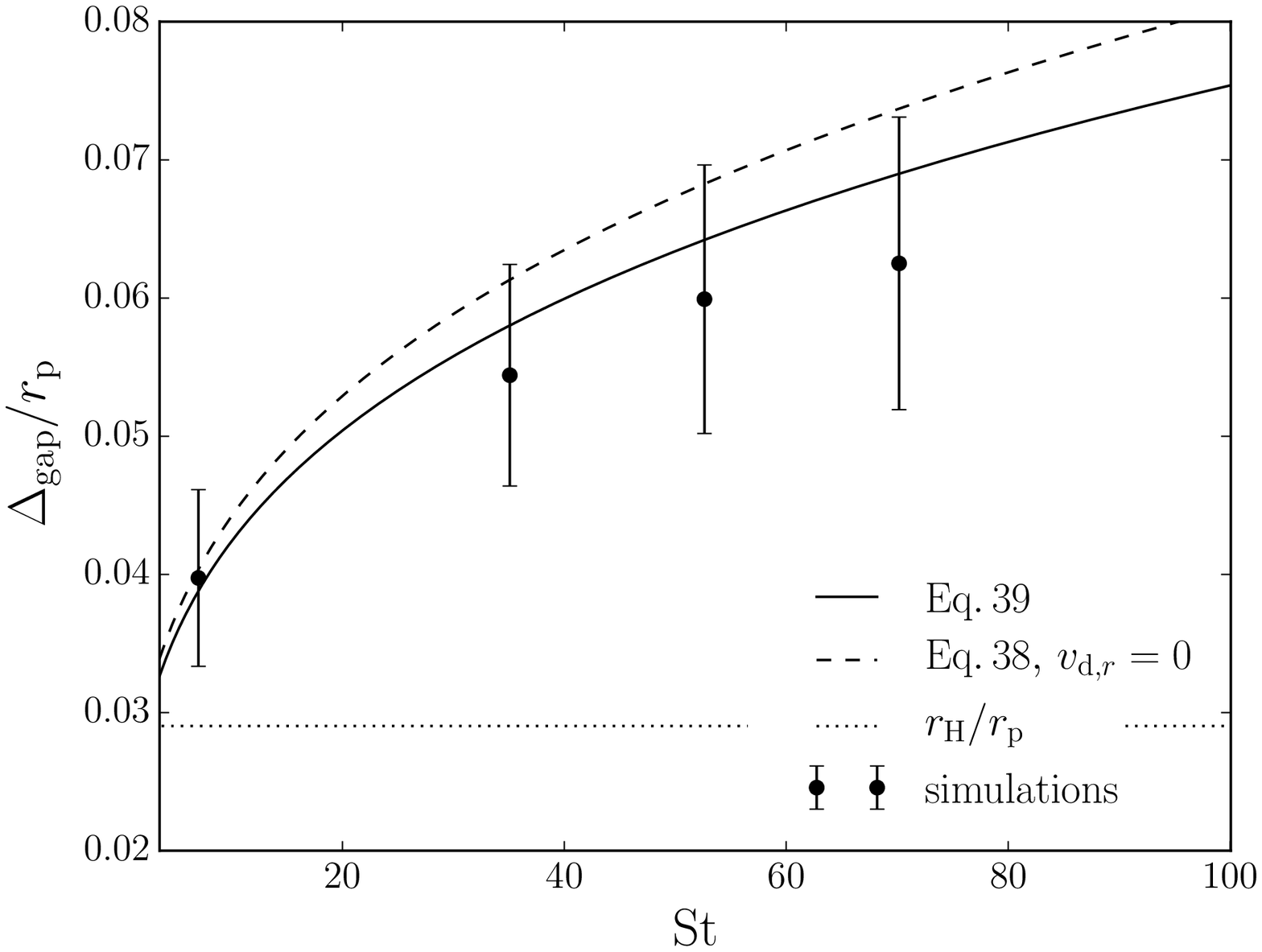}
\caption{(left) Azimuthally averaged dust surface density profile for a gap created by a planet of mass $0.1\, M_{\mathrm{J}}$ using grains with different sizes.
(right) Comparison between the outer gap edge estimated by simulation and theoretical predictions. Our theoretically predictions on the gap outer edge are consistent with the results of simulations.}
\label{fig:profilesstoke}
\end{center}
\end{figure*}

%\subsection{Smoothing length}

\subsection{Initial conditions}
\label{subsect:initialcond}
The disc is setup in \textsc{phantom} by following the procedure outlined in \citet{lodato10a}. The system consists in a central star of mass $1.3\, M_{\odot}$ surrounded by a gaseous disc of $5\times10^{5}$ SPH gas and $3\times10^{5}$ SPH dust particles extending from $r_{\mathrm{in}}$ = 1 to $r_{\mathrm{out}}$ = 120 au.  We model the initial surface density profiles of the discs using power-laws of the form $\Sigma(r) = \Sigma_{\rm in} (r/r_{\rm in})^{-p}$, where $\Sigma_{\rm in}$ is set such as the total gas mass contained between $r_{\rm in}$ and $r_{\rm out}$ is 0.0002 M$_{\odot}$ and a dust-to-gas ratio of 0.01. %We simulate only the inner part of the disc since this is what can be observed with ALMA, e.g. in HL Tau. If the gas phase were to extend to $r_{\mathrm{out}}= 1000 $ au, the total mass of the system is $\simeq 0.01 $ M$_{\odot}$, adopting $p=0.1$.
We adopt a power law exponent of the gas surface density profile $p=0.1$, and the aspect ratio of the disc is assumed to be $\sim0.07$ at the planet location (corresponding to $0.04$ at $r_{\rm in}$). 
We assume a vertically isothermal equation of state $P=c_{\rm s}^{2} \rho_{\mathrm{g}}$ with $c_{\rm s}(r) = c_{{\rm s,in}} (r/r_{\rm in})^{-0.35}$.  The exponent of the power-law profile of the pressure is therefore $\zeta=-1.95$. 
We set an SPH viscosity parameter $\alpha_{\rm AV} = 0.1$ which ensures an effective \citet{shakura73a} viscosity $\alpha_{\rm SS} \approx 0.004$. We study the evolution of the dust density resulting from the tides of one embedded planet located at a distance of 40 au from the central star. We perform a series of simulations varying the planet mass, the aspect ratio of the disc and the size of grains in order to test our criterion over a wide range of disc models.
%In addition, we perform simulations using dust grains with size 1 cm with the aim to show the dependence of the gravitational and drag torque on the particle coupling. %Moreover, we explore the disc-planet tidal interaction by varying the accretion radius of the embedded protoplanet in order to investigate if the choice of the sink properties affect the gap opening process. 

\subsection{Results}
\label{sec:results}
\subsubsection{Planet mass}
\label{sec:planet_mass}

%\citetalias{dipierro16a}, and vary the planet mass around the value predicted by Eq.~\ref{eq:sufficientcondition}. The 
We simulate the evolution of 1 mm sized dust grains over 40 planetary orbits, which leaves enough time for the dust to settle and for the gap to form. The dust grains in our model have an initial Stokes number of $\mathrm{St} \sim 7$ at the disc midplane. Those large grains settle efficiently to the midplane of the disc in a stable dust layer with dust-to-gas scale height ratio of $\sim\sqrt{\alpha_{\rm SS}/\mathrm{St}}\sim 0.02$, consistent with the \cite{dubrulle95a}'s model and SPH simulations of dusty discs \citep{laibe08a}. A dust-to-gas ratio in density $\epsilon$ of $\sim 0.5$ is achieved in the midplane of the disc. %This effect reduces the Stokes number to a value of $\sim 5$ compared to the case where the dust mass would be neglected (Eq.~\ref{eq:ts}), and reinforces the gap-closing (opening) effect induced by the drag torque outside (inside) the orbit of the planet. 
However, as long as dust back-reaction remains weak enough to not affect the local pressure profile, our analysis shows that the minimum mass for dust gap opening and the location of the outer gap edge do not depend on the local dust-to-gas ratio. Therefore, we do not need to know the exact shape of the dust-to-gas ratio profile to determine the outer gap edge.

Our analytic criterion (Eq.~\ref{eq:sufficientcondition}) predicts for the minimum mass able to carve a gap in the dust disc to be $M_{\rm p} \simeq 0.052 \,M_{\rm J}$ for grains with $\mathrm{St}> \mathrm{St_{crit}}=3.7$, assuming the nominal value of $f=0.4$ \citep{goldreich79a}. Above this mass, we expect that planets of mass $\gtrsim M_{\rm p,lim}\simeq0.13 \,M_{\rm J}$ (Eq.~\ref{eq:mplim}) are expected to perturb the local pressure profile, weakening the gap closing effect induced by drag in the outer orbit \citep{rosotti16a}.
Moreover, we expect to see gaps only in the dust for planets up to $ M_{\rm p,gap} \simeq 0.27\,M_{\rm J}$, according to our estimation of minimum mass to create a pressure maxima (Eq.~\ref{eq:mlimgas}). 
We therefore vary the planet mass in the range $0.01 - 0.25 \,M_{\mathrm{J}}$ and look at the eventual structure of dust gap. 

Fig.~\ref{fig:changemass} shows rendered images of the dust surface density of the disc hosting a planet with different masses. A planet of mass $M_{\rm p }\lesssim 0.09 M_{\mathrm{J}}$ causes a local depletion of grains in its close neighbourhood. This local void is permanently replenished by an incoming flux of drifting particles after the planet transit, leading to the formation of a non-axisymmetric depletion of dust.
%Interestingly, the outer edge of the gap carved by massive planets shows excitation of particles eccentricities, leading to the formation of narrow ridges just outside the orbit of the planet \citep{ayliffe12a}.
Fig.~\ref{fig:profiles} shows that planets of mass $0.09 M_{\mathrm{J}}\lesssim M_{\mathrm{p}}\lesssim 0.125 M_{\mathrm{J}}$ are not able to disturb the local pressure profile. For $M_{\rm p}\gtrsim 0.13 M_{\mathrm{J}}$, the pressure gradient is still always negative (bottom-right panel of Fig.~\ref{fig:profiles}), but the weakening of the local pressure profile produces a decrease in the radial drift velocity and therefore a deeper and wider gap compared to low-mass case. The surface density profiles of the dust shows that the width and the depth of the gap change sensitively at the transition between the two regimes (top left panel of Fig.~\ref{fig:profiles}). 
The minimum mass able to reshape the local pressure profile is $\sim 0.13 \,M_{\mathrm{J}}$, in excellent agreement with the value of $M_{\rm p,lim}$ found by \citet{rosotti16a}. However, we obtain deeper dust gaps for $M_{\mathrm{p}}\gtrsim M_{\rm p,lim}$ compared to \citet{rosotti16a}. This discrepancy originates from a different choice for the smoothing length used to soften the tidal torque in the dust. They use a length of order the gas scale height, whereas we use a smaller length of order the dust scale height $H_{\mathrm{d}}$. Hence, we obtain a better estimate for the tidal torque in the dust in the region $\max(r_{\mathrm{H}},H_{\mathrm{d}})<\left | r-r_{\mathrm{p}} \right| <H$. In this region, tides dominate and shape the density distribution of large grains around low-mass planets, for which $r_{\mathrm{H}}<H$.

As discussed in Sect.~\ref{subsec:suffcond}, we use the value of the critical mass obtained by our simulations to determine the value of the constant $f$ in front of Eq.~\ref{eq:sufficientcondapprox}. We obtain
\begin{equation}
f_{\rm sim} \simeq 0.28 \pm 0.01,
\label{eq:simu}
\end{equation}
in agreement with theoretical estimates (see Sect.~\ref{sec:discplanetinteraction}). The low uncertainty of $0.01$ is due to the sensitivity of the critical mass with respect to $f$ (see Eq.~\ref{eq:defxi}). Hereafter, we adopt this \textit{sole} value of $f$ to compare the results of \textit{all} numerical simulation to our theoretical model. Using Eq.~\ref{eq:simu}, the critical Stokes number above which we expect to observe a gap only in the dust is $\mathrm{St_{crit}}=5.3$.

To estimate the location of the gap outer edge from the dust surface density profile, we follow the approach of \citet{dong16f}, which appears to work better for shallow gaps where the density does not drop below an empirical threshold. The gap outer edge $r_{\mathrm{gap}}$ is defined as the location outside the planet orbit where the dust surface density $\Sigma_{\mathrm{d}}(r_{\mathrm{gap}})$ reaches the geometric mean between its minimum value in the gap $\Sigma_{\mathrm{d}}(r_{\mathrm{min}})$ and its unperturbed value at the same location $\Sigma_{\mathrm{d,0}}(r_{\mathrm{min}})$, i.e.
\begin{equation}
\Sigma_{\mathrm{d}}(r_{\mathrm{gap}})\equiv \sqrt{ \Sigma_{\mathrm{d}}(r_{\mathrm{min}})  \times  \Sigma_{\mathrm{d,0}}(r_{\mathrm{min}})} .
\end{equation}
%The spatial resolution in the dust at the outer edge of the gap corresponds to the local dust smoothing length \citep{price12a}. 
Fig.~\ref{fig:reedgecommpl} shows that the position of the gap outer edge estimated from all the simulations is consistent with the value predicted by Eq.~\ref{eq:redgeapprox} for the entire range of masses where our analysis is valid, i.e. $0.09\, M_{\mathrm{J}}\lesssim M_{\rm p} \lesssim M_{\rm p,lim}$. 
For more massive planets, shallower pressure profiles reduce the drag torque outside the planet location, which translates the outer edge of the gap further away from the planet compared to our analytical predictions. In this case, our model underestimates the value found in numerical simulations by $\sim 20\%$. 
We also evaluate the location of the gap outer edge from the zero dust-velocity condition of Eq.~\ref{eq:outedge}, using consistent radial profiles of surface density and temperature across the gap obtained from numerical simulations (dashed line in Fig.~\ref{fig:reedgecommpl}). We note that the analytical approximation given by Eq.~\ref{eq:redgeapprox} works fairly well -- the discrepancy being of order a few per cent -- due to the low values of $\Delta_{\mathrm{gap}}/r_{\mathrm{p}}$. More massive planets create denser dust gaps outer edges. Since resolution in SPH increases with density, errors on the location of the gap decrease with the mass of the planet.

It is worth remarking that the limiting case where the outer gap edge is equal to the Hill radius (Fig.~\ref{fig:reedgecommpl}, dotted line) is obtained for a planet of mass $\sim 0.017 M_{\mathrm{J}}$, where our simulations does not show a gap-like structure. 
%We verify that the outer gap edge can not be closer than the Hill radius for the planet (Fig.~\ref{fig:reedgecommpl}, dotted line). 
Finally, in the low-mass planet regime, no gap forms in the gas. We test this condition by running a simulation where the gravity is switched off in the dust, to discriminate the mechanism at the origin of the dust gap opening. We find that %for the range of planet mass presented here, the planet is not massive enough to affect the pressure structure and 
the gap is indeed carved by the mechanism explained in this paper, as expected (see lower right panel in Fig. 1 of \citealt{dipierro16a}).

\subsubsection{Aspect ratio of the disc}
Eq.~\ref{eq:sufficientcondition} shows that dust gap opening depends on the ratio $H/r_{\mathrm{p}}$ since radial drift is triggered by the radial pressure gradient of the gas. We test the criterion derived above by performing a simulation with a disc satisfying $H/r_{\mathrm{p}} = 0.035$ at the planet location, a value two times lower than the one previously adopted. From Eq.~\ref{eq:sufficientcondition}, the minimum mass of the planet able to carve a dust gap in this disc model is expected to be $0.011\,M_{\mathrm{J}}$. For planets more massive than $M_{\mathrm{p,lim}}=0.017\,M_{\mathrm{J}}$, the local pressure profile is perturbed (Eq.~\ref{eq:mlimgas}). We perform simulations using the higher adequate resolution to ensure the viscosity is the same as for the other models. We verify that in this simulation suit, the local pressure profile remains monotonic (see Fig.~\ref{fig:profileshonr}). Fig. \ref{fig:changemasshr002} shows that a planet of mass $\grtsim 0.012\, M_{\mathrm{J}}$ ($3\, M_{\oplus}$) is able to carve a gap in the dust, a result consistent with our predictions. Fig.~\ref{fig:reedgecomhonr} shows that the location of the outer edge of the gap predicted by our model reproduces well the results of simulations. From Eq.~\ref{eq:sufficientcondapprox}, the minimum planet mass scales as $\left(H/r_{\mathrm{p}}\right)^{-3}$ and is therefore eight times less massive than the one in our model of reference. Moreover, Eq.~\ref{eq:redgeapprox} indicates that the outer edge of the gap carved by the planet with the minimum mass expressed in Eq.~\ref{eq:sufficientcondapprox} scales with $\left(H/r_{\mathrm{p}}\right)^{2}$. Thus, the gap outer edge of the minimum mass planet is four times lower that the one in the reference model. This can be verified by comparing Figs.~\ref{fig:reedgecommpl} and \ref{fig:reedgecomhonr}.

\subsubsection{Grain sizes}
\label{sect:grainsize}
When the size of dust grains increases, the drag torque weakens and gap closing caused by drag becomes less efficient (see the first term in Eq.~\ref{eq:draggd}). Hence, planets of very low masses can open dust gaps as long as locally, the grains are large enough. %The formalism developed in this paper applies to this situation since the planet leave the gap density profile unperturbed. 
Moreover, we expect the outer edge to be further away from the planet for larger Stokes number (Sect.~\ref{subsec:suffcond}). To test these predictions, we perform a series of simulations using the disc model studied in Sect.~\ref{subsect:initialcond}, a planet of mass $0.1\, M_{\mathrm{J}}$, and varying the initial size of the grains. The simulations are evolved over 100 planetary orbits for the grains to relax in a steady state outside of the planet orbit. Fig.~\ref{fig:changestoke} and the left panel of Fig.~\ref{fig:profilesstoke} show that dust gaps of large grains present the asymmetric W-shape evidenced by \citet{ayliffe12a} and \citet{dipierro16a} around the orbit of the planet. In this region, the drag torque is too weak to prevent the formation of a large and stable population of dust grains in the corotation region. % millimetre grains are depleted towards the inner pressure maximum close to the central star.
The left panel of Fig.~\ref{fig:profilesstoke} compares the location of the gap outer edge obtained in SPH simulations and the one derived from our analytic model. The agreement between the theory and numerical results is very good (5-10 \%). The moderate errors between our theoretical estimate and numerical simulations are due to the peculiar shape of the dust gap.
%This can be due to the local approach adopted in our analysis. In detail, our analysis

%\subsection{A more accurate criterion}
%
%By combining our various numerical tests, our final criterion for gap opening in dusty discs is given by
%%
%\begin{equation}
%\frac{M_{\rm p}}{M_{\rm \star}} \grtsim \,\chi(p)  \left(\frac{-\zeta}{1 + \epsilon}\right)^{3/2}  \mathrm{St}^{-3/2}\left(\frac{H}{r_{\rm p}}\right)^3 ,
%\label{eq:sufficientconditionok}
%\end{equation}
%%
%%where 
%%
%%\begin{equation}
%%\chi(p) = -1.8p+3 ,
%%\end{equation}
%%
%a value inferred from a set of simulations with different power-law exponent of the gas surface density profile \textit{in the range} $[0.1,1]$ and different aspect ratio.
%
%%The simulations reveal that the constant is $\sim$ 2.8 for simulation with $p=0.1$ and decrease to a value of $\sim$ 2 for $p=0.5$ and $\sim$ 1.2 for $p=1$ (not shown in the paper). 
%Interestingly, the value of the constant in front of the criterion does not change with the local aspect ratio. %We found difficult to refine the criterion beyond this range of values. 
%Planets with masses larger than the limit given by Eq.~\ref{eq:sufficientconditionok} but lower than the one given by Eq.~\ref{eq:mlimgas} shall carve a deep gap in the dust without affecting the gas structure.

\subsection{Summary}

\begin{figure}
\begin{center}
\includegraphics[width=0.49\textwidth]{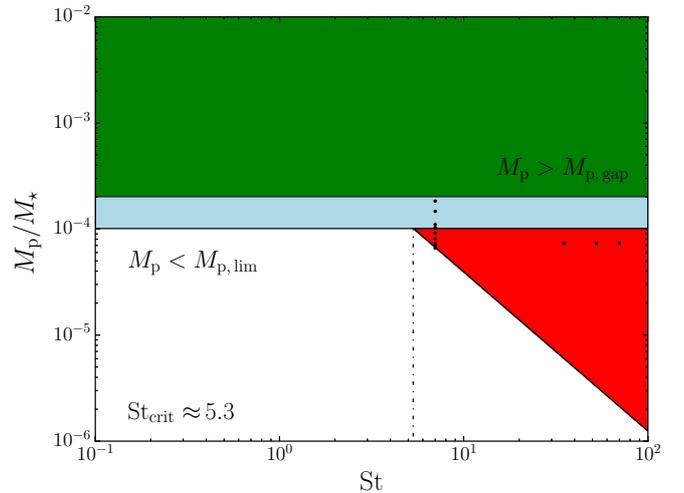}
 \caption{Same as Fig.~\ref{fig:criterion} for $p = 0.1$ and $q = 0.35$, which corresponds to the disc parameters fixed in our numerical simulations, and $f = 0.28$.  Dots corresponds to the different masses and Stokes numbers tested in the simulation suit described in Sect.~\ref{sec:planet_mass} and \ref{sect:grainsize}.}
\label{fig:criterion_num}
\end{center}
\end{figure}

We have considered a disc hosting a low-mass planet which does not disturb the local pressure profile of the gas. 
We obtained two analytic criteria for the minimum mass of the planet required to i) stop the inflow of dust particles (axisymmetric mechanism) and ii) ensure that drift can not refill the inner regions of the disc in dust (non-axisymmetric mechanism). These two criteria represent the necessary and sufficient conditions for dust gap opening, respectively.
The exact value of the minimum masses predicted by these conditions depends on the proportionality constant in front of Eq.~\ref{eq:torquegrav} (planet migration is neglected). By combining our various numerical tests, our final sufficient condition for gap opening in dusty discs is given by
%
%\begin{equation}
%\frac{M_{\rm p}}{M_{\rm \star}} \gtrsim 0.44 \, \left(  \frac{-\zeta}{1 + \epsilon} \right)^{3/2} \mathrm{St}^{-3/2}  \left(\frac{H}{r_{\rm p}}\right)^3 ,
%\label{eq:sum1}
%\end{equation}
%
%
\begin{equation}
\frac{M_{\rm p}}{M_{\rm \star}} \gtrsim 1.38 \, \left(  \frac{-\zeta}{1 + \epsilon} \right)^{3/2} \mathrm{St}^{-3/2}  \left(\frac{H}{r_{\rm p}}\right)^3 ,
\label{eq:sum2}
\end{equation}
for grains with 
\begin{equation}
\mathrm{St} \ge\mathrm{St_{crit}} \simeq   2.76\left(  \frac{-\zeta}{1 + \epsilon} \right)= \mathcal{O}(1)
\label{eq:critstokes}
\end{equation}
We predict the outer edge of the dust gap to be located at a distance $\Delta_{\mathrm{gap}}$ from the planet, where
\begin{equation}
\frac{\Delta_{\mathrm{gap}}}{r_{\rm p}} \simeq 0.87 \left(  \frac{-\zeta}{1 + \epsilon} \right) ^{-1/4} \, \mathrm{St}^{1/4} \left(\frac{H}{r_{\rm p}}\right)^{-1/2}  \left(\frac{M_{\rm p}}{M_{\star}} \right)^{1/2} .
\label{eq:sum3}
\end{equation}
Planets with masses larger than the limit given by Eq.~\ref{eq:sum2} but lower than $\sim M_{\mathrm{p,lim}}$ (Eq.~\ref{eq:mplim}) shall carve a deep gap in the dust without affecting the gas structure. More massive planet in the range $M_{\mathrm{p,lim}} \lesssim M_{\mathrm{p}} \lesssim M_{\mathrm{p,gap}}$ (Eq.~\ref{eq:mlimgas}) are expected to slightly perturb the local pressure profile, leading to the formation of a dust gap due to the combined action of tidal torque and the weakening of the drag. More massive planet, $M_{\mathrm{p}} \gtrsim M_{\mathrm{p,gap}}$, carve a gap both in the gas and dust phase \citep{lambrechts14a,rosotti16a}. 
Fig.~\ref{fig:criterion_num} shows that numerical simulations corroborate the different dust gap opening regimes predicted in Sect.~\ref{sec:criterion} for low-mass planets.

%----------------------------------------------------------------------------------
\section{Using the criterion}
\label{sect:discussion}

\subsection{Interpreting observations of gaps}

\subsubsection{Gap detectability}
Relating the morphology of dust gaps to the properties of the planet and the disc gives insights about the planet formation process. Ideally, multiple-wavelengths observations should be combined to infer density distributions of grains experiencing different aerodynamical regimes. Scattered light emission at optical and near-infrared frequencies trace small dust grains ($\simeq 0.1-10 ~\mu$m) at the surface of the disc, where stellar photons are absorbed or scattered \citep{watson07a}. The scattering emission intensity probes the gas structure at the surface of the disc, since these grains are efficiently coupled with the gas ($\mathrm{St}\aplt\alpha$). Emission at (sub)-millimetre wavelengths probes surface density of large grains in the mid-plane of the disc ($\simeq 0.1-10$ mm), since discs are usually optically thin at these wavelengths in the vertical direction \citep{dullemond07a,williams11a}.  A narrow beam is required to resolve the gap, together with a large signal-to-noise ratio to discriminate its weak emission. The gap depth can then be extrapolated, assuming that the weak emission in the gap is solely due to a low dust surface density.

\subsubsection{Estimating the Stokes number}

Eq.~\ref{eq:sum2} involves the Stokes number of the grains. $\mathrm{St}$ can be estimated directly if the gas surface density is known. Unfortunately, hydrogen density is a quantity which is not directly measurable in a disc. Gas masses are therefore usually estimated by processing the (sub-)millimetre continuum or line measurements of CO isotopologues. This requires to model fractions of isotopologues abundances, dust grains opacities and local gas-to-dust mass ratios \citep{williams11a,williams14a,miotello14b}. 
An estimate of the value of the Stokes number $\mathrm{St}$ can also be inferred indirectly via the ratio $\mathrm{St} / \alpha$ coming from the thickness of the dust layer \citep[e.g.][]{dubrulle95a}, assuming a fixed value for the turbulence parameter $\alpha$. Similarly, assuming a fixed value of the Stokes number, it is possible to infer the level of gas turbulence by analyzing the dust settling.
As an example, \citet{pinte16a} measured a dust scale height of $\sim1$ au at $r \sim100$ au in the disc around HL Tau. Assuming coupled grains (e.g. $\mathrm{St} \simeq 0.01$), this thickness implies $\alpha \simeq10^{-4}$. However, i) this value is $1-2$ orders of magnitude smaller than the value consistent with typical accretion rates of protostars, ii) such low $\mathrm{St}$ implies typical Minimum Mass Solar Nebula gas density, which would make the planet migrate very quickly onto the star, and iii) when planets interact with discs of such low viscosity, vortices develop via the Rossby wave instabilities, trap grains and produce non-axisymmetric structures detectable by ALMA \citep{lyra13a}, not detected in HL Tau. 
Assuming $\mathrm{St} \simeq 1$, the value of the measured thickness implies $\alpha \simeq 10^{-2}$, more consistent with the expected value of accretion rate of protostar.

\subsubsection{Gap in the gas}
The mass of the hypothetical planet and the local properties of the disc can be estimated when observing a gap in optical or NIR scattered light emission, given a degeneracy over the ratio $M_{\rm p} / \alpha$ \citep{fung14a,kanagawa16a,rosotti16a}. If the signal-to-noise ratio is high enough, the depth of the gap \citep{fung14a,kanagawa15b} or its shape \citep{kanagawa15b,kanagawa16a} can also be used to the same purpose. These methods suffer large systematic errors, mostly due to uncertainties on the local disc geometries which heavily affect the surface brightness around the gap \citep{jang-condell12a}. Scattering emission may additionally reveal spiral structures, whose morphologies may be related to the mass of the planet and the aspect ratio of the disc \citep{zhu15a,juhasz15a,dong15a}. 

An additional method to probe gaps in the gas is the detection of line emissions of CO isotopologues, such as $^{12}$CO, $^{13}$CO and C$^{18}$O. Those might be optically thin at the corresponding wavelengths and trace the gas down to the disc midplane \citep{williams14a,miotello14b}. \citet{isella16a} claim evidence of a decrement in
the density of CO isotopologues within the middle and outer continuum gaps in the disc around HD163296. Yet, the decreased emission of the CO molecular lines might also be produced by a reduced density of large grains. In detail, since photodissociation by ultraviolet radiation (UV) is the primary process that regulates the abundance of gas phase CO in the emitting layer of discs, a reduced dust density around the planet location might induce a less efficient absorption of the UV photons. The decreased shielding of the CO molecules by dust lead the UV photons to penetrate into the disc and become optically thick at higher column densities, i.e. closer to the midplane. The higher efficiency of UV photodissociation at the planet orbit with respect to the adjacent regions might therefore produce a decreased emission of CO molecular lines, that can be misinterpreted as a real gas gap.  However, since small grains are much more efficient in absorbing UV radiation, a gap in only large dust grain might not affect remarkably the shielding of CO molecules since larger grains have less opacity in the UV and do not shield CO strongly \citep{visser09a}. Connecting the variation of CO isotopologues emission lines with real gas density variations still remains an open question.

\subsubsection{Applying the criterion}
A good use of the criterion starts with two preliminary remarks. Firstly, Fig.~\ref{fig:criterion} shows that $M_{\mathrm{p}}\grtsim M_{\mathrm{p, gap}}$ is the only condition required for planets to open gaps in both the gas and the dust and this, independently of the Stokes number. Thus, if a gap is detected in NIR scattering \textit{and} thermal-mm emission, no information can be extracted on $\mathrm{St}$ with the criterion derived in this study. Indeed, grains of all sizes tend to drift towards the pressure maximum at the outer edge of the gas gap, as long as $\mathrm{St} > \alpha$. Further analysis of the morphological details of the gap should be conducted to infer the properties of the system. Secondly, the absence of \textit{any} gap in both gas and dust does not necessarily reflect the absence of any gravitational body in the disc. Grains may replenish the orbit of the planet as they drift inwards if $\mathrm{St} \lesssim \mathrm{St_{crit}}$. However, if any additional detection limit is given for the maximum planet mass a disc can embed, Eq.~\ref{eq:critstokes} provides a condition on the minimum Stokes number compatible with the eventual existence of the planet.

We now focus on the non-trivial case, i.e. a gap detected only at millimetre wavelengths. This work shows that any low mass planet can create this structure as long as the Stokes number is large enough. As an example, for $\mathrm{St} \simeq 20$, the necessary mass of the planet required to open the dust gap is $\simeq 100$ lower than the one required to open a gap in the gas due to the factor $\mathrm{St}^{3/2}$ in Eq.~\ref{eq:sum2}. We therefore expect observations of dust only gaps to be more frequent in the outer disc, where grains of a given sizes have Stokes numbers much larger than unity. At least, $\mathrm{St} \gtrsim 1$, which constrains the \textit{maximum} local density of the gas. Eq.~\ref{eq:sum3} gives the expression of the distance between the planet mass and the outer edge of the dust gap. Even if the Stokes number is only roughly approximated, the weak sensitivity brought by the factor $\mathrm{St}^{1/4}$ allows to determine the planet mass relatively precisely (assuming that the aspect ratio of the disc is known). The absence of any gap in the gas additionally implies that $M_{\mathrm{p}} < M_{\mathrm{p, lim}}$. Combined with the roughly known value of $M_{\rm p}$, this condition provides a minimum value for $\alpha$ in the disc. Comparing this value and the degree of dust settling may give a way to infer if the seminal diffusive description of turbulence in disc is relevant or not. Interestingly, if several dust only gaps are detected in the same disc, the degeneracy over the constant in front of the criterion can be broken, helping to determine the masses of the planets more precisely.

\subsection{Limitations}
\label{sec:limits}

\begin{figure}
\begin{center}
\includegraphics[width=0.47\textwidth]{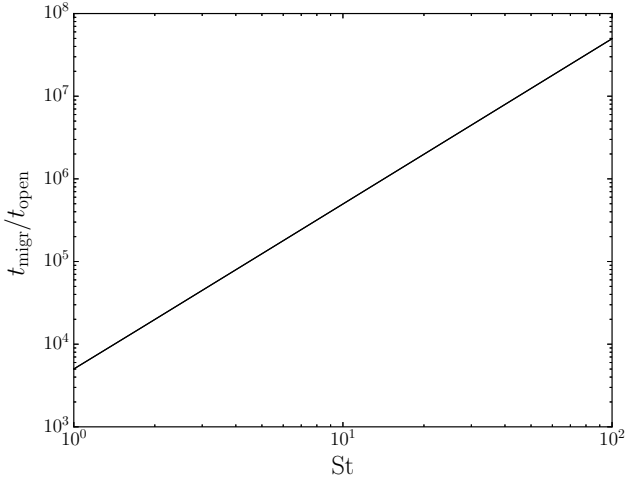}
\caption{Ratio between the type I migration time scale and the gap opening time in dusty discs  by a planet with mass $0.1 M_{\mathrm{J}}$. The gap is carved in the dusty disc faster than the planet migration.}
\label{fig:tmigration}
\end{center}
\end{figure}

In this study, we have restricted our analysis to planets on fixed orbits. Migration may strongly affect the ability of a planet to carve gaps in the gas phase \citep{malik15a} and can appreciably change the density structure around the planet \citep{rafikov02b}. On the contrary, Fig.~\ref{fig:tmigration} shows that the dust gap opening time scale obtained from Eq.~\ref{eq:topengap} is much shorter than the migration time scale, estimated from the differential Lindblad torque derived in \citet{tanaka02a} in our reference disc model with an embedded planet of mass $0.1 M_{\mathrm{J}}$. Hence, planet migration is not expected to affect the ability of a planet to carve gaps in our mechanism. Note that the ratio between both timescales depends indirectly on the Stokes number through the local gas density. 

We have also assumed dust grains of constant sizes. It is known that dust coagulation or fragmentation may strongly affect the dust dynamics \citep[e.g.][]{laibe08a,birnstiel10a}. However, we can safely neglect the grains size evolution over the small time required to open the dust gap. Finally, we have assumed that back-reaction is weak enough and does not affect the gas surface density significantly as grains are repelled outside of the planet orbit.  However, large grains with $\mathrm{St}>1$ are not able to affect significantly the gas structure (see Eq.~\ref{eq:vrad}). Whether back-reaction can trigger the formation of gaps in the gas would be worth investigating, but remains beyond the scope of this study. Hence, the criterion has not been proven to work at dust-to-gas ratio larger than unity.

\begin{figure}
\begin{center}
\includegraphics[width=0.47\textwidth]{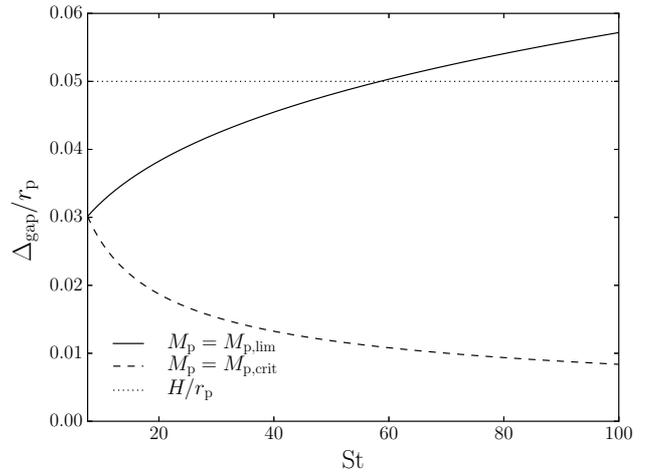}
\caption{Location of the outer edge of the dust gap for different Stokes number as predicted by Eq.~\ref{eq:sum3}, in a typical disc of local aspect ratio 0.05 (dotted line) and $\zeta=-2.75$. The dashed and solid lines delimit the range of planet masses for which our analysis is valid. The two lines intersect at $\mathrm{St}=\mathrm{St_{crit}}\sim 7$ (Eq.~\ref{eq:critstokes}), where $M_{\mathrm{p,crit}}=M_{\mathrm{p,lim}}$ (Sect.~\ref{subsec:examples}).}
\label{fig:redgelimit}
\end{center}
\end{figure}

In Sect.~\ref{sec:discplanetinteraction} we rationalised the use of Eq.~\ref{eq:torquegrav} to model the tidal torque on dust by advocating that gaps open in a region where high-order Lindblad resonances are highly concentrated and degenerate into a continuum, i.e. $r_{\mathrm{H}}\lesssim\Delta_{\mathrm{gap}}\lesssim H$. We test this assumption by comparing the value of $\Delta_{\mathrm{gap}}$ predicted by our theory with the scale height of the disc, for planet masses within the range $\left[M_{\mathrm{p,crit}} , M_{\mathrm{p, lim}} \right]$ and various Stokes number. We first note that
\begin{equation}
\frac{\Delta_{\mathrm{gap}}}{r_{\rm p}} =   
\begin{cases}
\displaystyle   \left(  \frac{-\zeta}{1 + \epsilon} \right) ^{1/2} \, \mathrm{St}^{-1/2} \left(\frac{H}{r_{\rm p}}\right), & \text{$M_{\mathrm{p}}=M_{\mathrm{p,crit}}$,} \\ 
\displaystyle  0.5 \left(  \frac{-\zeta}{1 + \epsilon} \right) ^{-1/4} \, \mathrm{St}^{1/4} \left(\frac{H}{r_{\rm p}}\right), & \text{$M_{\mathrm{p}}=M_{\mathrm{p,lim}}$.}
\end{cases}
\end{equation}
In both cases, the outer edge of the gap increases linearly with the aspect ratio of the disc. Fig.~\ref{fig:redgelimit} shows the range of theoretical locations of the gap outer edge as a function of the Stokes numbers. For $M_{\mathrm{p}}=M_{\mathrm{p, lim}}$, the gap outer edge is smaller than the local gas scale height if
\begin{equation}
\mathrm{St}\lesssim 20 \left(  \frac{-\zeta}{1 + \epsilon} \right).
\end{equation}
For a typical disc with $H / r_{\mathrm{p}}  = 0.05$ and $\zeta=-2.75$, the Stokes number above which the gap is larger than the disc scale height is $\sim 55$. Hence, the location of the outer gap edge is smaller the local gas scale height for a large range of disc model parameters, which supports our initial assumption.

%------------------------------------------------------------------------------
\section{Conclusion}
\label{sec:conclusion}

We derived an analytic criterion that predicts the minimum mass required for a planet to open a gap in the dust phase of a viscous protoplanetary disc in the case where the planet do not perturb the local pressure profile of the disc. In this regime, a gap opens in the dust if the tidal torque overpowers the drag torque outside the planet orbit. We generalised the approach of \citetalias{nakagawa86a} to include the disc-planet tidal interaction and the viscous forces in the equations of motion. Gas and dust velocities in steady state were obtained analytically (Eq.~\ref{eq:vrad}-\ref{eq:vthetad}). From there, assuming that the planet is not able to affect the local pressure structure, we derived a relation between the minimum mass required to open a gap in the dust, and the key parameters of the dust motion: the Stokes number, the aspect ratio of the disc and the dust-to-gas ratio (Eq.~\ref{eq:sufficientcondapprox}). 
We benchmarked the value of the scaling constant in front of the tidal torque density formula using 3D dust-and-gas SPH simulations of various discs.

Our final opening criterion for dust gaps is given by Eq.~\ref{eq:sum2}.  We found that low mass planets are able to carve dust gaps when the Stokes number $\mathrm{St}\geq \mathrm{St_{crit}}\simeq 1$ (Eq.~\ref{eq:critstokes}). We also derived an analytic formula for the radial extension of the outer dust gap edge (Eq.~\ref{eq:sum3}). The criterion and the location of the gap outer edge estimated by our analysis have been tested through 3D SPH simulations of a variety of dusty disc models with an embedded planet. The numerical results appear consistent with our analysis.

%The dependency of the criterion with respect to the different parameters is fully consistent with the analytic model. 
Observations in near-infrared scattering or millimetre thermal dust emission might reveal gaps, in both phases or in the dust only. This criterion can be used to constrain the mass of the planet embedded in the disc, or the Stokes number of the grains. 

\section*{Acknowledgements}
GD thanks Monash for CPU time on NeCTaR. 
We wish to thank Giuseppe Lodato, Daniel J. Price and Richard Alexander for fruitful discussions and the referee for an insightful report of the manuscript. 
GD acknowledge funding via PRIN MIUR 2010-2011 `Chemical and dynamical evolution of the Milky Way and Local Group Galaxies'. This project has received funding from the European Research Council (ERC) under the European Union's Horizon 2020 research and innovation programme (grant agreement No 681601).
GL is funded by European Research Council FP7 advanced grant ECOGAL. We used SPLASH \citep{price07a}.

%, prot. 2010LY5N2T.
\label{lastpage}

\bibliography{biblio}

\begin{thebibliography}{}
\makeatletter
\relax
\def\mn@urlcharsother{\let\do\@makeother \do\$\do\&\do\#\do\^\do\_\do\%\do\~}
\def\mn@doi{\begingroup\mn@urlcharsother \@ifnextchar [ {\mn@doi@}
  {\mn@doi@[]}}
\def\mn@doi@[#1]#2{\def\@tempa{#1}\ifx\@tempa\@empty \href
  {http://dx.doi.org/#2} {doi:#2}\else \href {http://dx.doi.org/#2} {#1}\fi
  \endgroup}
\def\mn@eprint#1#2{\mn@eprint@#1:#2::\@nil}
\def\mn@eprint@arXiv#1{\href {http://arxiv.org/abs/#1} {{\tt arXiv:#1}}}
\def\mn@eprint@dblp#1{\href {http://dblp.uni-trier.de/rec/bibtex/#1.xml}
  {dblp:#1}}
\def\mn@eprint@#1:#2:#3:#4\@nil{\def\@tempa {#1}\def\@tempb {#2}\def\@tempc
  {#3}\ifx \@tempc \@empty \let \@tempc \@tempb \let \@tempb \@tempa \fi \ifx
  \@tempb \@empty \def\@tempb {arXiv}\fi \@ifundefined
  {mn@eprint@\@tempb}{\@tempb:\@tempc}{\expandafter \expandafter \csname
  mn@eprint@\@tempb\endcsname \expandafter{\@tempc}}}

\bibitem[\protect\citeauthoryear{{ALMA Partnership} et~al.,}{{ALMA Partnership}
  et~al.}{2015}]{alma-partnership15a}
{ALMA Partnership} et~al., 2015, \mn@doi [\apjl] {10.1088/2041-8205/808/1/L3},
  \href {http://adsabs.harvard.edu/abs/2015ApJ...808L...3A} {808, L3}

\bibitem[\protect\citeauthoryear{{Adachi}, {Hayashi}  \& {Nakazawa}}{{Adachi}
  et~al.}{1976}]{adachi76a}
{Adachi} I.,  {Hayashi} C.,   {Nakazawa} K.,  1976, \mn@doi [Progress of
  Theoretical Physics] {10.1143/PTP.56.1756}, \href
  {http://adsabs.harvard.edu/abs/1976PThPh..56.1756A} {56, 1756}

\bibitem[\protect\citeauthoryear{{Andrews} \& {Williams}}{{Andrews} \&
  {Williams}}{2007}]{andrews07a}
{Andrews} S.~M.,  {Williams} J.~P.,  2007, \mn@doi [\apj] {10.1086/522885},
  \href {http://adsabs.harvard.edu/abs/2007ApJ...671.1800A} {671, 1800}

\bibitem[\protect\citeauthoryear{{Andrews} et~al.,}{{Andrews}
  et~al.}{2016}]{andrews16a}
{Andrews} S.~M.,  et~al., 2016, \mn@doi [\apjl] {10.3847/2041-8205/820/2/L40},
  \href {http://adsabs.harvard.edu/abs/2016ApJ...820L..40A} {820, L40}

\bibitem[\protect\citeauthoryear{{Armitage} \& {Natarajan}}{{Armitage} \&
  {Natarajan}}{2002}]{armitage02a}
{Armitage} P.~J.,  {Natarajan} P.,  2002, \mn@doi [\apjl] {10.1086/339770},
  \href {http://adsabs.harvard.edu/abs/2002ApJ...567L...9A} {567, L9}

\bibitem[\protect\citeauthoryear{{Artymowicz}}{{Artymowicz}}{1993}]{artymowicz93a}
{Artymowicz} P.,  1993, \mn@doi [\apj] {10.1086/173470}, \href
  {http://adsabs.harvard.edu/abs/1993ApJ...419..166A} {419, 166}

\bibitem[\protect\citeauthoryear{{Ayliffe}, {Laibe}, {Price}  \&
  {Bate}}{{Ayliffe} et~al.}{2012}]{ayliffe12a}
{Ayliffe} B.~A.,  {Laibe} G.,  {Price} D.~J.,   {Bate} M.~R.,  2012, \mn@doi
  [\mnras] {10.1111/j.1365-2966.2012.20967.x}, \href
  {http://adsabs.harvard.edu/abs/2012MNRAS.423.1450A} {423, 1450}

\bibitem[\protect\citeauthoryear{{Bai}}{{Bai}}{2016}]{bai16a}
{Bai} X.-N.,  2016, \mn@doi [\apj] {10.3847/0004-637X/821/2/80}, \href
  {http://adsabs.harvard.edu/abs/2016ApJ...821...80B} {821, 80}

\bibitem[\protect\citeauthoryear{{Balbus} \& {Hawley}}{{Balbus} \&
  {Hawley}}{1991}]{balbus91a}
{Balbus} S.~A.,  {Hawley} J.~F.,  1991, \mn@doi [\apj] {10.1086/170270}, \href
  {http://adsabs.harvard.edu/abs/1991ApJ...376..214B} {376, 214}

\bibitem[\protect\citeauthoryear{{Baruteau} et~al.,}{{Baruteau}
  et~al.}{2014}]{baruteau14a}
{Baruteau} C.,  et~al., 2014, \mn@doi [Protostars and Planets VI]
  {10.2458/azu_uapress_9780816531240-ch029}, \href
  {http://adsabs.harvard.edu/abs/2014prpl.conf..667B} {pp 667--689}

\bibitem[\protect\citeauthoryear{{Bate}, {Bonnell}  \& {Price}}{{Bate}
  et~al.}{1995}]{bate95a}
{Bate} M.~R.,  {Bonnell} I.~A.,   {Price} N.~M.,  1995, \mn@doi [\mnras]
  {10.1093/mnras/277.2.362}, \href
  {http://adsabs.harvard.edu/abs/1995MNRAS.277..362B} {277, 362}

\bibitem[\protect\citeauthoryear{{Bate}, {Lubow}, {Ogilvie}  \&
  {Miller}}{{Bate} et~al.}{2003}]{bate03a}
{Bate} M.~R.,  {Lubow} S.~H.,  {Ogilvie} G.~I.,   {Miller} K.~A.,  2003,
  \mn@doi [\mnras] {10.1046/j.1365-8711.2003.06406.x}, \href
  {http://adsabs.harvard.edu/abs/2003MNRAS.341..213B} {341, 213}

\bibitem[\protect\citeauthoryear{{B{\'e}thune}, {Lesur}  \&
  {Ferreira}}{{B{\'e}thune} et~al.}{2016}]{bethune16a}
{B{\'e}thune} W.,  {Lesur} G.,   {Ferreira} J.,  2016, \mn@doi [\aap]
  {10.1051/0004-6361/201527874}, \href
  {http://adsabs.harvard.edu/abs/2016A%26A...589A..87B} {589, A87}

\bibitem[\protect\citeauthoryear{{Birnstiel}, {Dullemond}  \&
  {Brauer}}{{Birnstiel} et~al.}{2010}]{birnstiel10a}
{Birnstiel} T.,  {Dullemond} C.~P.,   {Brauer} F.,  2010, \mn@doi [\aap]
  {10.1051/0004-6361/200913731}, \href
  {http://adsabs.harvard.edu/abs/2010A%26A...513A..79B} {513, A79}

\bibitem[\protect\citeauthoryear{{Bryden} et~al.}{{Bryden}
  et~al.}{1999}]{bryden99a}
{Bryden} G.,  et~al., 1999, \mn@doi [\apj] {10.1086/306917}, \href
  {http://adsabs.harvard.edu/abs/1999ApJ...514..344B} {514, 344}

\bibitem[\protect\citeauthoryear{{Canovas}, {Caceres}, {Schreiber}, {Hardy},
  {Cieza}, {M{\'e}nard}  \& {Hales}}{{Canovas} et~al.}{2016}]{canovas16a}
{Canovas} H.,  {Caceres} C.,  {Schreiber} M.~R.,  {Hardy} A.,  {Cieza} L.,
  {M{\'e}nard} F.,   {Hales} A.,  2016, \mn@doi [\mnras]
  {10.1093/mnrasl/slw006}, \href
  {http://adsabs.harvard.edu/abs/2016MNRAS.458L..29C} {458, L29}

\bibitem[\protect\citeauthoryear{{Crida}, {Morbidelli}  \& {Masset}}{{Crida}
  et~al.}{2006}]{crida06a}
{Crida} A.,  {Morbidelli} A.,   {Masset} F.,  2006, \mn@doi [\icarus]
  {10.1016/j.icarus.2005.10.007}, \href
  {http://adsabs.harvard.edu/abs/2006Icar..181..587C} {181, 587}

\bibitem[\protect\citeauthoryear{{D'Angelo} \& {Lubow}}{{D'Angelo} \&
  {Lubow}}{2008}]{dangelo08a}
{D'Angelo} G.,  {Lubow} S.~H.,  2008, \mn@doi [\apj] {10.1086/590904}, \href
  {http://adsabs.harvard.edu/abs/2008ApJ...685..560D} {685, 560}

\bibitem[\protect\citeauthoryear{{Dipierro}, {Price}, {Laibe}, {Hirsh},
  {Cerioli}  \& {Lodato}}{{Dipierro} et~al.}{2015}]{dipierro15a}
{Dipierro} G.,  {Price} D.,  {Laibe} G.,  {Hirsh} K.,  {Cerioli} A.,   {Lodato}
  G.,  2015, \mn@doi [\mnras] {10.1093/mnrasl/slv105}, \href
  {http://adsabs.harvard.edu/abs/2015MNRAS.453L..73D} {453, L73}

\bibitem[\protect\citeauthoryear{{Dipierro}, {Laibe}, {Price}  \&
  {Lodato}}{{Dipierro} et~al.}{2016}]{dipierro16a}
{Dipierro} G.,  {Laibe} G.,  {Price} D.~J.,   {Lodato} G.,  2016, \mn@doi
  [\mnras] {10.1093/mnrasl/slw032}, \href
  {http://adsabs.harvard.edu/abs/2016MNRAS.459L...1D} {459, L1}

\bibitem[\protect\citeauthoryear{{Dong} \& {Fung}}{{Dong} \&
  {Fung}}{2017}]{dong16f}
{Dong} R.,  {Fung} J.,  2017, \apj, \href
  {http://adsabs.harvard.edu/abs/2016arXiv161204821D} {835, 146}

\bibitem[\protect\citeauthoryear{{Dong}, {Rafikov}, {Stone}  \&
  {Petrovich}}{{Dong} et~al.}{2011}]{dong11a}
{Dong} R.,  {Rafikov} R.~R.,  {Stone} J.~M.,   {Petrovich} C.,  2011, \mn@doi
  [\apj] {10.1088/0004-637X/741/1/56}, \href
  {http://adsabs.harvard.edu/abs/2011ApJ...741...56D} {741, 56}

\bibitem[\protect\citeauthoryear{{Dong}, {Zhu}, {Rafikov}  \& {Stone}}{{Dong}
  et~al.}{2015}]{dong15a}
{Dong} R.,  {Zhu} Z.,  {Rafikov} R.~R.,   {Stone} J.~M.,  2015, \mn@doi [\apjl]
  {10.1088/2041-8205/809/1/L5}, \href
  {http://adsabs.harvard.edu/abs/2015ApJ...809L...5D} {809, L5}

\bibitem[\protect\citeauthoryear{{Dubrulle}, {Morfill}  \&
  {Sterzik}}{{Dubrulle} et~al.}{1995}]{dubrulle95a}
{Dubrulle} B.,  {Morfill} G.,   {Sterzik} M.,  1995, \mn@doi [\icarus]
  {10.1006/icar.1995.1058}, \href
  {http://adsabs.harvard.edu/abs/1995Icar..114..237D} {114, 237}

\bibitem[\protect\citeauthoryear{{Duffell} \& {Dong}}{{Duffell} \&
  {Dong}}{2015}]{duffell15b}
{Duffell} P.~C.,  {Dong} R.,  2015, \mn@doi [\apj]
  {10.1088/0004-637X/802/1/42}, \href
  {http://adsabs.harvard.edu/abs/2015ApJ...802...42D} {802, 42}

\bibitem[\protect\citeauthoryear{{Duffell} \& {MacFadyen}}{{Duffell} \&
  {MacFadyen}}{2012}]{duffell12a}
{Duffell} P.~C.,  {MacFadyen} A.~I.,  2012, \mn@doi [\apj]
  {10.1088/0004-637X/755/1/7}, \href
  {http://adsabs.harvard.edu/abs/2012ApJ...755....7D} {755, 7}

\bibitem[\protect\citeauthoryear{{Duffell} \& {MacFadyen}}{{Duffell} \&
  {MacFadyen}}{2013}]{duffell13a}
{Duffell} P.~C.,  {MacFadyen} A.~I.,  2013, \mn@doi [\apj]
  {10.1088/0004-637X/769/1/41}, \href
  {http://adsabs.harvard.edu/abs/2013ApJ...769...41D} {769, 41}

\bibitem[\protect\citeauthoryear{{Dullemond}, {Hollenbach}, {Kamp}  \&
  {D'Alessio}}{{Dullemond} et~al.}{2007}]{dullemond07a}
{Dullemond} C.~P.,  {Hollenbach} D.,  {Kamp} I.,   {D'Alessio} P.,  2007,
  Protostars and Planets V, \href
  {http://adsabs.harvard.edu/abs/2007prpl.conf..555D} {pp 555--572}

\bibitem[\protect\citeauthoryear{{Fedele} et~al.,}{{Fedele}
  et~al.}{2017}]{fedele17a}
{Fedele} D.,  et~al., 2017, \aap, \href
  {http://adsabs.harvard.edu/abs/2017arXiv170202844F} {600, A72}

\bibitem[\protect\citeauthoryear{{Flock} et~al.}{{Flock}
  et~al.}{2015}]{flock15a}
{Flock} M.,  et~al., 2015, \mn@doi [\aap] {10.1051/0004-6361/201424693}, \href
  {http://adsabs.harvard.edu/abs/2015A%26A...574A..68F} {574, A68}

\bibitem[\protect\citeauthoryear{{Follette} et~al.,}{{Follette}
  et~al.}{2013}]{follette13a}
{Follette} K.~B.,  et~al., 2013, \mn@doi [\apj] {10.1088/0004-637X/767/1/10},
  \href {http://adsabs.harvard.edu/abs/2013ApJ...767...10F} {767, 10}

\bibitem[\protect\citeauthoryear{{Fouchet}, {Maddison}, {Gonzalez}  \&
  {Murray}}{{Fouchet} et~al.}{2007}]{fouchet07a}
{Fouchet} L.,  {Maddison} S.~T.,  {Gonzalez} J.-F.,   {Murray} J.~R.,  2007,
  \mn@doi [\aap] {10.1051/0004-6361:20077586}, \href
  {http://adsabs.harvard.edu/abs/2007A%26A...474.1037F} {474, 1037}

\bibitem[\protect\citeauthoryear{{Fouchet}, {Gonzalez}  \&
  {Maddison}}{{Fouchet} et~al.}{2010}]{fouchet10a}
{Fouchet} L.,  {Gonzalez} J.-F.,   {Maddison} S.~T.,  2010, \mn@doi [\aap]
  {10.1051/0004-6361/200913778}, \href
  {http://adsabs.harvard.edu/abs/2010A%26A...518A..16F} {518, A16}

\bibitem[\protect\citeauthoryear{{Fung} \& {Chiang}}{{Fung} \&
  {Chiang}}{2017}]{fung17a}
{Fung} J.,  {Chiang} E.,  2017, preprint, \href
  {http://adsabs.harvard.edu/abs/2017arXiv170108161F} {} (\mn@eprint {arXiv}
  {1701.08161})

\bibitem[\protect\citeauthoryear{{Fung}, {Shi}  \& {Chiang}}{{Fung}
  et~al.}{2014}]{fung14a}
{Fung} J.,  {Shi} J.-M.,   {Chiang} E.,  2014, \mn@doi [\apj]
  {10.1088/0004-637X/782/2/88}, \href
  {http://adsabs.harvard.edu/abs/2014ApJ...782...88F} {782, 88}

\bibitem[\protect\citeauthoryear{{Ginski} et~al.,}{{Ginski}
  et~al.}{2016}]{ginski16a}
{Ginski} C.,  et~al., 2016, \aap, \href
  {http://adsabs.harvard.edu/abs/2016arXiv160904027G} {595, A112}

\bibitem[\protect\citeauthoryear{{Goldreich} \& {Tremaine}}{{Goldreich} \&
  {Tremaine}}{1979}]{goldreich79a}
{Goldreich} P.,  {Tremaine} S.,  1979, \mn@doi [\apj] {10.1086/157448}, \href
  {http://adsabs.harvard.edu/abs/1979ApJ...233..857G} {233, 857}

\bibitem[\protect\citeauthoryear{{Goldreich} \& {Tremaine}}{{Goldreich} \&
  {Tremaine}}{1980}]{goldreich80a}
{Goldreich} P.,  {Tremaine} S.,  1980, \mn@doi [\apj] {10.1086/158356}, \href
  {http://adsabs.harvard.edu/abs/1980ApJ...241..425G} {241, 425}

\bibitem[\protect\citeauthoryear{{Gonzalez}, {Pinte}, {Maddison}  \&
  {M{\'e}nard}}{{Gonzalez} et~al.}{2012}]{gonzalez12a}
{Gonzalez} J.-F.,  {Pinte} C.,  {Maddison} S.~T.,   {M{\'e}nard} F.,  2012,
  \mn@doi [\aap] {10.1051/0004-6361/201218806}, \href
  {http://adsabs.harvard.edu/abs/2012A%26A...547A..58G} {547}

\bibitem[\protect\citeauthoryear{{Gonzalez}, {Laibe}, {Maddison}, {Pinte}  \&
  {M{\'e}nard}}{{Gonzalez} et~al.}{2015}]{gonzalez15a}
{Gonzalez} J.-F.,  {Laibe} G.,  {Maddison} S.~T.,  {Pinte} C.,   {M{\'e}nard}
  F.,  2015, \mn@doi [\mnras] {10.1093/mnrasl/slv120}, \href
  {http://adsabs.harvard.edu/abs/2015MNRAS.454L..36G} {454, L36}

\bibitem[\protect\citeauthoryear{{Gonzalez}, {Laibe}  \& {Maddison}}{{Gonzalez}
  et~al.}{2017}]{gonzalez17a}
{Gonzalez} J.-F.,  {Laibe} G.,   {Maddison} S.~T.,  2017, \mnras, \href
  {http://adsabs.harvard.edu/abs/2017arXiv170101115G} {467, 1984}

\bibitem[\protect\citeauthoryear{{Goodman} \& {Rafikov}}{{Goodman} \&
  {Rafikov}}{2001}]{goodman01a}
{Goodman} J.,  {Rafikov} R.~R.,  2001, \mn@doi [\apj] {10.1086/320572}, \href
  {http://adsabs.harvard.edu/abs/2001ApJ...552..793G} {552, 793}

\bibitem[\protect\citeauthoryear{{Henon} \& {Petit}}{{Henon} \&
  {Petit}}{1986}]{henon86a}
{Henon} M.,  {Petit} J.-M.,  1986, \mn@doi [Celestial Mechanics]
  {10.1007/BF01234287}, \href
  {http://adsabs.harvard.edu/abs/1986CeMec..38...67H} {38, 67}

\bibitem[\protect\citeauthoryear{Hill}{Hill}{1878}]{hill78a}
Hill G.~W.,  1878, American Journal of Mathematics, 1, 5

\bibitem[\protect\citeauthoryear{Isella et~al.,}{Isella
  et~al.}{2016}]{isella16a}
Isella A.,  et~al., 2016, \mn@doi [Phys. Rev. Lett.]
  {10.1103/PhysRevLett.117.251101}, 117, 251101

\bibitem[\protect\citeauthoryear{{Jang-Condell} \& {Turner}}{{Jang-Condell} \&
  {Turner}}{2012}]{jang-condell12a}
{Jang-Condell} H.,  {Turner} N.~J.,  2012, \mn@doi [\apj]
  {10.1088/0004-637X/749/2/153}, \href
  {http://adsabs.harvard.edu/abs/2012ApJ...749..153J} {749, 153}

\bibitem[\protect\citeauthoryear{{Jin}, {Li}, {Isella}, {Li}  \& {Ji}}{{Jin}
  et~al.}{2016}]{jin16a}
{Jin} S.,  {Li} S.,  {Isella} A.,  {Li} H.,   {Ji} J.,  2016, \mn@doi [\apj]
  {10.3847/0004-637X/818/1/76}, \href
  {http://adsabs.harvard.edu/abs/2016ApJ...818...76J} {818, 76}

\bibitem[\protect\citeauthoryear{{Juh{\'a}sz}, {Benisty}, {Pohl}, {Dullemond},
  {Dominik}  \& {Paardekooper}}{{Juh{\'a}sz} et~al.}{2015}]{juhasz15a}
{Juh{\'a}sz} A.,  {Benisty} M.,  {Pohl} A.,  {Dullemond} C.~P.,  {Dominik} C.,
   {Paardekooper} S.-J.,  2015, \mn@doi [\mnras] {10.1093/mnras/stv1045}, \href
  {http://adsabs.harvard.edu/abs/2015MNRAS.451.1147J} {451, 1147}

\bibitem[\protect\citeauthoryear{{Kanagawa}, {Muto}, {Tanaka}, {Tanigawa},
  {Takeuchi}, {Tsukagoshi}  \& {Momose}}{{Kanagawa} et~al.}{2015}]{kanagawa15b}
{Kanagawa} K.~D.,  {Muto} T.,  {Tanaka} H.,  {Tanigawa} T.,  {Takeuchi} T.,
  {Tsukagoshi} T.,   {Momose} M.,  2015, \mn@doi [\apjl]
  {10.1088/2041-8205/806/1/L15}, \href
  {http://adsabs.harvard.edu/abs/2015ApJ...806L..15K} {806, L15}

\bibitem[\protect\citeauthoryear{{Kanagawa}, {Muto}, {Tanaka}, {Tanigawa},
  {Takeuchi}, {Tsukagoshi}  \& {Momose}}{{Kanagawa} et~al.}{2016}]{kanagawa16a}
{Kanagawa} K.~D.,  {Muto} T.,  {Tanaka} H.,  {Tanigawa} T.,  {Takeuchi} T.,
  {Tsukagoshi} T.,   {Momose} M.,  2016, \mn@doi [\pasj] {10.1093/pasj/psw037},
  \href {http://adsabs.harvard.edu/abs/2016PASJ...68...43K} {68, 43}

\bibitem[\protect\citeauthoryear{{Kley} \& {Nelson}}{{Kley} \&
  {Nelson}}{2012}]{kley12a}
{Kley} W.,  {Nelson} R.~P.,  2012, \mn@doi [\araa]
  {10.1146/annurev-astro-081811-125523}, \href
  {http://adsabs.harvard.edu/abs/2012ARA%26A..50..211K} {50, 211}

\bibitem[\protect\citeauthoryear{{Kwok}}{{Kwok}}{1975}]{kwok75a}
{Kwok} S.,  1975, \mn@doi [\apj] {10.1086/153637}, \href
  {http://adsabs.harvard.edu/abs/1975ApJ...198..583K} {198, 583}

\bibitem[\protect\citeauthoryear{{Laibe} \& {Price}}{{Laibe} \&
  {Price}}{2011}]{laibe11a}
{Laibe} G.,  {Price} D.~J.,  2011, \mn@doi [\mnras]
  {10.1111/j.1365-2966.2011.19291.x}, \href
  {http://adsabs.harvard.edu/abs/2011MNRAS.418.1491L} {418, 1491}

\bibitem[\protect\citeauthoryear{{Laibe} \& {Price}}{{Laibe} \&
  {Price}}{2012}]{laibe12a}
{Laibe} G.,  {Price} D.,  2012, \mn@doi [\mnras]
  {10.1111/j.1365-2966.2011.20202.x}, \href
  {http://adsabs.harvard.edu/abs/2012MNRAS.420.2345L} {420, 2345}

\bibitem[\protect\citeauthoryear{{Laibe}, {Gonzalez}, {Fouchet}  \&
  {Maddison}}{{Laibe} et~al.}{2008}]{laibe08a}
{Laibe} G.,  {Gonzalez} J.-F.,  {Fouchet} L.,   {Maddison} S.~T.,  2008,
  \mn@doi [\aap] {10.1051/0004-6361:200809522}, \href
  {http://adsabs.harvard.edu/abs/2008A%26A...487..265L} {487, 265}

\bibitem[\protect\citeauthoryear{{Lambrechts}, {Johansen}  \&
  {Morbidelli}}{{Lambrechts} et~al.}{2014}]{lambrechts14a}
{Lambrechts} M.,  {Johansen} A.,   {Morbidelli} A.,  2014, \mn@doi [\aap]
  {10.1051/0004-6361/201423814}, \href
  {http://adsabs.harvard.edu/abs/2014A%26A...572A..35L} {572, A35}

\bibitem[\protect\citeauthoryear{{Laughlin} \& {Lissauer}}{{Laughlin} \&
  {Lissauer}}{2015}]{laughlin15a}
{Laughlin} G.,  {Lissauer} J.~J.,  2015, preprint, \href
  {http://adsabs.harvard.edu/abs/2015arXiv150105685L} {} (\mn@eprint {arXiv}
  {1501.05685})

\bibitem[\protect\citeauthoryear{{Lin} \& {Papaloizou}}{{Lin} \&
  {Papaloizou}}{1979}]{lin79a}
{Lin} D.~N.~C.,  {Papaloizou} J.,  1979, \mn@doi [\mnras]
  {10.1093/mnras/186.4.799}, \href
  {http://adsabs.harvard.edu/abs/1979MNRAS.186..799L} {186, 799}

\bibitem[\protect\citeauthoryear{{Lin} \& {Papaloizou}}{{Lin} \&
  {Papaloizou}}{1986}]{lin86a}
{Lin} D.~N.~C.,  {Papaloizou} J.,  1986, \mn@doi [\apj] {10.1086/164426}, \href
  {http://adsabs.harvard.edu/abs/1986ApJ...307..395L} {307, 395}

\bibitem[\protect\citeauthoryear{{Lin} \& {Papaloizou}}{{Lin} \&
  {Papaloizou}}{1993}]{lin93a}
{Lin} D.~N.~C.,  {Papaloizou} J.~C.~B.,  1993, in {Levy} E.~H.,  {Lunine}
  J.~I.,  eds, Protostars and Planets III. pp 749--835

\bibitem[\protect\citeauthoryear{{Lodato} \& {Price}}{{Lodato} \&
  {Price}}{2010}]{lodato10a}
{Lodato} G.,  {Price} D.~J.,  2010, \mn@doi [\mnras]
  {10.1111/j.1365-2966.2010.16526.x}, \href
  {http://adsabs.harvard.edu/abs/2010MNRAS.405.1212L} {405, 1212}

\bibitem[\protect\citeauthoryear{{Lodato} \& {Rice}}{{Lodato} \&
  {Rice}}{2004}]{lodato04a}
{Lodato} G.,  {Rice} W.~K.~M.,  2004, \mn@doi [\mnras]
  {10.1111/j.1365-2966.2004.07811.x}, \href
  {http://adsabs.harvard.edu/abs/2004MNRAS.351..630L} {351, 630}

\bibitem[\protect\citeauthoryear{{Lodato}, {Nayakshin}, {King}  \&
  {Pringle}}{{Lodato} et~al.}{2009}]{lodato09a}
{Lodato} G.,  {Nayakshin} S.,  {King} A.~R.,   {Pringle} J.~E.,  2009, \mn@doi
  [\mnras] {10.1111/j.1365-2966.2009.15179.x}, \href
  {http://adsabs.harvard.edu/abs/2009MNRAS.398.1392L} {398, 1392}

\bibitem[\protect\citeauthoryear{{Lor{\'e}n-Aguilar} \&
  {Bate}}{{Lor{\'e}n-Aguilar} \& {Bate}}{2016}]{loren-aguilar16a}
{Lor{\'e}n-Aguilar} P.,  {Bate} M.~R.,  2016, \mn@doi [\mnras]
  {10.1093/mnrasl/slv206}, \href
  {http://adsabs.harvard.edu/abs/2016MNRAS.457L..54L} {457, L54}

\bibitem[\protect\citeauthoryear{{Lynden-Bell} \& {Pringle}}{{Lynden-Bell} \&
  {Pringle}}{1974}]{lynden-bell74a}
{Lynden-Bell} D.,  {Pringle} J.~E.,  1974, \mn@doi [\mnras]
  {10.1093/mnras/168.3.603}, \href
  {http://adsabs.harvard.edu/abs/1974MNRAS.168..603L} {168, 603}

\bibitem[\protect\citeauthoryear{{Lyra} \& {Lin}}{{Lyra} \&
  {Lin}}{2013}]{lyra13a}
{Lyra} W.,  {Lin} M.-K.,  2013, \mn@doi [\apj] {10.1088/0004-637X/775/1/17},
  \href {http://adsabs.harvard.edu/abs/2013ApJ...775...17L} {775, 17}

\bibitem[\protect\citeauthoryear{{Malik}, {Meru}, {Mayer}  \& {Meyer}}{{Malik}
  et~al.}{2015}]{malik15a}
{Malik} M.,  {Meru} F.,  {Mayer} L.,   {Meyer} M.,  2015, \mn@doi [\apj]
  {10.1088/0004-637X/802/1/56}, \href
  {http://adsabs.harvard.edu/abs/2015ApJ...802...56M} {802, 56}

\bibitem[\protect\citeauthoryear{{Miotello}, {Bruderer}  \& {van
  Dishoeck}}{{Miotello} et~al.}{2014}]{miotello14b}
{Miotello} A.,  {Bruderer} S.,   {van Dishoeck} E.~F.,  2014, \mn@doi [\aap]
  {10.1051/0004-6361/201424712}, \href
  {http://adsabs.harvard.edu/abs/2014A%26A...572A..96M} {572, A96}

\bibitem[\protect\citeauthoryear{{Nakagawa}, {Sekiya}  \& {Hayashi}}{{Nakagawa}
  et~al.}{1986}]{nakagawa86a}
{Nakagawa} Y.,  {Sekiya} M.,   {Hayashi} C.,  1986, \mn@doi [\icarus]
  {10.1016/0019-1035(86)90121-1}, \href
  {http://adsabs.harvard.edu/abs/1986Icar...67..375N} {67, 375}

\bibitem[\protect\citeauthoryear{{Okuzumi}, {Momose}, {Sirono}, {Kobayashi}  \&
  {Tanaka}}{{Okuzumi} et~al.}{2016}]{okuzumi16a}
{Okuzumi} S.,  {Momose} M.,  {Sirono} S.-i.,  {Kobayashi} H.,   {Tanaka} H.,
  2016, \mn@doi [\apj] {10.3847/0004-637X/821/2/82}, \href
  {http://adsabs.harvard.edu/abs/2016ApJ...821...82O} {821, 82}

\bibitem[\protect\citeauthoryear{{Paardekooper} \& {Mellema}}{{Paardekooper} \&
  {Mellema}}{2004}]{paardekooper04a}
{Paardekooper} S.-J.,  {Mellema} G.,  2004, \mn@doi [\aap]
  {10.1051/0004-6361:200400053}, \href
  {http://adsabs.harvard.edu/abs/2004A%26A...425L...9P} {425, L9}

\bibitem[\protect\citeauthoryear{{Paardekooper} \& {Mellema}}{{Paardekooper} \&
  {Mellema}}{2006}]{paardekooper06a}
{Paardekooper} S.-J.,  {Mellema} G.,  2006, \mn@doi [\aap]
  {10.1051/0004-6361:20054449}, \href
  {http://adsabs.harvard.edu/abs/2006A%26A...453.1129P} {453, 1129}

\bibitem[\protect\citeauthoryear{{Petit} \& {Henon}}{{Petit} \&
  {Henon}}{1987a}]{petit87a}
{Petit} J.-M.,  {Henon} M.,  1987a, \aap, \href
  {http://adsabs.harvard.edu/abs/1987A%26A...173..389P} {173, 389}

\bibitem[\protect\citeauthoryear{{Petit} \& {Henon}}{{Petit} \&
  {Henon}}{1987b}]{petit87b}
{Petit} J.-M.,  {Henon} M.,  1987b, \aap, \href
  {http://adsabs.harvard.edu/abs/1987A%26A...188..198P} {188, 198}

\bibitem[\protect\citeauthoryear{{Picogna} \& {Kley}}{{Picogna} \&
  {Kley}}{2015}]{picogna15a}
{Picogna} G.,  {Kley} W.,  2015, \mn@doi [\aap] {10.1051/0004-6361/201526921},
  \href {http://adsabs.harvard.edu/abs/2015A%26A...584A.110P} {584, A110}

\bibitem[\protect\citeauthoryear{{Pinilla}, {de Juan Ovelar}, {Ataiee},
  {Benisty}, {Birnstiel}, {van Dishoeck}  \& {Min}}{{Pinilla}
  et~al.}{2015}]{pinilla15a}
{Pinilla} P.,  {de Juan Ovelar} M.,  {Ataiee} S.,  {Benisty} M.,  {Birnstiel}
  T.,  {van Dishoeck} E.~F.,   {Min} M.,  2015, \mn@doi [\aap]
  {10.1051/0004-6361/201424679}, \href
  {http://adsabs.harvard.edu/abs/2015A%26A...573A...9P} {573, A9}

\bibitem[\protect\citeauthoryear{{Pinte}, {Dent}, {M{\'e}nard}, {Hales},
  {Hill}, {Cortes}  \& {de Gregorio-Monsalvo}}{{Pinte} et~al.}{2016}]{pinte16a}
{Pinte} C.,  {Dent} W.~R.~F.,  {M{\'e}nard} F.,  {Hales} A.,  {Hill} T.,
  {Cortes} P.,   {de Gregorio-Monsalvo} I.,  2016, \mn@doi [\apj]
  {10.3847/0004-637X/816/1/25}, \href
  {http://adsabs.harvard.edu/abs/2016ApJ...816...25P} {816, 25}

\bibitem[\protect\citeauthoryear{{Price}}{{Price}}{2007}]{price07a}
{Price} D.~J.,  2007, \mn@doi [\pasa] {10.1071/AS07022}, \href
  {http://adsabs.harvard.edu/abs/2007PASA...24..159P} {24, 159}

\bibitem[\protect\citeauthoryear{{Price} \& {Laibe}}{{Price} \&
  {Laibe}}{2015}]{price15a}
{Price} D.~J.,  {Laibe} G.,  2015, \mn@doi [\mnras] {10.1093/mnras/stv996},
  \href {http://adsabs.harvard.edu/abs/2015MNRAS.451..813P} {451, 813}

\bibitem[\protect\citeauthoryear{{Price} et~al.,}{{Price}
  et~al.}{2017}]{price17a}
{Price} D.~J.,  et~al., 2017, preprint, \href
  {http://adsabs.harvard.edu/abs/2017arXiv170203930P} {} (\mn@eprint {arXiv}
  {1702.03930})

\bibitem[\protect\citeauthoryear{{Rafikov}}{{Rafikov}}{2001}]{rafikov01a}
{Rafikov} R.~R.,  2001, \mn@doi [\aj] {10.1086/323451}, \href
  {http://adsabs.harvard.edu/abs/2001AJ....122.2713R} {122, 2713}

\bibitem[\protect\citeauthoryear{{Rafikov}}{{Rafikov}}{2002a}]{rafikov02a}
{Rafikov} R.~R.,  2002a, \mn@doi [\apj] {10.1086/339399}, \href
  {http://adsabs.harvard.edu/abs/2002ApJ...569..997R} {569, 997}

\bibitem[\protect\citeauthoryear{{Rafikov}}{{Rafikov}}{2002b}]{rafikov02b}
{Rafikov} R.~R.,  2002b, \mn@doi [\apj] {10.1086/340228}, \href
  {http://adsabs.harvard.edu/abs/2002ApJ...572..566R} {572, 566}

\bibitem[\protect\citeauthoryear{{Rafikov}}{{Rafikov}}{2015}]{rafikov15a}
{Rafikov} R.~R.,  2015, \mn@doi [\apj] {10.1088/0004-637X/804/1/62}, \href
  {http://adsabs.harvard.edu/abs/2015ApJ...804...62R} {804, 62}

\bibitem[\protect\citeauthoryear{{Rafikov}}{{Rafikov}}{2017}]{rafikov17a}
{Rafikov} R.~R.,  2017, \apj, \href
  {http://adsabs.harvard.edu/abs/2017arXiv170102352R} {837, 163}

\bibitem[\protect\citeauthoryear{{Rafikov} \& {Petrovich}}{{Rafikov} \&
  {Petrovich}}{2012}]{rafikov12a}
{Rafikov} R.~R.,  {Petrovich} C.,  2012, \mn@doi [\apj]
  {10.1088/0004-637X/747/1/24}, \href
  {http://adsabs.harvard.edu/abs/2012ApJ...747...24R} {747, 24}

\bibitem[\protect\citeauthoryear{{Rice}. et~al.}{{Rice}.
  et~al.}{2006}]{rice06a}
{Rice}. W. K.~M.,  et~al., 2006, \mn@doi [\mnras]
  {10.1111/j.1365-2966.2006.11113.x}, \href
  {http://adsabs.harvard.edu/abs/2006MNRAS.373.1619R} {373, 1619}

\bibitem[\protect\citeauthoryear{{Rosotti}, {Juhasz}, {Booth}  \&
  {Clarke}}{{Rosotti} et~al.}{2016}]{rosotti16a}
{Rosotti} G.~P.,  {Juhasz} A.,  {Booth} R.~A.,   {Clarke} C.~J.,  2016, \mn@doi
  [\mnras] {10.1093/mnras/stw691}, \href
  {http://adsabs.harvard.edu/abs/2016MNRAS.459.2790R} {459, 2790}

\bibitem[\protect\citeauthoryear{{Shakura} \& {Sunyaev}}{{Shakura} \&
  {Sunyaev}}{1973}]{shakura73a}
{Shakura} N.~I.,  {Sunyaev} R.~A.,  1973, \aap, \href
  {http://adsabs.harvard.edu/abs/1973A%26A....24..337S} {24, 337}

\bibitem[\protect\citeauthoryear{{Takahashi} \& {Inutsuka}}{{Takahashi} \&
  {Inutsuka}}{2016}]{takahashi16a}
{Takahashi} S.~Z.,  {Inutsuka} S.-i.,  2016, \apj, \href
  {http://adsabs.harvard.edu/abs/2016arXiv160405450T} {152, 184}

\bibitem[\protect\citeauthoryear{{Takeuchi}, {Miyama}  \& {Lin}}{{Takeuchi}
  et~al.}{1996}]{takeuchi96a}
{Takeuchi} T.,  {Miyama} S.~M.,   {Lin} D.~N.~C.,  1996, \mn@doi [\apj]
  {10.1086/177013}, \href {http://adsabs.harvard.edu/abs/1996ApJ...460..832T}
  {460, 832}

\bibitem[\protect\citeauthoryear{{Taki}, {Fujimoto}  \& {Ida}}{{Taki}
  et~al.}{2016}]{taki16a}
{Taki} T.,  {Fujimoto} M.,   {Ida} S.,  2016, \mn@doi [\aap]
  {10.1051/0004-6361/201527732}, \href
  {http://adsabs.harvard.edu/abs/2016A%26A...591A..86T} {591, A86}

\bibitem[\protect\citeauthoryear{{Tanaka}, {Takeuchi}  \& {Ward}}{{Tanaka}
  et~al.}{2002}]{tanaka02a}
{Tanaka} H.,  {Takeuchi} T.,   {Ward} W.~R.,  2002, \mn@doi [\apj]
  {10.1086/324713}, \href {http://adsabs.harvard.edu/abs/2002ApJ...565.1257T}
  {565, 1257}

\bibitem[\protect\citeauthoryear{{Varni{\`e}re}, {Quillen}  \&
  {Frank}}{{Varni{\`e}re} et~al.}{2004}]{varniere04a}
{Varni{\`e}re} P.,  {Quillen} A.~C.,   {Frank} A.,  2004, \mn@doi [\apj]
  {10.1086/422542}, \href {http://adsabs.harvard.edu/abs/2004ApJ...612.1152V}
  {612, 1152}

\bibitem[\protect\citeauthoryear{{Visser}, {van Dishoeck}  \& {Black}}{{Visser}
  et~al.}{2009}]{visser09a}
{Visser} R.,  {van Dishoeck} E.~F.,   {Black} J.~H.,  2009, \mn@doi [\aap]
  {10.1051/0004-6361/200912129}, \href
  {http://adsabs.harvard.edu/abs/2009A%26A...503..323V} {503, 323}

\bibitem[\protect\citeauthoryear{{Watson}, {Stapelfeldt}, {Wood}  \&
  {M{\'e}nard}}{{Watson} et~al.}{2007}]{watson07a}
{Watson} A.~M.,  {Stapelfeldt} K.~R.,  {Wood} K.,   {M{\'e}nard} F.,  2007,
  Protostars and Planets V, \href
  {http://adsabs.harvard.edu/abs/2007prpl.conf..523W} {pp 523--538}

\bibitem[\protect\citeauthoryear{{Williams} \& {Best}}{{Williams} \&
  {Best}}{2014}]{williams14a}
{Williams} J.~P.,  {Best} W.~M.~J.,  2014, \mn@doi [\apj]
  {10.1088/0004-637X/788/1/59}, \href
  {http://adsabs.harvard.edu/abs/2014ApJ...788...59W} {788, 59}

\bibitem[\protect\citeauthoryear{{Williams} \& {Cieza}}{{Williams} \&
  {Cieza}}{2011}]{williams11a}
{Williams} J.~P.,  {Cieza} L.~A.,  2011, \mn@doi [\araa]
  {10.1146/annurev-astro-081710-102548}, \href
  {http://adsabs.harvard.edu/abs/2011ARA%26A..49...67W} {49, 67}

\bibitem[\protect\citeauthoryear{{Youdin} \& {Lithwick}}{{Youdin} \&
  {Lithwick}}{2007}]{youdin07a}
{Youdin} A.~N.,  {Lithwick} Y.,  2007, \mn@doi [\icarus]
  {10.1016/j.icarus.2007.07.012}, \href
  {http://adsabs.harvard.edu/abs/2007Icar..192..588Y} {192, 588}

\bibitem[\protect\citeauthoryear{{Zhang}, {Blake}  \& {Bergin}}{{Zhang}
  et~al.}{2015}]{zhang15a}
{Zhang} K.,  {Blake} G.~A.,   {Bergin} E.~A.,  2015, \mn@doi [\apjl]
  {10.1088/2041-8205/806/1/L7}, \href
  {http://adsabs.harvard.edu/abs/2015ApJ...806L...7Z} {806, L7}

\bibitem[\protect\citeauthoryear{{Zhu}, {Stone}  \& {Rafikov}}{{Zhu}
  et~al.}{2013}]{zhu13a}
{Zhu} Z.,  {Stone} J.~M.,   {Rafikov} R.~R.,  2013, \mn@doi [\apj]
  {10.1088/0004-637X/768/2/143}, \href
  {http://adsabs.harvard.edu/abs/2013ApJ...768..143Z} {768, 143}

\bibitem[\protect\citeauthoryear{{Zhu}, {Stone}, {Rafikov}  \& {Bai}}{{Zhu}
  et~al.}{2014}]{zhu14a}
{Zhu} Z.,  {Stone} J.~M.,  {Rafikov} R.~R.,   {Bai} X.-n.,  2014, \mn@doi
  [\apj] {10.1088/0004-637X/785/2/122}, \href
  {http://adsabs.harvard.edu/abs/2014ApJ...785..122Z} {785, 122}

\bibitem[\protect\citeauthoryear{{Zhu}, {Dong}, {Stone}  \& {Rafikov}}{{Zhu}
  et~al.}{2015}]{zhu15a}
{Zhu} Z.,  {Dong} R.,  {Stone} J.~M.,   {Rafikov} R.~R.,  2015, \mn@doi [\apj]
  {10.1088/0004-637X/813/2/88}, \href
  {http://adsabs.harvard.edu/abs/2015ApJ...813...88Z} {813, 88}

\bibitem[\protect\citeauthoryear{{de Boer} et~al.,}{{de Boer}
  et~al.}{2016}]{de-boer16a}
{de Boer} J.,  et~al., 2016, \aap, \href
  {http://adsabs.harvard.edu/abs/2016arXiv161004038D} {595, A114}

\bibitem[\protect\citeauthoryear{{van Boekel} et~al.,}{{van Boekel}
  et~al.}{2017}]{van-boekel16a}
{van Boekel} R.,  et~al., 2017, \apj, \href
  {http://adsabs.harvard.edu/abs/2016arXiv161008939V} {837, 132}

\bibitem[\protect\citeauthoryear{{van der Plas} et~al.,}{{van der Plas}
  et~al.}{2017}]{van-der-plas16a}
{van der Plas} G.,  et~al., 2017, \aap, \href
  {http://adsabs.harvard.edu/abs/2016arXiv160902488V} {597, A32}

\makeatother
\end{thebibliography}

\end{document}